\documentclass[12pt,preprint]{aastex6}

\usepackage{graphicx}
\usepackage{amssymb}
\usepackage{amsmath}
\usepackage{fixltx2e}
\usepackage{float}
\usepackage{natbib}
\usepackage{color} 

\newcommand{\eTi}{\mbox{$\epsilon^{50}{\rm Ti}$}} 
\newcommand{\eCr}{\mbox{$\epsilon^{54}{\rm Cr}$}} 
\newcommand{\dOx}{\mbox{$\Delta^{17}{\rm O}$}} 
\newcommand{\alratio}{\mbox{${}^{26}{\rm Al}/{}^{27}{\rm Al}$}}

\begin{document} 

\title{The Effect of Jupiter's Formation on the Distribution of Refractory Elements and Inclusions in Meteorites}

\author{Steven J.\ Desch}
\affil{School of Earth and Space Exploration, Arizona State University, PO Box 871404, Tempe, AZ 85287-1404}

\author{Anusha Kalyaan} 
\affil{School of Earth and Space Exploration, Arizona State University, PO Box 871404, Tempe, AZ 85287-1404}

\author{Conel M.\ O'D.\ Alexander}
\affil{Department of Terrestrial Magnetism, Carnegie Institution of Washington, 5241 Broad Branch Rd NW, 
 Washington DC 20015} 

\begin{abstract}

We present a comprehensive evolutionary model of the Sun's protoplanetary disk, constructed to resolve the 
``CAI Storage" problem of meteoritics.  
We predict the abundances of calcium-rich, aluminum-rich inclusions (CAIs) and refractory 
lithophile elements under the central assumption that Jupiter's $\sim 30 \, M_{\oplus}$ core forms at about 3 AU at around 0.6 Myr and opened a gap 
CAIs are trapped in the pressure maximum beyond Jupiter; carbonaceous chondrites formed there. 
Inside Jupiter's orbit, CAIs were depleted by aerodynamic drag; ordinary and enstatite chondrites formed there. For 16 chondrites and achondrites, we review meteoritic data on their CAI and refractory abundances and their times of formation, constrained by radiometric dating and thermal models.
We predict their formation locations, finding excellent consistency with other location information (water content, asteroid spectra and parent bodies). We predict the size of particle concentrated by turbulence for each chondrite, finding excellent matches to each chondrites's mean chondrule diameter.
These consistencies imply meteorite parent bodies assembled quickly from local materials concentrated by turbulence, and usually did not migrate far. We predict CI chondrites are depleted in refractory lithophile elements relative to the Sun, by about 12\% (0.06 dex). We constrain the variation of turbulence parameter $\alpha$ in the disk, and find a limited role for magnetorotational instability, favoring hydrodynamical instabilities in the outer disk, plus magnetic disk winds in the inner disk. Between 3 and 4 Myr at least, gas persisted outside Jupiter but was depleted inside it, and the solar nebula was a transition disk.
\end{abstract} 


%
%
\section{Introduction}

Meteorites are tight-lipped witnesses to the beginning of the solar system and the formation of its planets.
Meteorites offer the opportunity to conduct chemical, isotopic, and other analyses on rock samples that were
present in the protoplanetary disk, but sometimes the sheer quantity and precision of meteoritic data is 
overwhelming and complicates rather than clarifies the issues (Kerr 2001).
The key to unlocking the data from meteorites and unraveling the secrets of our protoplanetary disk is to use
the data to constrain and help build predictive astrophysical models of the disk.

Such models are needed especially to resolve the ``CAI Storage" problem of meteoritics, a long-standing mystery
described by many authors (Wood 1996, 2005; Chambers 2006; Russell et al.\ 2006; Ciesla 2007).
As we review below (\S 2.2.1), CAIs (calcium-rich, aluminum-rich inclusions) are the first solids to form in the solar system, 
and the minerals they contain are rich in refractory elements (Ca, Al, Ti) indicating they formed at high tempertures 
$> 1400$ K obtained only near the Sun.
CAIs are typically hundreds of microns in size, and can even reach diameters $\sim 1$ cm in some meteorite types (CV chondrites).
From a wide variety of evidence (\S 2.4), it is known that the meteorites containing CAIs assembled at least 
2 - 3 Myr after CAIs. 
This is difficult to understand, because objects the size of CAIs are widely accepted to spiral in to the Sun by 
aerodynamic drag forces at rates $> 1$ AU/Myr, and therefore should not have been present in the protoplanetary disk when 
the meteorites formed (Weidenschilling 1977a; Cuzzi et al.\ 2003). 

The physics of this inspiral of particles is simple and has been understood for decades 
(Whipple 1972; Adachi et al.\ 1976; Weidenschilling 1977a). 
Whereas particles balance centrifugal force against the gravitational force of the star and orbit the Sun at the 
Keplerian velocity $v_{\rm K}$, gas is partially supported against gravity by the pressure gradient force, which 
generally acts outward (because usually pressure decreases with distance from the Sun).
Gas therefore orbits at a speed slightly slower (typically $\sim 30 \, {\rm m} \, {\rm s}^{-1}$) than $v_{\rm K}$.
Particles see a constant headwind as they orbit the Sun and feel a drag force opposite to their motion that robs
them of angular momentum at a rate that depends on the ratio of the particles' mass and area.
While $\mu$m-sized particles remain effectively well mixed with the gas, cm-sized particles spiral into the Sun on
timescales $< 10^4$ yr.

It is a mystery how abundant CAIs could remain in the disk to find their way into {\it any} meteorites,
but it is doubly puzzling that CAIs are most common in those meteorites that formed farthest from the Sun.
Whereas CAIs typically comprise 0.5 - 3 vol\% of carbonaceous chondrites that formed far from the Sun (as suggested by
volatile content and spectral match to C-type asteroids), they comprise $< 0.1$ vol\% of enstatite and ordinary chondrites 
that formed much closer to the Sun (as suggested by volatile contents and spectral match to E- and S-type asteroids).
Previous models of CAI distribution often address the first point but not the second.
Cuzzi et al.\ (2003) pointed out that CAIs produced closer to the Sun may diffuse outward through the disk by turbulent diffusion, 
an idea expanded on by Yang \& Ciesla (2012).
Boss (2012) similarly suggested gravitational instabilities may disperse CAIs. 
These ideas help explain how CAIs can be spread through the disk, but not why CAIs are so much more abundant in the 
meteorites formed farthest from the Sun.
The mystery of how CAIs remained in the disk and preferentially accreted into carbonaceous chondrites is the CAI storage problem.

We have constructed a detailed astrophysical model of the protoplanetary disk designed to resolve the CAI storage problem.
Our hypothesis for CAI production and redistribution builds on the ``CAI Factory" model of Cuzzi et al.\ (2003), as follows. 
First, accretional heating led to regions of the disk hot enough ($> 1400$ K) to form CAIs (the ``CAI Factory"), inside about 1 AU,
for the first few $\times 10^5$ yr of the disk. 
CAIs formed there were transported out of this region, both into the Sun (by accretion and aerodynamic drag) and outward through the 
disk, aided by turbulent diffusion and the outward motion of gas at the disk midplane known as meridional flow. 
The loss of CAIs from this region depleted it in refractory elements. 
CAIs remained concentrated inside 1 AU, but some CAIs diffused past 3 AU from the Sun. 
Following the conclusion of Kruijer et al.\ (2017), we hypothesize that Jupiter's core of mass $\sim 30 \, M_{\oplus}$ formed very 
rapidly, at $\sim$ 0.6 Myr into disk evolution. We assume it formed at about 3.0 AU from the Sun.
Jupiter rapidly accreted gas and grew over $\sim 5$ Myr, opening a gap in the disk by $\sim 1$ Myr.
CAIs interior to this gap spiraled in toward the Sun and were depleted from the 2-3 AU region, where enstatite and ordinary 
chondrites formed. 
The gap caused a pressure maximum exterior to Jupiter, and aerodynamic drag concentrated CAIs and other large 
($> 100 \, \mu{\rm m}$) particles in this ``pressure bump," where carbonaceous chondrites formed.
Trapping of large particles in such a pressure maximum has been invoked to explain why Jupiter's core did not grow by 
pebble accretion indefinitely (Lambrechts et al.\ 2014), and why the inner solar system does not contain abundant water
despite being cold enough to condense ice (Morbidelli et al.\ 2016). 
Trapping of CAIs in this same pressure bump, we argue, fundamentally resolves the CAI storage problem.
At sometime between 4 and 5 Myr, Jupiter migrated outward from 3.0 AU to its current location at 5.2 AU, scattering some
asteroids in the process to produce the current distribution of S-type and C-type asteroids in the asteroid belt. 

To model this sequence of events and test this hypothesis, we have constructed a 1-D hydrodynamics code that calculates 
the evolution of disk surface density $\Sigma(r,t)$ and temperature $T(r,t)$, as well as abundances of refractory elements 
and abundances of CAIs, as functions of distance from the Sun, $r$, and time, $t$.
Temperatures including accretional heating are calculated self-consistently. 
We introduce a treatment by which starting material is thermally processed in zones $> 1400$ K in temperature, thereby forming CAIs. 
We also introduce a treatment to approximate the growth of Jupiter and its ability to open a gap. 
The radial flows of the gas are computed, and the dynamical response of particles of various sizes are calculated. 
A key input is the radial variation of turbulent viscosity, parameterized by the standard parameter $\alpha(r)$.
We calculate the concentrations of CAIs and other refractory-bearing materials at different times $t$ and locations $r$ in the nebula.

Assuming meteorite parent bodies are ``snapshots" of the solar nebula at the time and place they formed, we calculate where and 
when 11 chondrite (unmelted meteorites) and 5 achondrite (melted meteorites) types of meteorites formed. 
We review information about chondrite compositions, especially of refractory lithophile elements (Ca, Al, Ti, Hf, Sc, rare earth
elements, etc.), and CAI volume fractions to constrain refractory and CAI abundances in each of these different meteorites.
We review meteoritic data from radiometric dating and thermal modeling of meteorite parent bodies to constrain the time $t$ at 
which they formed.
With these two pieces of information and the outputs of the disk model, we predict where each meteorite parent body formed.

In every case, the predicted range of $r$ closely matches expectations for the meteorites' formation locations, based on 
matches to asteroid spectra, volatile content, or other information. 
In some cases we find great consistency between the model predictions and the locations of suggested asteroid parent bodies
(4 Vesta for the HED meteorites, 6 Hebe for H chondrites, and 8 Flora for LL chondrites).
The model predicts the physical conditions where each meteorite parent body formed and, in the case of achondrites, predicts
the starting abundances of CAIs. 
We also use the model to predict the size of particle that would be concentrated by turbulence in the location where each 
chondrite parent body formed. We find an excellent match to the mean diameters of chondrules in most chondrites. 

Our model is detailed, with many inputs that can be varied. It can be characterized as ``finely tuned," but it can also be
argued that it fits many more data (locations of 16 meteorite types and mean chondrule diameters of 11 chondrites) than there
are input parameters.
The model robustly predicts that CI chondrites are depleted in refractories relative to the Sun by measureable amounts,
and makes a host of other predictions about meteorites that can be used to test our hypothesis.
To the extent the model is verified, it places restrictive constraints on $\alpha(r)$ and therefore the nature of turbulence 
and angular momentum transport in disks, as well as the lifetime and dispersal of the disk, and other disk properties. 
Our modeling and synthesis of the existing data thus provides unique, vital insights into the evolution of our
protoplanetary disk that can be extrapolated to other disks, and the compositions and architectures of 
exoplanetary systems.

The outline of this paper is as follows.
We first review in \S 2 the background of meteorites, especially their refractory and CAI abundances, and the times of 
accretion of their parent bodies. 
In \S 3 we describe the numerical code we have written to calculate the formation and distribution of CAIs and refractory elements
in an evolving protoplanetary disk.
We present the results of our calculations for our favored case in \S 4, and discuss the sensitivity of our results to different
input parameters. 
In \S 5 we analyze the results to describe the time and place of formation of the parent bodies of 11 types of chondrites and 
5 types of achondrites, comparing to meteoritic constraints.
In \S 6 we discuss directions for future research, and summarize our results in \S 7.

%
%
%
\section{Background on Meteorites}

Because the model we present is comprehensive, and because we aim to communicate to a broad audience including astronomers,
we first briefly review the different types of meteorites and their components, including CAIs.
We then compile data from the literature on the abundances of CAIs and refractory elements in different meteorite classes,
and the time of accretion of different meteorite parent bodies. 

%
%

\subsection{Classes of Meteorites}

For a discussion of meteorite classification, we refer the reader to reviews by Scott \& Krot (2005, 2014) and by Weisberg et al.\ (2006). 
Meteorites are classified by whether they derive from asteroidal material that did or did not melt.
Achondrites are melted meteorites, although the term usually refers to meteorites of rocky composition.
Iron meteorites usually derive from the cores of melted asteroids in which metal-silicate differentiation took place.
Pallasites are mixtures of core and mantle materials within asteroids, and rocky achondrites formed from the melted crustal
material.
Primitive achondrites are meteorites from asteroidal material that only partially melted, for example losing
metal that melted but leaving behind only barely melted rock. 
In this paper we concentrate on chondrites, which derive from asteroidal material that never melted. 
Three main classes of chondrite are recognized, comprising almost all of the thousands of chondrites in our
collections: ordinary chondrites (OCs), enstatite chondrites (ECs), and carbonaceous chondrites (CCs).
Rarer, Rumuruti-class chondrites (RCs) also exist. 
Chondrites are categorized into these classes based on bulk rock composition, bulk oxygen isotopic
composition, bulk ${}^{54}{\rm Cr}$ and ${}^{50}{\rm Ti}$ and other isotopic abundances, and petrologic features. 
Syntheses of isotopic data by Trinquier et al.\ (2009), Warren (2011), and Kruijer et al.\ (2017) show 
that OCs, ECs, RCs, and most achondrites have similar isotopic ratios and probably formed in a common 
reservoir in the inner solar system, while CCs probably formed in a distinct reservoir. 

{\it Ordinary, Enstatite, Rumuruti Chondrites}: 
By far the most common type of chondrite, comprising 85\% of all observed meteorite falls, are OCs, which 
number in the thousands. 
OCs are subdivided into H, L and LL, with high, low, and very low abundances of metal.
OCs are predominantly pyroxenes, olivines and metal, contain 0.1 - 1 wt\% water 
(structurally bound in clays), and show evidence for aqueous alteration (Hutchison et al.\ 1987; Alexander et al.\ 1989;
Grossman et al.\ 2000, 2002).
OCs are linked based on spectra and other factors to S-type asteroids (Chapman 2004), found between 2 and 3 AU, 
and which predominate from 2 to 2.5 AU.

There are roughly 200 ECs, subdivided into types EH and EL, with high and low abundances of metal.
ECs contain minerals that are remarkably reduced chemically, including abundant sulfides like oldhamite (CaS).
They formed with possibly elevated levels of sulfur (Lehner et al.\ 2013), and with essentially no water, 
as they show no evidence for aqueous alteration. 
Baedecker \& Wasson (1975) noted that the chemical compositions of ECs demands a depletion of both 
refractory elements and of water. 
ECs are spectroscopically linked with E-type asteroids that are concentrated at 1.9 - 2.1 AU from the Sun 
(Scott \& Krot 2005).

Chemically related to OCs, but distinct from them, are the rarer (56 total) RCs.
These often are highly metamorphosed and brecciated.
Many show evidence of aqueous alteration: they are often oxidized, and one subgroup of RCs contains mica and amphibole, 
minerals formed by reaction of water with silicates at high temperature (McCanta et al.\ 2008). 
In addition, very rare (2 total) Kakangari-class chondrites, plus a handful of other ungrouped chondrites exist.

{\it Carbonaceous Chondrites}:
CCs are subdivided into multiple classes: CB, CI, CH, CK, CM, CO, CR, and CV. 
The well-studied meteorite Allende is a CV3 chondrite (the number following the class denoting the degree of 
thermal or aqueous alteration of the meteorite on its asteroidal parent body).
CV and CK chondrites show many similarities to each other (Wasson et al.\ 2013), suggesting they either formed 
in the same time and place of the solar nebula, or even on the same parent body (Elkins-Tanton et al.\ 2011). 
CB and CH chondrites have unusual properties, and the meteorite Isheyevo combines elements of both types; they 
probably formed following an impact between two asteroids, at least one of them a carbonaceous chondrite parent body 
(Krot et al. 2005).
CI chondrites are remarkable for having elemental abundances that are an excellent match to the abundances of elements 
in the Sun's photosphere (e.g., Lodders et al.\ 2009), and CI compositions justifiably are assumed to be the starting 
composition of the solar system (``solar composition"), or at least close to it. 
This is true even for very volatile species, implying CI chondrites accreted in a relatively cold 
region of the solar nebula.
Likewise, CIs contain the highest abundance of water, $\approx 12-13$ wt\% (Alexander et al.\ 2013), of any chondrites, 
consistent with accreting an initial water-to-rock mass ratio 
nearer the primordial value in the solar system, $\sim 1.2$ 
(Krot et al.\ 2015), again suggesting an origin far from the Sun. 
CCs are associated with C-type asteroids found over a range of distances from 2 to 4 AU, and which 
predominate beyond 2.5 AU (Gradie \& Tedesco 1982; Gaffey 1993; Carry \& DeMeo 2013).


{\it Rocky Achondrites}: 
In addition to chondrites, about 8\% of all meteorites (by number) are achondrites, meteorites from parent 
bodies that did melt.  
Most achondrites are HED (howardite, eucrite or diogenite) meteorites spectroscopically linked to the asteroid
4 Vesta; these record a complex history of Vesta's differentiation, of crystallization within plutons, and eruption 
onto the surface.
Aubrites are another achondrite type. 
A few classes of achondrites are classified as primitive, meaning that they are meteorites containing rock that 
melted but did not see their composition change much, e.g, by movement of the melt. 
These are acapulcoites, lodranites (these first two are strongly associated with each other and may derive
from the same parent body), and winonaites.
Ureilites are sometimes classified as primitive. 
All achondrites have seen removal of some melt, meaning detailed models must be developed to infer the composition
of the parent body as it accreted. 
We consider those achondrites for which the composition of the parent body can be inferred, and for which either the
location or time of formation can be inferred.
We therefore consider only HED meteorites, aubrites, ureilites, and acapulcoites/lodranites and winonaites. 
We were unable to obtain the information we needed about the angrite or brachinite achondrites.


{\it Iron Meteorites}: 
A large variety of iron meteorites with different chemical compositions are known and inferred to 
derive from dozens of different asteroidal parent bodies.
Most are thought to derive from the cores of asteroids, formed when high temperatures caused FeNi metal to melt and sink. 
Iron meteorites of type I AB, and possibly also type III CD, chemically resemble the winonaite primitive achondrites, 
suggesting they are from the core of the winonaite parent body (Weisberg et al.\ 2006; Mittlefehldt 2014 and references therein). 
Pallasites are iron meteorites with silicate inclusions that appear to include material from both 
an iron core and rocky mantle, from at least four different parent bodies.
The subgroup known as Eagle Station pallasites appear be consistent with melting
of CV chondrite material, and they are sometimes inferred to derive from the same parent body 
(Humayun \& Weiss 2011).
Further information about iron meteorites and pallasites is outside the scope of this paper, which is focused 
on the refractory lithophile elements (Ca, Al, and Ti) and on CAIs.
We defer discussion of iron meteorites to a future paper.


%
%
%
\subsection{Meteoritic components}

\subsubsection{Components of chondrites}

{\it Chondrules}:
The defining characteristic of chondrites is that they contain chondrules, millimeter-sized igneous 
inclusions of ferromagnesian silicates that were melted while freely floating in the solar nebula. 
The volume fraction of chondrules varies among chondrites but is typically 15-60\% in CCs, 60-80\% in 
OCs and ECs, and 20-40\% in R and K chondrites (Scott \& Krot 2014). 
For the most part CI chondrites (alone among chondrites) do not contain chondrules, although they do
contain rare chondrule fragments amounting to $<5$ vol\% (Leshin et al.\ 1997). 
The mechanism that melted the chondrules was necessarily very energetic and an important mechanism
in the protoplanetary disk, but its identification remains contentious. 
Among the most successful models for chondrule formation (because of their adherence to meteoritic 
constraints; Desch et al.\ 2012) are passage through large-scale nebular shocks
(Wood 1963; Connolly \& Love 1998; Iida et al.\ 2001; Desch \& Connolly 2002; Ciesla \& Hood 2002; 
Morris \& Desch 2010) and passage through bow shocks around large planetary embryos (Morris et al.\ 2012;
Boley et al.\ 2013; Mann et al.\ 2016). 
Formation of chondrules as impact melt droplets during collisions between asteroids might also be consistent
with meteoritics constraints (Sanders \& Scott 2012; Johnson et al.\ 2015; Hasegawa et al.\ 2016; 
Lichtenberg et al.\ 2017).  
We defer a detailed discussion of chondrule formation mechanisms to a future paper.  For the purposes of
the present paper it suffices that chondrules may form by a variety of mechanisms and be found in the disk from very early times.

A significant fact about chondrules is that they have similar but distinct mean sizes in different chondrites: 
chondrule diameters range from 0.15 mm (in CO chondrites) to 0.9 mm (in CV chondrites).
Chondrules appear to have been collected into chondrites by 
an aerodynamic sorting mechanism based on particle stopping times $t_{\rm stop} = \rho_{\rm s} a / (\rho_{\rm g} C)$,
where $\rho_{\rm s}$ and $a$ are the internal density and radius of particles, and $\rho_{\rm g}$ and $C$ are the 
density and sound speed of the gas (e.g., Cuzzi \& Weidenschilling 2005).
Within a chondrite, the sizes of chondrules and CAIs correlate with each other (CVs have among the largest of all
of these objects, COs the smallest), and the metal grains are typically smaller than chondrules by factors consistent with their
internal densities (Skinner \& Leenhouts 1993; Keubler et al.\ 1999).

Streaming instability is a mechanism that would collect particles in a local region in the disk into an asteroid-sized 
chondrite parent body on short timescales, a few orbits (Johansen et al.\ 2007; Lambrechts \& Johansen 2012; Lambrechts 
et al.\ 2014). 
It would sort particles according to their aerodynamic stopping time, as required; however, the size of particle it sorts
tends to be too large ($\sim 10$ cm) to match chondrules.
Cuzzi et al.\ (2017) and Simon et al.\ (2018) have pointed out that the mechanism of turbulent concentration (Cuzzi et al.\ 2001)
would naturally create such aggregates of chondrule-sized particles, which could then be rapidly assembled into chondrites by 
the streaming instability, wherever that instability is triggered.
According to Cuzzi et al.\ (2001), the particles that are sorted by turbulence have 
\begin{equation}
t_{\rm stop} = \Omega^{-1} \, \frac{\pi}{2} \, \frac{ \rho_{\rm s} a }{ \Sigma }.
 = \Omega^{-1} \, {\rm Re}^{-1/2} = \Omega^{-1} \, \left( \frac{ \alpha C H }{ \nu_{\rm m} } \right)^{-1/2},
\end{equation}
where ${\rm Re} \gg 1$ is the Reynolds number and $\nu_{\rm m} = \eta(T) / \rho_{\rm g}$ is the molecular
viscosity.
Desch (2007) discussed the calculation of molecular viscosity of ${\rm H}_{2}$-He mixtures, deriving 
$\eta(T) \approx 40 \, (T / 100 \, {\rm K}) \, \mu{\rm P}$.
From this we derive the diameter of optimally concentrated particles:
\begin{equation}
d = 0.56 \, \left( \frac{ \rho_{\rm s} }{ 3 \, {\rm g} \, {\rm cm}^{-3} } \right)^{-1} \,
 \left( \frac{ T }{ 100 \, {\rm K} } \right)^{1/4} \,
 \left( \frac{ \Sigma }{ 1000 \, {\rm g} \, {\rm cm}^{-2} } \right)^{1/2} \,
 \left( \frac{ \alpha }{ 1 \times 10^{-4} } \right)^{-1/2} \, {\rm mm}
\end{equation}
If turbulence is responsible for the size distributions of chondrules in chondrites, this offers a way to probe the
surface density of the disk at the time and place each chondrite class formed, something we do in \S 5.3. 

{\it CAIs}: 
All chondrites, with the exception of the CI chondrites, contain CAIs.
To date only one CAI has been found in any CI chondrite (Frank et al.\ 2011).
The mineralogy of CAIs is dominated by minerals like 
corundum (${\rm Al}_{2}{\rm O}_{3}$),
hibonite (${\rm CaAl}_{12}{\rm O}_{19}$), 
grossite (${\rm CaAl}_{4}{\rm O}_{7}$), perovskite (${\rm CaTiO}_{3}$), 
melilite (${\rm Ca}_{2}({\rm Al}_{2},{\rm MgSi}){\rm SiO}_{7}$),
spinel ($({\rm FeMg,Cr,Al,Ti})_{3}{\rm O}_{4}$), and 
anorthite (${\rm CaAl}_{2}{\rm Si}_{2}{\rm O}_{8}$), that are rich in Ca and Al (MacPherson et al.\ 2005).
These minerals are either those predicted to condense first from a cooling solar-composition gas,
at temperatures of about 1400 K at total pressures $\sim 10^{-4}$ bar
(Grossman \& Larimer 1974; Wood \& Hashimoto 1993; Ebel \& Grossman 2000),
or they are the results of melting and alteration of condensates. 
CAIs are enriched in all the refractory lithophiles (Ca, Al, Ti, Sc, Y, Zr, Hf, rare earth elements, etc.), 
those rock-forming elements that condense at temperatures 
above $\approx 1350$ K at which the dominant rock-forming minerals in the solar nebula (olivine, 
pyroxene, FeNi metal) condense (Lodders 2003).
At a temperature $\approx 1400$ K, essentially only the minerals found in CAIs can condense. 

Many CAIs also show textural evidence for condensation from a gas.
The most common textural type is known as Fluffy Type A CAIs, with distinctly non-spheroidal shapes that resemble 
irregular snowflakes.
They are composed of aggregations of small, chemically zoned spheroids (Wark \& Lovering 1977), 
that appear to have coagulated.
Other CAI textures exist, including Compact CAIs of types A, B, or C, depending on their mineral compositions.
These have igneous textures consistent with the melting of pre-existing CAIs (MacPherson 2014).

The sizes of CAIs also support the hypothesis of coagulation-driven growth of condensates. 
CAIs span a range of diameters from micron to centimeter sizes (Kunihiro et al.\ 2005). 
It is generally accepted that the largest, cm-sized CAIs are found only in CV and CK chondrites, with CAI
sizes in all other chondrites being smaller (e.g., MacPherson et al.\ 2005). 
May et al.\ (1999) reported mean radii of CAIs of $100 \, \mu{\rm m}$ in CO chondrites, but $160 \, \mu{\rm m}$ 
in CV chondrites.
Very few data exist on the sizes of CAIs in any other chondrites other than CV and CK chondrites, and even
in these chondrite types, only a few studies exist of the size distribution of CAIs 
(Chaumard et al.\ 2014; Fisher et al.\ 2014; Charnoz et al.\ 2015).
Charnoz et al.\ (2015) measured the sizes of CAIs in thin sections of four CV and CK chondrites, and 
(after making necessary corrections for the distribution of objects in thin section vs.\ whole rock), found 
that CAIs formed a power-law cumulative size distribution, with the total number of CAIs with radii greater
than $r$ scaling as $N(> r) \propto r^{-s+1}$, with $1.54 < s < 1.83$ in the size range 0.1 - 1 mm.
This implies a differential size distribution $dn / dr \propto r^{-s}$, and while the vast majority of CAIs
are at the small end of the size distribution ($R_{\rm min} \sim$ tens of microns), the mass is overwhelmingly 
carried by particles at the large end, in mm-sized particles. 
In the dataset of Charnoz et al. (2015), maximum CAI radii ranged from 1.3 mm (Allende), to 1.7 mm (NWA 779),
to 4 mm (NWA 2900) to 10 mm (TNZ 057), implying a typical maximum radius $\sim 3$ mm.

Since our goal is to model the distribution of refractory elements, we are concerned first and foremost
with the aerodynamic properties of the CAIs, which scale with their radius.
For the sake of simplicity, we model the distribution of CAIs with a single size.
For the distribution above, the mass-weighted mean radius is $\approx (4-s)/(5-s) \, R_{\rm max} \approx 0.7 \, R_{\rm max}$,
and we choose a single radius of CAIs of 2.5 mm.
This provides a decent approximation to CAIs in CK/CV chondrites, but the sizes of CAIs in other chondrites 
(e.g., CO and EC and OC chondrites) are typically much smaller (tens of $\mu{\rm m}$ or less).
Also, the size distribution we assume yields essentially zero CAIs in ECs and OC, whereas they are observed to have 
small abundances ($\sim 0.1$ wt\%) of small CAIs.
Two effects that we do not model could match this observation.
If a few percent of all CAIs were simply formed small, with radii $< 100 \, \mu{\rm m}$, they would readily be distributed 
by diffusion throughout the disk, providing a baseline abundance of small CAIs.
We do not consider this likely, though, as only a very small fraction $< 0.01$ wt\% of the CAI mass is initially in such
small particles; but measurements of the size distribution do not extend to sizes smaller than $100 \, \mu{\rm m}$, so the total 
fraction of mass in this size range is unconstrained, and small CAIs may simply be produced at a $\sim 0.1$ wt\% level.
A second possibility is that CAIs are born large, are transported out of the CAI Factory, and then shattered into smaller
particles in the different chondrite-forming regions.
Charnoz et al.\ (2015) modeled the growth of CAIs and showed that they are consistent with coagulation /
fragmentation starting with submicron monomers (the size of particle expected to condense from the gas), 
but with a sharp cessation in growth above about 1 cm.
Charnoz et al.\ (2015) interpreted this size cutoff as being due to shattering of larger particles in 
the CAI-forming region. 
If large CAIs were transported out of the CAI Factory and then shattered before they were melted and transformed to igneous CAIs, 
they would break apart into their smaller aggregates of their constituent submicron monomers.
Again, we do not model such processes, choosing to fix a single radius for all CAIs.

In summary, a single CAI size will not capture the rich diversity of the CAI distribution, in particular the 
small but non-zero abundance of CAIs in the inner disk. 
This is easily explained away, though. The greater challenge is to simultaneously explain the near-zero abundances 
in the inner disk where ECs and OCs formed, and the $> 3$ wt\% abundances in the outer disk where CCs formed.
Equally important is explaining the existence of large (almost cm-sized) CAIs in CV chondrites.
Explaining these CAIs and explaining the distribution of mass is why we choose a single CAI radius of 2.5 mm.

{\it AOAs:}
Chondrites also contain varying portions of another type of refractory inclusion, amoeboid olivine aggregates,
or AOAs (Grossman \& Steel 1976; Krot et al.\ 2004).
AOAs appear to have condensed directly from the gas, in the same environment as CAIs (Krot et al.\ 2004), and 
have been radiometrically dated to have formed at the same time as most CAIs (Itoh \& Yurimoto 2003).
However, AOAs are dominated by olivine, and are not as refractory-rich as CAIs. 
They exhibit a wide range of sizes, $1 - 250 \, \mu{\rm m}$ (Scott \& Krot 2005).
We include AOAs in our model and arbitrarily assume a uniform radius of $600 \, \mu{\rm m}$.
%
%

\subsubsection{Formation times of chondritic components}

An integral feature of CAIs is the evidence that they contained the short-lived radionuclide 
${}^{26}{\rm Al}$ ($t_{1/2} = 0.71$ Myr) when they formed. 
Most, but not all, measured CAIs are consistent with a uniform 
``canonical" level $\alratio \sim 5 \times 10^{-5}$ (MacPherson et al.\ 1995), 
a value now refined to $\alratio \approx 5.23 \times 10^{-5}$ (Jacobsen et al.\ 2008). 
It appears no meteoritic objects had initial $\alratio$ ratio greater than the canonical value,
and since CAIs and AOAs are made of the minerals predicted to condense first in a cooling solar composition gas,
CAIs and AOAs are widely accepted to be the first objects formed in the solar nebula.
It should be noted that the initial ratio $\alratio \approx 5 \times 10^{-5}$ is most precisely measured in
the large type B CAIs found in CV chondrites; the spread of initial $\alratio$ ratios in small CAIs,
e.g., from ordinary chondrites, is not as precisely known.

The uniformity of the initial ${}^{26}{\rm Al}/{}^{27}{\rm Al}$ ratios in those CAIs with canonical 
${}^{26}{\rm Al}$ abundance implies a very short formation duration. 
Estimates of this duration range from the very short, $\sim 2 \times 10^4$ years (Thrane et al.\ 2006; 
Mishra \& Chaussidon 2014), to $\sim 0.3$ Myr (MacPherson et al.\ 2012; Kita et al.\ 2013; MacPherson et al.\ 2017; 
Ushikubo et al.\ 2017).
The time of formation of CAIs with canonical ${}^{26}{\rm Al}$ abundance is taken as 
$t = 0$ for the solar system, and in the simulations we present here.
While most CAIs show evidence for initial ${}^{26}{\rm Al} / {}^{27}{\rm Al} \approx 5 \times 10^{-5}$, not all do.
Many CAIs show reduced initial $\alratio$ ratio, consistent with either late formation, or resetting of their Al-Mg 
isotopic clock by heating, at 2 - 3 Myr after $t = 0$ (MacPherson et al.\ 2003) 
A large fraction ($\sim 1/4$) have $\alratio \approx 0$ (MacPherson et al.\ 1995; Sajipal \& Goswami 1998). 
We concur with the assessment of Wood (2005), that the data speak to an initial pulse of CAI formation,
followed by continued CAI creation at a reduced rate, potentially for several Myr. 

Based on their ${}^{26}{\rm Al}$ content, the majority of chondrules {\it in chondrites}
(not necessarily all chondrules) appear to have formed roughly 
1.5 - 3.0 Myr after CAIs (Mostefaoui et al.\ 2002; Rudraswami \& Goswami 2007; Rudraswami et al.\ 2008; Kurahashi et al.\ 2008; 
Hutcheon et al.\ 2009; Villeneuve et al.\ 2009; Nagashima et al.\ 2015; Schrader et al.\ 2017).
This is corroborated by Pb-Pb dating (Amelin et al.\ 2002; Connelly et al.\ 2017), although it is currently debated whether 
some fraction of chondrules formed contemporaneously with CAIs, i.e. at time $t = 0$ (Connelly et al.\ 2012).
Because meteorite parent bodies formed in the first 2 Myr after CAIs would be expected to contain a sufficiently large 
inventory of radioactive ${}^{26}{\rm Al}$ that they would melt (Sanders \& Taylor 2005), it is not surprising 
that no chondrites are found with only early-formed chondrules.
The primitive acapulcoite-lodranite and winonaite achondrites appear to be examples of chondrule-rich chondrite parent bodies 
that had just enough live ${}^{26}{\rm Al}$ to melt.
Significantly, as we review below, the ages of chondrules differ in different chondrite classes, a fact that can be used to 
date the time of accretion of chondrite parent bodies. 

\subsection{Refractory and CAI Abundances in Chondrites}

Our model is developed to explain the abundances of refractory elements and CAIs in chondrites, so it is essential to review
the values measured in various meteorites. 
Drawing on several resources (Brearley \& Jones 1998; Scott \& Krot 2005, 2014; Rubin 2011) we summarize in Table 1 the 
trends among  different chondrites of the abundances of CAIs and refractory elements. 
In the first two columns, for each chondrite class, are the average bulk-composition molar ratios of refractory lithophile
(Ca, Al, Ti, etc.) atoms relative to Mg atoms, normalized to the same ratio in CI chondrites,
drawn respectively from Scott \& Krot (2014) and Rubin (2011) and references therein.
Mg is selected as a proxy for the abundance of most meteoritic material, which is predominantly Mg-rich silicate.
There is good agreement between these two sources, and it is generally agreed that CCs are enriched in 
refractory elements and ECs and OCs depleted, relative to CI chondrites. 
But the degree of depletion depends on which refractory lithophile elements are included in the average abundance,
and which element is used as the proxy for the bulk.
For example, EC and OC refractory depletions of 20\% are typical when normalized to Si and CI composition,
(Wood 2005), whereas Table 1 makes clear that refractory depletions are only half as much when normalized to Mg.
We will generally try to compare our model against the refractory abundances reported by Rubin (2011). 

The next two columns in Table 1 list estimates of the CAIs volume fractions in chondrites, again from 
Scott \& Krot (2014) and Rubin (2011). 
These estimates are derived from multiple analyses (usually visual or tomographic estimates of areal extent of the CAIs)
reported in the literature.
As is clear, the agreement between these two compilations is not always good. 
CAI abundances are not particularly well known or agreed upon, and often are difficult to disentangle visually from the 
abundances of AOAs, as pointed out by Hezel et al. (2008) and Hezel \& Russell (2008).
AOA modal abundances in CCs often exceed those of CAIs (Hezel \& Russell 2008 and references therein).
In CV chondrites, the chondrites with the most CAIs, estimates for the modal abundance of CAIs range from 
0.65 - 1.89 vol\% (May et al.\ 1999), to 2.52 vol\% (Kornacki et al.\ 1984), to 2.5 - 9.4 vol\% (McSween 1977a),
to 15 vol\% (Chaumard et al.\ 2014).
Hezel et al.\ (2008) estimate an abundance $3.0_{-0.1}^{+0.3}$ vol\% in CV chondrites. 
[Hezel \& Russell (2008) point out that based on the Al abundance alone, the CV3 chondrite Allende must have far less
than 9.3 vol\% CAIs.]
Likewise, in CO chondrites, reported abundances range from 0.63 - 1.5 vol\% (Russell et al.\ 1998), to 
1.0 - 3.6 vol\% (Rubin et al.\ 1985), to 1.2 - 3.5 vol\% (McSween 1977b).
In the last column we list our adopted CAI abundances, against which the model predictions should be compared. 
(We report weight percent, assuming that the densities of CAIs and their host chondrites do not differ significantly.)

%
%
\begin{table}
\begin{centering}
\caption{Chondrite refractory and CAI abundances}
\vspace{0.2in} 
\begin{tabular}{c|cc|ccc}
Type &  (X/Mg)         & (X/Mg)         & CAIs        &  CAIs           & Adopted CAIs    \\
     &   / CI$^{a}$    & / CI$^{b}$     & (vol\%)$^{a}$ & (vol\%)$^{b}$ & (wt\%)  \\
\hline
EH   &   0.87          &  {\bf 0.884}   & $< 0.1$     &  0.01           & \boldmath$< 0.1$ \\
EL   &   0.83          &  {\bf 0.871}   & $< 0.1$     &  0.01           & \boldmath$< 0.1$ \\
\hline
R    &   0.95          &  {\bf 0.974}   & $< 0.1$     &  0.04           & \boldmath$< 0.1$ \\
\hline
H    &   0.93          &  {\bf 0.899}   & 0.01 - 0.2  &  0.02           & \boldmath$< 0.2$ \\
L    &   0.94          &  {\bf 0.904}   & $< 0.1$     &  0.02           & \boldmath$< 0.1$ \\
LL   &   0.90          &  {\bf 0.890}   & $< 0.1$     &  0.02           & \boldmath$< 0.1$ \\
\hline
CK   &   1.21          &  {\bf 1.24}    & 0.2         &  4              & \boldmath$< 4$   \\  
CV   &   1.35          &  {\bf 1.35}    & 3.0         &  3.0            & {\bf 3}          \\ 
CO   &   1.13          &  {\bf 1.11}    & 1.0         &  1.0            & {\bf 1}          \\
CM   &   1.15          &  {\bf 1.13}    & 1.2         &  1.2            & {\bf 1-2}        \\ 
CR   &   1.03          &  {\bf 1.02}    & 0.12        &  0.6            & {\bf 0.5-1}      \\ 
CI   &  $\equiv$ 1.00  &  {\bf 1.00}    & 0.00        &  0.0            & {\bf 0.0}        \\
\hline 
\end{tabular}
\end{centering} 

\noindent 
a. Scott \& Krot (2014);
b. Rubin (2011) and references therein;
\end{table}

Table 1 makes evident that ECs and OCs are depleted in refractories and almost devoid of CAIs,
whereas CCs are enriched in refractories and have must higher CAI abundances.
The abundances of CAIs and refractory elements are related but not identical:
Figure 2 of Rubin (2011) makes clear that among CCs, the CAI abundance is linearly correlated with refractory abundance, 
but among OCs and ECs, varying degrees of refractory depletions occur despite the uniform lack of CAIs.
Even in CCs, most ($\approx 70\%$) of refractory elements do not reside in the CAIs. 

\subsection{Time of Accretion of Meteorite Parent Bodies} 

Equally important to our model is constraining the time at which various achondrite and chondrite parent bodies accreted.
Since the abundances of refractory elements and CAIs vary in the disk over time, what material ends up in a meteorite
depends on when its parent body formed in the solar nebula. 
This can be constrained in three general ways. 
The time of formation of components like chondrules and CAIs, that formed in the solar nebula before accretion into 
meteorites, can be radiometrically dated using isotopic analyses. 
Or, minerals that formed on the parent body after its formation (e.g., carbonates formed by aqueous alteration of rock 
on a chondrite parent body, or solidification of basaltic lava on the surface of an achondrite parent body) 
can also be radiometrically dated. 
The time of accretion necessarily lies between these dates.
An additional, powerful constraint comes from estimates of the peak temperature reached by the parent body (as indicated by 
products of thermal metamorphism), which constrains the amount of live ${}^{26}{\rm Al}$ incorporated and the time of accretion,
subject to model-based assumptions.

In Table 2 we list all the available data we could find in each category, for all chondrite classes and several achondrites.
In the first column are the radiometric dates of formation of chondrules in various chondrites, culled from the literature.
Chondrites must accrete after the last chondrule within them forms. 
Where a range of chondrule ages was measured, we report the latest-formed chondrules. 
In the second column we list the time of accretion for each achondrite and chondrite, based on thermal models.
We especially draw on the thermal models of Sugiura \& Fujiya (2014; hereafter SF14), who include the parent bodies'
observed Al abundances in their calculations to match parent bodies' peak metamorphic temperatures against time of accretion.
These are simplified models and subject to uncertainties in key inputs (degree of serpentinization, initial Al abundances, 
surface temperatures, etc.). Most of these uncertainties have been quantified by SF14, leading to the listed uncertainties in 
the time of accretion.
It is possible that other uncertainties may alter these predicted times of accretion, but the models of SF14 are generally 
consistent with the radiometric dating.
In the third column we list other estimates of date of accretion, usually found by combining
radiometric dating with thermal models.
For example, Doyle et al.\ (2015) used ${}^{53}{\rm Mn}-{}^{53}{\rm Cr}$ dating of carbonates formed 
by aqueous alteration on the L parent body to infer it accreted before a time $2.4_{-1.3}^{+1.8}$ Myr.
In the last column of Table 2 we list our adopted times of accretion of different meteorite parent bodies, 
against which the model should be compared. 
For all chondrites and some achondrites we use the model results of SF14, which satisfactorily fit the radiometrically 
measured constraints of forming after chondrules but before post-accretionary events.
An exception is the CV chondrites, which we suspect may have experienced higher peak temperatures of metamorphism than 
modeled by SF14, implying an earlier time of accretion.
We use the detailed models of Goodrich et al.\ (2015) for ureilites and Hunt et al.\ (2017) for winonaites.

%
%
\begin{table}
\begin{centering}
\caption{Chondrite and achondrite accretion ages}
\vspace{0.2in} 
\begin{tabular}{c|c|c|c|c} 
Type          & Chondrule               & Thermal model         & Post-accretion                 & {\bf Adopted Time of} \\ 
              & ages (Myr)              & accretion times (Myr) & event (Myr)                    & {\bf Accretion (Myr)} \\ 
\hline 
ureilites     &                         & $1.0 \pm 0.3{}^{a}$   & $1.9_{-0.7}^{+2.2}{}^{b}$,     & $\sim$ {\bf 0.6}      \\
              &                         & $\sim 1.6{}^{c}$      & $3.8_{-1.3}^{+1.3}{}^{c}$      &                       \\ 
              &                         & $\approx 0.6{}^{d}$   & $3.3_{-0.7}^{+0.7}{}^{e}$      &                       \\ 
HEDs          &                         & $0.8 \pm 0.3{}^{a}$   & $2.2_{-1.1}^{+1.1}{}^{f}$      & $\sim$ {\bf 0.8}      \\
              &                         & $< 1{}^{g}$           & $0.6_{-0.5}^{+0.4}{}^{h}$      &                       \\
acapulcoites- &                         & $1.3 \pm 0.3{}^{a}$   &                                & $\sim$ {\bf 1.3}      \\
lodranites    &                         & $1.5 - 2.0{}^{i}$     &                                &                       \\
aubrites      &                         & $1.5 \pm 0.1{}^{a}$   &                                & $\sim$ {\bf 1.5}      \\
winonaites    &                         & $\sim 1.8^{j}$        & $3.6_{-2.0}^{+2.3}{}^{k}$      & $\sim$ {\bf 1.8}      \\
\hline
EH            &                         & $1.8  \pm 0.1{}^{a}$  & $\sim 2^{l}$                   &  {\bf 1.7 - 1.9}      \\
EL            &                         & $1.8  \pm 0.1{}^{a}$  &                                &  {\bf 1.7 - 1.9}      \\
\hline 
R             &                         & $2.1 \pm 0.1{}^{a}$   &                                &  {\bf 2.0 - 2.2}      \\
\hline
H             & $1.7 \pm 0.7^{m}$       & $2.1 \pm 0.1{}^{a}$   &                                &  {\bf 2.0 - 2.2}      \\
              &                         & $1.8 - 2.7{}^{n}$     &                                &                       \\
              &                         & $2.05 - 2.25{}^{o}$   &                                &                       \\
L             & $1.0-2.2^{p}$           & $2.1 \pm 0.1^{a}$     & $2.4_{-1.3}^{+1.8}{}^{q}$      &  {\bf 2.0 - 2.2}      \\
              & $3.3 \pm 0.5{}^{r}$     &                       &                                &                       \\
              &                         & $2.05 - 2.25{}^{o}$   &                                &                       \\
LL            & $\sim 1.0-2.5{}^{s}$    & $2.1 \pm 0.1^{a}$     & $4.0_{-1.1}^{+1.4}{}^{q}$      &  {\bf 2.0 - 2.2}      \\
              & $2.4_{-0.4}^{+0.7}{}^{t}$ &                     &                                &                       \\ 
              & $2.5 \pm 0.3(?)^{u}$    &                       &                                &                       \\
              & $2.5 \pm 0.4^{v}$       &                       &                                &                       \\
              & $1.8^{w}$               &                       &                                &                       \\
\hline
CK            &                         & $2.6 \pm 0.2^{a}$     &                                &  {\bf 2.4 - 2.8}      \\  
CV            & $2.2 \pm 0.8{}^{x}$     & $3.0 \pm 0.2^{a}$     & $4.2_{-0.7}^{+0.8}{}^{q}$      &  {\bf 2.4 - 3.0}      \\
              & $2.5 \pm 0.4{}^{y}$     &                       &                                &                       \\
              & $3.0 \pm 0.4{}^{z}$     &                       &                                &                       \\
CO            & $2.5 \pm 0.3{}^{v}$     & $2.7 \pm 0.2^{a}$     & $5.1_{-0.4}^{+0.5}{}^{q}$      &  {\bf 2.5 - 2.9}      \\
CM            &                         & $3.5_{-0.5}^{0.7}{}^{a}$ & $4.4 - 5.7{}^{aa}$          &  {\bf 3.0 - 4.2}      \\
CR            & $3.7 \pm 0.6^{ab}$      & $3.5 \pm 0.5^{a}$     &                                &  {\bf 3.7 - 4.0}      \\ 
              & $3.7_{-0.2}^{+0.3}{}^{ac}$ & $4.0_{-0.3}^{+0.5}{}^{ac}$ &                        &                       \\
              & $3.6 \pm 0.6^{ad}$      &                       &                                &                       \\
              & $4.0 \pm 0.6^{r}$       &                       &                                &                       \\ 
CI            &                         & $3.6 \pm 0.5^{a}$     & $4.4 - 5.7{}^{aa}$             &  {\bf 3.1 - 4.1}      \\
\hline 
\end{tabular}
\end{centering} 

\noindent 
a. Sugiura \& Fujiya (2014);
b. Melting event, Al-Mg: Baker et al. (2012);
c. Impact event, Al-Mg and Hf-W: van Kooten et al.\ (2017);
d. Wilson \& Goodrich (2016); Goodrich et al. (2015); 
e. Silicate melting, Hf-W: Budde et al.\ (2015); 
f. Differentiation, Mn-Cr: Trinquier et al.\ (2008);
g. Neumann et al.\ (2014); 
h. Magma ocean crystallization, Al-Mg: Schiller et al. (2011);
i. Models plus Hf-W: Touboul et al.\ (2009); 
j. Hunt et al.\ (2017); 
k. Differentiation, Hf-W: Kruijer et al.\ (2014);
l. Mn-Cr: Shukolyukov \& Lugmair (2004); 
m. Hf-W: Kleine et al.\ (2008);
n. Henke et al. (2013); Gail et al.\ (2014) and references therein; 
o. Blackburn et al.\ (2017);
p. Al-Mg: Rudraswami \& Goswami (2007); 
q. Mn-Cr: Doyle et al.\ (2016);
r. U-Pb: Bollard et al.\ (2017);
s. Al-Mg: Rudraswami et al.\ (2008);
t. Al-Mg: Mostefaoui et al.\ (2002); 
u. Al-Mg: Villeneuve et al.\ (2009);
v. Al-Mg: Kurahashi et al.\ (2008);
w. Al-Mg: Kita et al.\ (2000);
x. Hf-W: Budde et al.\ (2016);
y. Al-Mg: Nagashima et al.\ (2015);
z. Al-Mg: Hutcheon et al.\ (2009); 
aa. Mn-Cr: Fujiya et al.\ (2013); 
ab. Pb-Pb: Amelin et al.\ (2002), reanalyzed by Schrader et al.\ (2017);
ac. Al-Mg: Schrader et al.\ (2017);
ad. Metal-silicate fractionation during chondrule formation, Hf-W: Budde et al.\ (2018).


\end{table} 

There is generally good but not perfect agreement among various sources for many meteorite types.
Table 2 makes evident that achondrites probably accreted early, in the first $\sim 1$ Myr, both 
from radiometric dating and the need to accrete sufficient live ${}^{26}{\rm Al}$ to melt.
Chondrite parent bodies generally accreted later, with most OCs probably accreting slightly earlier 
($\approx 2 - 3$ Myr) than most CCs. The CR parent body appears to have accreted latest, at about 4 Myr.

%
%
%

\subsection{Origin of chondrite diversity} 

The cause of the diversity in chondrite refractory and CAI abundances has been intepreted as reflecting 
either spatial or temporal heterogeneities, or both.
All chondrites (except CI) contain CAIs formed at the birth of the solar system, and 
chondrules that formed over many Myr; 
but chondrites may sample these different components at different times and places in the nebula, and 
variations in the relative  proportions of each component could give rise to chondrite diversity. 
Rubin (2011) considered the solar nebula to have a spatial structure that varied relatively little in time, 
with the less oxidized ECs and OCs closer to the Sun, the CCs farther away.
Chambers (2006) considered the nebula to vary little in space, with chondrite diversity reflecting differences 
in time, with CCs being the oldest, and OCs and ECs forming later. 
Improvements in isotopic data and thermal models since 2006 (Table 2) refute a solely temporal interpretation,
but it is likely that the variations among chondrites are due to variations in both space and time. 

Great progress toward resolving these issues was made when Warren (2011) plotted different chondrites and other
meteorites according to their bulk isotopic anomalies, in particular $\eTi$, $\eCr$, and $\dOx$.
The first two refer to the excesses in the bulk composition of these chondrites in the  
${}^{50}{\rm Ti} / {}^{48}{\rm Ti}$ or ${}^{54}{\rm Cr} / {}^{52}{\rm Cr}$ isotopic ratios, above a terrestrial 
standard, measured in parts per 10,000.
Likewise, $\dOx$ refers to the excess in the ${}^{17}{\rm O} / {}^{16}{\rm O}$ isotopic ratio in the bulk 
composition, above the value expected for terrestrial samples with the same ${}^{18}{\rm O} / {}^{16}{\rm O}$ 
ratio, measured in parts per 1000. 
Warren (2011) showed that when plotted in a field of $\eTi$ vs. $\eCr$, or 
$\dOx$ vs. $\eCr$, meteorites show a striking bimodality, falling into one of two groups.
One group includes the Earth and Moon, Mars, ECs, OCs, and several achondrites (including angrites, Main Group 
pallasites, HEDs, and ureilites).
The other group includes CI chondrites and all other CCs.

This dichotomy was recently found to extend to Mo and W isotopes, and to iron meteorites as well 
(Kruijer et al.\ 2017). 
When plotted in a field of bulk $\epsilon^{95}{\rm Mo}$ vs. $\epsilon^{94}{\rm Mo}$, or 
bulk $\epsilon^{182}{\rm W}$ vs. $\epsilon^{183}{\rm W}$, meteorites again divide into same two groups. 
One group includes ECs, OCs, and several iron meteorites 
as well as many achondrites and Earth and Mars. 
The other group includes CCs and several different iron meteorites.
These data strongly imply that the solar nebula was divided into two chemical and isotopic reservoirs, the
one (`non-carbonaceous') associated with the inner disk and ECs and OCs, the other (`carbonaceous') associated 
with the outer disk and with CCs.
Based on the $\epsilon^{182}{\rm W}$ data, which is affected by the decay of the radiaoctive ${}^{182}{\rm Hf}$
($t_{1/2} = 9.3$ Myr), Kruijer et al.\ (2017) inferred that the reservoirs were at first identical, but began to 
evolve separately at a time $< 1$ Myr after CAIs.
What caused the two regions of the disk to evolve differently is unclear, but may be attributable to depletion of 
of $r$ process material from the inner disk.
Whatever the cause, something separated the two reservoirs, preventing them from mixing. 
Only much later in disk evolution were the parent bodies of chondrites dynamically scattered by Jupiter 
and commingled in the asteroid belt, as in the Grand Tack model of Walsh et al.\ (2011), or by Jupiter at 5.2 AU
(Raymond \& Izidoro 2017). 

Kruijer et al.\ (2017) suggested that Jupiter is what separated the solar nebula into two reservoirs that 
could not perfectly mix.
If Jupiter opens a gap in the disk, the surface density and pressure just beyond Jupiter must necessarily increase outward.
This reversal of the pressure gradient causes gas to rotate faster than Keplerian, and for particles to experience a 
tailwind. Instead of spiraling inward, they spiral outward, until they reach the pressure maximum beyond Jupiter.
If the isotopic anomalies like $\eTi$ and $\eCr$ are carried mostly by large ($> 100 \, \mu{\rm m}$) particles,
then Jupiter's formation can lead to the bimodality of meteorites with respect to these isotopes.
Kruijer et al.\ (2017) inferred that Jupiter grew to $> 20 \, M_{\oplus}$, large enough to reverse the pressure
gradient and direction of particle drift in the disk (Lambrechts et al.\ 2014), in $< 1$ Myr.

The trapping of particles in the pressure maximum beyond Jupiter has been invoked to explain other features of the solar system.
Jupiter's core is predicted to grow by pebble accretion until it opens a gap in the disk, at which point the the core is 
isolated from the source of pebbles and stops growing (Lambrechts et al.\ 2014). 
Formation of Jupiter has also been invoked to explain why the inner solar system is so dry, as measured by 
the lack of evidence for water in enstatite chondrites, or Earth's low ($< 0.1$ wt\%) water abundance 
(Mottl et al.\ 2007). 
Morbidelli et al.\ (2016) argued that Jupiter formed inside the snow line, and when the disk subsequently cooled,
and ice particles tried to drift into the inner solar nebula, Jupiter in like fashion prevented the inflow of 
large icy particles.

Our focus in this paper is to assess whether such a model might be also explain the abundances of refractory elements
and CAIs in meteorite types, as listed in Table 1.
Compared to CI compositions, CCs are clearly enriched in refractory lithophile elements, 
but to varying degrees ranging from 3\% to 35\%.
The reasons for this trend and variation are not completely clear. 
The enrichments are clearly correlated with the abundance of refractory-rich CAIs, but CAIs are not the sole 
or even main carriers of Ca, Al and Ti in chondrites (Rubin 2011).
Also clear from Table 1 is that OCs and ECs are {\it depleted} in refractories compared to a CI composition,
by about 10\%.
It is perplexing that the chondrites thought to form closest to the Sun should have the depletions in
refractory elements, while those that form farthest from the Sun should have enhancements.
Very much related to this is the CAI Storage problem and the mystery of why CCs contain the highest proportion of CAIs, 
while CAIs are practically absent in OCs and ECs.
It is paradoxical that CAIs created at high temperatures near the Sun should be more abundant in
those chondrites that formed farthest from the Sun.
Resolution of these paradoxes is the purpose behind our model.

\section{Methods}

\subsection{Overall Disk Dynamics}

We have written a numerical code that calculates the evolution of a protoplanetary disk and the radial distribution of 
varying components within it.
Our 1-D explicit code solves for properties as a function of heliocentric distance only, on a grid extending from 
0.06 AU to 60 AU, with 450 zones logarithmically spaced in most of the results we present. 
We initialize the surface density of gas using a self-similar profile:
\begin{equation}
\Sigma(r,t=0) = \frac{(2-\gamma) \, M_{\rm disk}}{2\pi R_1^2} \, \left( \frac{ r }{ R_1 } \right)^{-\gamma} \,  
 \exp \left( -\left( \frac{ r }{ R_1 } \right)^{2-\gamma} \right),
\end{equation}
(Hartmann et al.\ 1998), 
with $\gamma = 15/14$, $R_1 = 1 \, {\rm AU}$.
The total disk mass on the grid is $0.089 \, M_{\odot}$, about 7 times the minimum-mass solar nebula mass of
Weidenschilling (1977b),
but the disk is gravitationally stable: at $t = 0$, nowhere in the disk does the Toomre $Q$ parameter fall below about
$Q = C \Omega / (\pi G \Sigma) \approx 4$, and the minimum value of $Q$ at any time in the disk is $\approx 2.3$,
achieved between about 0.7 and 1.4 Myr, in the region around 5 AU. 
We note that the evolved disks in Taurus are better characterized by $R_1 = 10 \, {\rm AU}$ (Hartmann et al.\ 1998),
but our choice of $R_1 = 1$ AU is motivated by the need to have high densities in the inner disk, and resembles 
the initial state of the disk calculated by Yang \& Ciesla (2012). 

We track the surface density of gas a function of distance and time, $\Sigma(r,t)$, solving the equation 
\begin{equation}
\frac{\partial \Sigma}{\partial t} = \frac{1}{2\pi r} \, \frac{\partial \dot{M}}{\partial r},
\end{equation}
where 
\begin{equation}
\dot{M} = 6\pi r^{1/2} \, \frac{\partial}{\partial r} \left( r^{1/2} \Sigma \nu \right)
\end{equation}
is the mass flux at every location ($\dot{M} > 0$ if mass flow is inward).
In practice we calculate the (vertically mass-weighted average) gas velocity 
\begin{equation}
V_{\rm g,r} = -\frac{3}{2} \frac{\nu}{r} \, \left( 1 + \frac{\partial \ln \Sigma \nu}{\partial \ln r} \right)
\end{equation} 
($V_{\rm g,r} < 0$ if mass flow is inward) 
and then use the donor cell approximation to determine $\Sigma$ in the formula 
$\dot{M} = -2\pi r \, \Sigma \, V_{\rm g,r}$.
(While donor cell is a relatively diffusive advection scheme, it is appropriate for profiles like ours with
no sharp discontinuities, and we show that our results our numerically converged.) 
Here $\nu$ is the turbulent viscosity, which we parameterize as 
\begin{equation}
\nu = \alpha \, C \, H,
\end{equation}
were $H = C / \Omega$ is the disk scale height, $C = (k T / \bar{m})^{1/2}$ is the sound speed
($k$ is Boltzmann's constant and $\bar{m} = 2.33$ amu), and 
$\Omega = (G M_{\star} / r^3)^{1/2}$ is the Keplerian orbital frequency ($M_{\star} = 1 \, M_{\odot}$).
At the inner boundary we apply the standard zero-torque boundary condition.
We remove gas from the outer portions of the disk in a manner consistent with external photoevaporation 
with far-ultraviolet flux given by $G_0 \approx 30$, as described by Kalyaan et al.\ (2015).
In practice, we find very small mass loss rates at the outer boundary, $\ll 10^{-10} \, M_{\odot} \, {\rm yr}^{-1}$.

Equations 4 and 5 constitute a diffusion equation for $\Sigma$. 
We apply finite difference formulas to solve these, solving for $\Sigma$ at time $t + dt$ using spatial 
derivatives of $\dot{M}$, and therefore of $\Sigma$, at the current time $t$.
Because the code is an explicit code, to maintain numerical stability we are limited by the Courant condition
to a maximum timestep that scales as the square of the grid zone size. 
We did not use an adaptive timestep, but found that a fixed timestep of 
0.01 yr provided numerical stability for our standard case of 450 zones.

To determine the temperature $T$ we assume (for the purposes of calculating mass fluxes) a 
vertically uniform temperature profile.
For a passively heated disk, we adopt a profile using the approach of Chiang \& Goldreich (1997):
\begin{equation}
T_{\rm passive} = 163 \, \left( \frac{ L }{1 \, L_{\odot} } \right)^{2/7} \, 
 \left( \frac{ r }{ 1 \, {\rm AU} } \right)^{-3/7} \, {\rm K},
\end{equation} 
and adopt the luminosity $L(t) = 1.2 \, ( t / 1 \, {\rm Myr} )^{-0.5} \, L_{\odot}$ (Baraffe et al.\ 2002).
If the disk is actively accreting, its temperature can be higher.
We follow Min et al.\ (2011) in using the formula from Hubeny (1990) for the midplane temperature: 
\begin{equation}
\sigma T_{\rm acc}^4 = \frac{27}{128} \, \Sigma^2 \kappa \, \Omega^2 \, \nu,
\end{equation}
where $\kappa$ is the opacity per gram of gas and $\sigma$ is the Stefan-Boltzmann constant.
Substituting $\nu = \alpha (k T_{\rm acc} / \bar{m}) \Omega^{-1}$ yields
\begin{equation}
T_{\rm acc} = \left[ \frac{27}{128} \, \frac{k}{\sigma \bar{m}} \, \alpha \, \Sigma^2 \, \kappa \, \Omega \right]^{1/3} 
\end{equation}
For the opacity, we assume $\kappa = 5 \, {\rm cm}^{2} \, {\rm g}^{-1}$, a value chosen to approximate the Rosseland mean opacity 
of particles in the models presented by Semenov et al.\ (2003; their Figure 1), in particular their composite aggregates model 
(at least for temperatures greater than 100 K).
We assume this opacity up to the temperature at which silicates evaporate, which we set to 1400 K.
Most ferromagnesian silicates will evaporate at temperatures over 1350 K, above which the opacity drops drastically.
Additional accretion heating cannot lead to increases in temperature much above this level, or about 1400 K
(e.g., Lesniak \& Desch 2011; see also Yang \& Ciesla 2012).
We approximate the combined effects of passive and accretional heating by setting the (vertically uniform) temperature 
at each radius to be 
\begin{equation}
T = \left[ T_{\rm passive}^4 + T_{\rm acc}^4 \right]^{1/4}
\end{equation}
In practice, the temperature is close to either $T_{\rm acc}$ or $T_{\rm passive}$, which ever is greater. 

The cause of turbulence in disks is largely unknown and unconstrained. 
We do not attempt to calculate $\alpha$ from first principles, instead parameterizing it.
We set $\alpha = 5 \times 10^{-4}$ at $r < 1 \, {\rm AU}$, falling as a power law 
$\alpha = 5 \times 10^{-4} \, ( r / 1 \, {\rm AU})^{-1.699}$ between 1 AU and 10 AU, 
and $\alpha  = 1 \times 10^{-5}$ for $r \geq 10$ AU. 
This form and the specific values are arbitrary, but are motivated by our finding that high
$\alpha$ is needed in the inner disk to generate the temperatures $> 1400$ K to create CAIs, but
especially the high value and the negative slope $d \ln \alpha / d \ln r < 0$ are needed to help
transport CAIs out of the inner disk.
In the outer disk, low values are required to prevent the CI chondrite-forming region from mixing 
with the rest of the disk.  
In \S 4.4 below we discuss the effects of changing this $\alpha$ profile.

We assume Jupiter forms at 0.6 Myr at 3.0 AU.
In \S 4.4 we explore the effects of Jupiter forming at different times and locations in the disk.
To mimic Jupiter forming and opening a gap in the disk, we artificially increase the 
viscosity in its vicinity. 
Simulations by Lambrechts et al.\ (2014) show that once Jupiter grows to $\approx 20 - 30 \, M_{\oplus}$ at heliocentric 
distance $r_{\rm J}$, it decreases the density of gas in its vicinity and leads to a density maximum at about 
$0.8 \, r_{\rm J}$ and $1.2 \, r_{\rm J}$, and a super-Keplerian rotation at about $1.1 r_{\rm J}$. 
These effects are mimicked in our code by imposing a higher value, $\alpha'$, in the vicinity of Jupiter:
\begin{equation}
\alpha' = \alpha + \left( 10^{-2} - \alpha \right) \, e^{-x^2},
\end{equation}
where $x = (r - r_{\rm J}) / R_{\rm H}$, where $R_{\rm H} = r_{\rm J} \, \left( M_{\rm J} / 3 M_{\star} \right)^{1/3}$ is 
the Hill radius of the growing Jupiter, 
using Jupiter's instantanteous mass.
The maximum value $\alpha' \sim 0.01$ in the gap is comparable to the value of $\alpha$ found in the vicinity of Jupiter
in numerical simulations (Lyra et al.\ 2016). 

We simultaneously mimic the growth of Jupiter in a manner similar to that of Kley (1999), as follows.  
We assume Jupiter instantaneously acquires a mass $30 \, M_{\oplus}$ at time $t = t_{\rm J}$, after which it accretes gas.
At each timestep $dt$, we integrate the mass 
\begin{equation} 
dM = \int_{r} \, \Sigma(r) \, \exp(-x^2) \, 2\pi r \, dr,
\end{equation}
where $x$ is defined as above, and then let Jupiter accrete a fraction $dt / \tau$ of this mass, where 
$\tau = 7.5 \times 10^4$ yr is a growth rate.
That is, $dM_{\rm J} / dt = dM / \tau$. 
The same amount of gas is removed from the disk in the vicinity of Jupiter. 
Our approach is similar to that taken by Raymond \& Izidoro (2017) in their treatment of how Jupiter's growth
will scatter planetesimals; they parameterized Jupiter's growth as linear on a timescale $\sim 10^5$ yr.
While some models of Jupiter's formation envision a slow increase in mass until runaway gas accretion proceeds
(Pollack et al.\ 1996; Ikoma et al.\ 2000; Rice \& Armitage 2003), we allow Jupiter to accrete gas within its
Hill sphere rapidly, essentially its Kelvin-Helmholtz timescale (Thommes et al.\ 2008).
As modeled by Raymond \& Izidoro (2017), Jupiter carves a gap at the same rate it grows. 

While our treatments of Jupiter's growth and its opening of a gap are admittedly crude, they capture the main effects 
of Jupiter, which are that a gap and pressure bump beyond Jupiter are created, and that gas is removed from the disk, at a rate 
$dM_{\rm J} / dt \sim (318 \, M_{\oplus}) / (1 \, {\rm Myr}) \sim 1 \times 10^{-9} \, M_{\odot} \, {\rm yr}^{-1}$
that is a significant fraction of the disk accretion rate.
In our simulations, Jupiter reaches its final mass of $317.8 \, M_{\oplus}$ at about 4.5 Myr.

\subsection{Particle Dynamics} 

Into this disk we introduce solids.
We initialize the disk with small (radius $a \approx 1 \, \mu{\rm m}$) particles with 
``solar nebula" (SN) composition, uniformly mixed throughout the disk.
We assume CI chondrites have a composition close to SN composition, but actually are depleted in
refractories lithophiles (Ca, Al, Ti, etc.), by 5 - 10\%. 
In what follows, we assume CI $= 0.878 \times$ SN. 
We do not track volatiles, concentrating only on silicate / metal grains, so the combined mass fraction 
relative to the gas of the two populations is $5 \times 10^{-3}$ (Lodders 2003). 
Volatiles such as water ice and other organics are assumed to evolve separately, and we defer the calculation of 
their radial distribution to future work. Here, we compare the abundance of CAI to the abundance of silicates and 
metal only. 

When particles with these starting compositions enter a region with $T > 1400$ K, we assume they can be converted into CAIs.
We assume two populations of small grains: 
population \#1, with a mass fraction $1.5 \times 10^{-3}$, can be thermally processed into large CAIs, but 
population \#2, with mass fraction $3.5 \times 10^{-3}$, is not.
In any timestep, any population \#1 grains in a region with $T = 1400$ K (the CAI Factory) are destroyed and converted into
other grains: 89\% of the material is converted into small ($a = 1 \, \mu{\rm m}$) grains of indeterminate but 
refractory-depleted composition (population \#3); 3\% is turned into refractory-depleted, large ($a = 600 \, \mu{\rm m}$) 
grains that resemble AOAs (population \#4); 
and 8\% of the material is converted into CAIs with radii $a = 2500 \, \mu{\rm m}$ (population \# 5).
We calculate that starting with a CI composition of silicates and metal, production of melilite and spinel will consume 
8\% of the mass.
Grossman (2010) likewise calculated that condensation of a solar composition gas would yield a mass fraction of CAIs 
$\approx 5.7\%$, starting with a CI composition that includes volatiles (which we do not).
After removing water and organics from the CI composition, condensation would yield a mass fraction of CAIs 
$\approx 8\%$ (Jacquet et al.\ 2012).
The AOA fraction 3\% was arbitrarily chosen to yield the approximate AOA/CAI ratio in meteorites.

A shortcoming of our model is that we do not model chondrule formation, which will vary over time in the disk.
We therefore do not include abundant populations of particles with radii of a few hundred microns, which might 
drain from the inner disk, reducing the solids-to-gas ratio there, or which might collect in the pressure bump,
increasing the solids-to-gas ratio there.
When computing the mass fraction of a meteorite that is CAIs, we are comparing only to the well-mxed fraction of 
micron-sized solids.
This assumption probably does not affect our conclusions about the inner disk: chondrules would drain slowly out of the 
inner disk, and chondrules would be continuously formed from dust. Our predicted CAI mass fraction is small enough ($< 0.01$ wt\%)
that even a factor of 10 depletion in solids would not violate abundance constraints.
In the outer disk, in contrast, it is possible that chondrules throughout the outer disk could aerodynamically collect 
in the pressure bump, lowering the mass fraction of CAIs in CCs below our predicted amounts, by an unknown factor.
Our results would sill hold if chondrules formed throughout the disk accreted into chondrites immediately, before radially
drifting; or if chondrules only formed in the outer disk in the pressure bump region.
We note that because of uncertainties in CAI abundances, a factor of 2 increase in the solids-to-gas ratio, diluting the
mass fraction of solids that is CAIs by a factor of 2 in the pressure bump, would 
still be consistent with the meteoritic constraints. 
We defer a calculation of chondrule concentrations in the disk to a future paper. 

Each of these particles moves through the disk according to the three processes of advection, diffusion and drift. 
The surface density of a tracer species, $\Sigma_{\rm c}$, evolves according to 
\begin{equation}
\frac{\partial \Sigma_{\rm c}}{\partial t} = \frac{1}{2\pi r} \, \frac{\partial \dot{M}_{\rm c}}{\partial r},
\end{equation}
with $\dot{M}_{\rm c}$ containing three terms corresponding to each process.
Advection is accounted for by assuming a tracer particles with mass fraction $c = \Sigma_{\rm c} / \Sigma$ have a mass flux 
$\dot{M}_{\rm c,adv} = c \dot{M}$.
Diffusion is accounted for (e.g., Desch et al.\ 2017) by adding a term
\begin{equation}
\dot{M}_{\rm c,diff} = 2\pi r \, \Sigma \, {\cal D} \, \frac{\partial}{\partial r} \left( \frac{\Sigma_{\rm c}}{\Sigma} \right),
\end{equation}
where 
\begin{equation}
{\cal D} = \frac{ \nu / {\rm Sc} }{ 1 + {\rm St}^2 },
\end{equation} 
and we assume a Schmidt number ${\rm Sc} = 0.7$.
Here the Stokes number is 
\begin{equation}
{\rm St} = \Omega \, t_{\rm stop} = \Omega \, \frac{\rho_{\rm s} a}{\rho_{\rm g} v_{\rm th}}
\end{equation}
and we take a grain internal density $\rho_{\rm s} = 3 \, {\rm g} \, {\rm cm}^{-3}$, and calculate
$v_{\rm th} = \left( 8 k T / \pi \bar{m} \right)^{1/2}$ and midplane gas density 
$\rho_{\rm g} = \Sigma / (\sqrt{2\pi} H)$ using local conditions in each cell.

To this we add $\dot{M}_{\rm c,drift} = -2\pi r \, \Sigma_{\rm c} \, \Delta u$ to account for the fact that particles 
drift with speed $\Delta u$ with respect to the gas, where 
\begin{equation}
\Delta u = -\frac{ {\rm St} }{ 1 + {\rm St}^2 } \, \left[ {\rm St} \, V_{\rm gr} -\eta \, r \Omega \right],
\end{equation}
and 
\begin{equation}
\eta = -\frac{1}{\Omega^2 r} \, \frac{1}{\rho_{\rm g}} \, \frac{\partial P}{\partial r}
\end{equation}
(Takeuchi \& Lin 2002; see Desch et al.\ 2017). 

For large particles we make one further adjustment to account for meridional flow.
It is well known that in $\alpha$ disks the gas velocity tends to be outward at the midplane even if the net accretion
is inward (Urpin 1984; Takeuchi \& Lin 2002; Ciesla 2009; Philippov \& Rafikov 2017;
but see objections by Jacquet 2017). 
Assuming a vertically isothermal disk with $\rho(r,z) = \rho(r,0) \, \exp( - z^2 / 2 H^2 )$, the 
formulation of Philippov \& Rafikov (2017) yields a gas velocity at the midplane 
\begin{equation}
u_{\rm g,r} = \frac{\nu}{r} \, \left[ -\frac{3}{2} - 3 \, \frac{\partial \ln (\Sigma \alpha)}{\partial \ln r} 
 -\frac{1}{2} \, \frac{\partial \ln T}{\partial \ln r} \right]
\end{equation}
We calculate this in every zone and find that it is generally outward ($> 0$). 
We assume that small grains are vertically well mixed, and therefore experience on average the same net radial flow 
$V_{\rm g,r}$ as the gas; but large particles are concentrated at the midplane and experience only this midplane
radial flow. 
For those particles we accordingly add a mass flux
\begin{equation}
\dot{M}_{\rm c,merid} = -2\pi r \, \Sigma_{\rm c} \, \frac{ u_{\rm g,r} - V_{\rm g,r} }{ 1 + {\rm St}^2 },
\end{equation}
to account for the fact that we should have used $u_{\rm g,r}$ instead of $V_{\rm g,r}$ in the 
drift equation. 
We have used the formulation of Philippov \& Rafikov (2017) to calculate the gas velocity as a function of height above
the midplane $z$, and the mass flux of particles if they follow a vertical density distribution proportional to 
$\exp( - z^2 / 2 H_{\rm p}^2 )$.
If $H_{\rm p} / H = 1$, the particles will be well mixed with the gas, but for $H_{\rm p} / H < 1$, they are 
concentrated at the midplane.
We find that if $H_{\rm p} / H < 0.92$ even, then particles are concentrated enough at the midplane that their net outflow
is outward. 
We use the following formula from Youdin \& Lithwick (2007), with the assumption that $\xi = 1$ (because
$\Omega \, t_{\rm eddy} \ll 1$), to calculate the particle scale height:
\begin{equation}
H_{\rm p} = H \, \left( \frac{ \alpha }{ \alpha + {\rm St} } \right)^{1/2}
\end{equation}
(where ${\rm St} = \Omega \, t_{\rm stop}$ as above).
The requirement that $H_{\rm p} < 0.92 \, H$ therefore requires ${\rm St} > 0.18 \, \alpha$.
If particles in any zone have ${\rm St} > 0.18 \, \alpha$, we assume they experience meridional flow as above; if 
${\rm St} < 0.18 \, \alpha$, we assume they are vertically well mixed and experience only the average velocity of the gas.
The mass flux of particles is 
$\dot{M}_{\rm c} = \dot{M}_{\rm c,adv} + \dot{M}_{\rm c,diff} + \dot{M}_{\rm c,drift} + \dot{M}_{\rm c,merid}$.

\section{Results}

\subsection{Canonical Case}

We first present results using the methodology and parameters described above.
Figure 1 shows the evolution of $\Sigma(r,t)$ in the disk.
The black dashed curve shows the surface density at $t = 0$, and subsequent curves, colored maroon, red, ... violet, 
denote $\Sigma(r,t)$ at 0.5, 1.0, ... 5.0 Myr.
After 0.6 Myr, Jupiter opens a gap at 3.0 AU.
Through the combined effects of increased viscosity near Jupiter and accretion of nebular gas by Jupiter, this 
region is cleared of gas.
Interior to about 3 AU, the surface density tends to follow power-law profiles in $\Sigma(r)$, because the mass accretion 
rate is close to uniform and $\Sigma(r) \approx \dot{M} / (3\pi \nu)$, or $\Sigma \propto \alpha^{-1} T^{-1} r^{-3/2}$.
Because $\alpha$ is decreasing as $r^{-1.7}$ beyond 1 AU, and temperature is also decreasing, these terms nearly cancel
the radial dependence from $r^{-3/2}$, so that $\Sigma(r)$ is essentially flat inside of several AU. 
Throughout the simulation, mass continues to spread from the inner disk to the outer disk, increasing the surface density
%
%
\begin{centering}
\begin{figure}[ht]
 \centering
 \includegraphics[width=.8\linewidth]{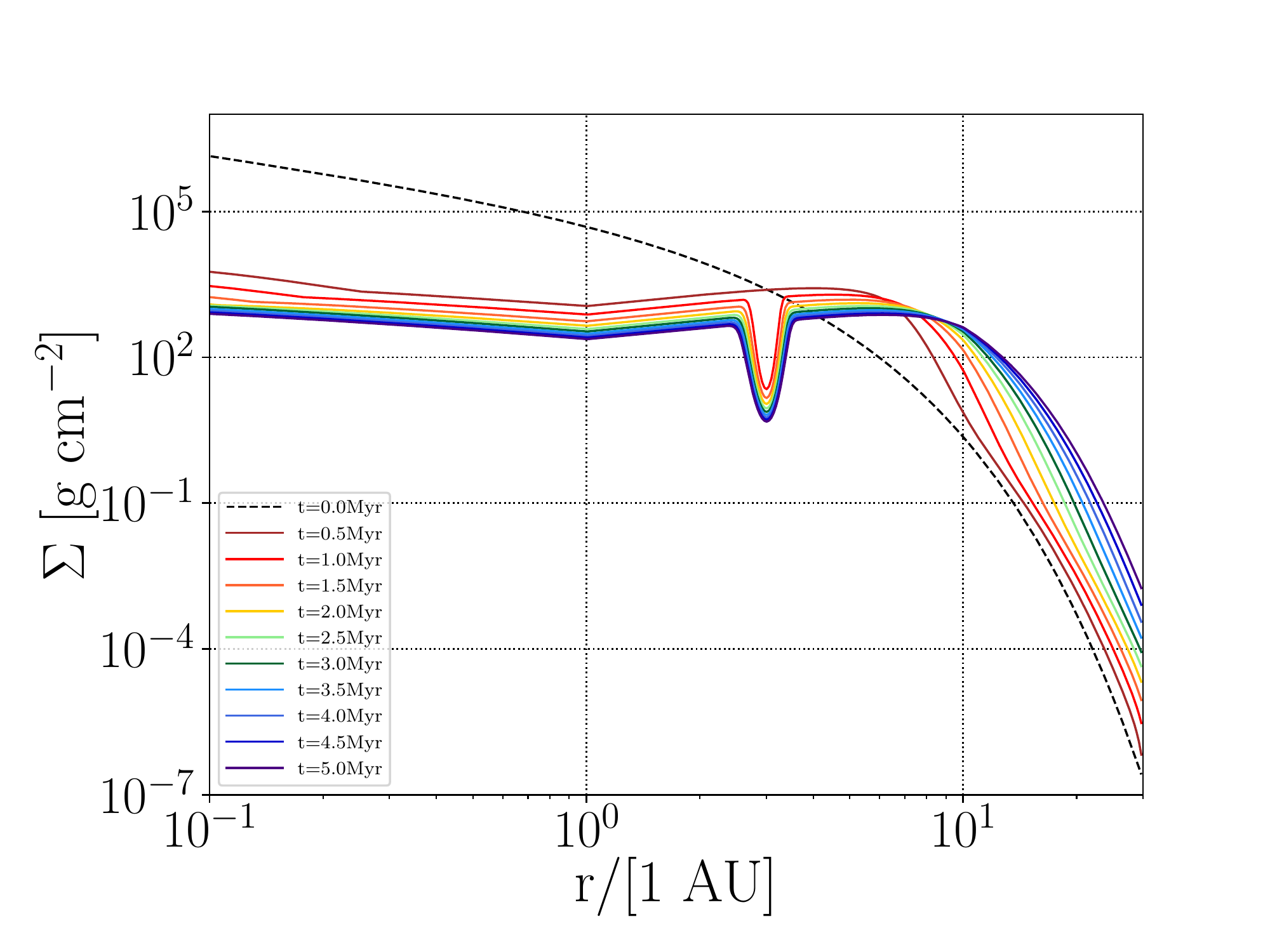}
 \caption{Surface densities at various times in the disk's evolution (black dashed curve at $t = 0$ Myr; maroon,
 red, ... violet curves at $t = 0.5$, 1.0, ... 5.0 Myr). The inner disk ($< 1$ AU) is marked by $\alpha = 5 \times 10^{-4}$,
 the outer disk ($> 10$ AU) by $\alpha = 1 \times 10^{-5}$, with $\alpha$ falling as a power law in between. 
 Jupiter is assumed to form at 0.6 Myr and open a gap at 3 AU.  }
 \label{fig:Sigma}
\end{figure}
\end{centering}

%
%
\begin{centering}
\begin{figure}[ht]
 \centering
 \includegraphics[width=.8\linewidth]{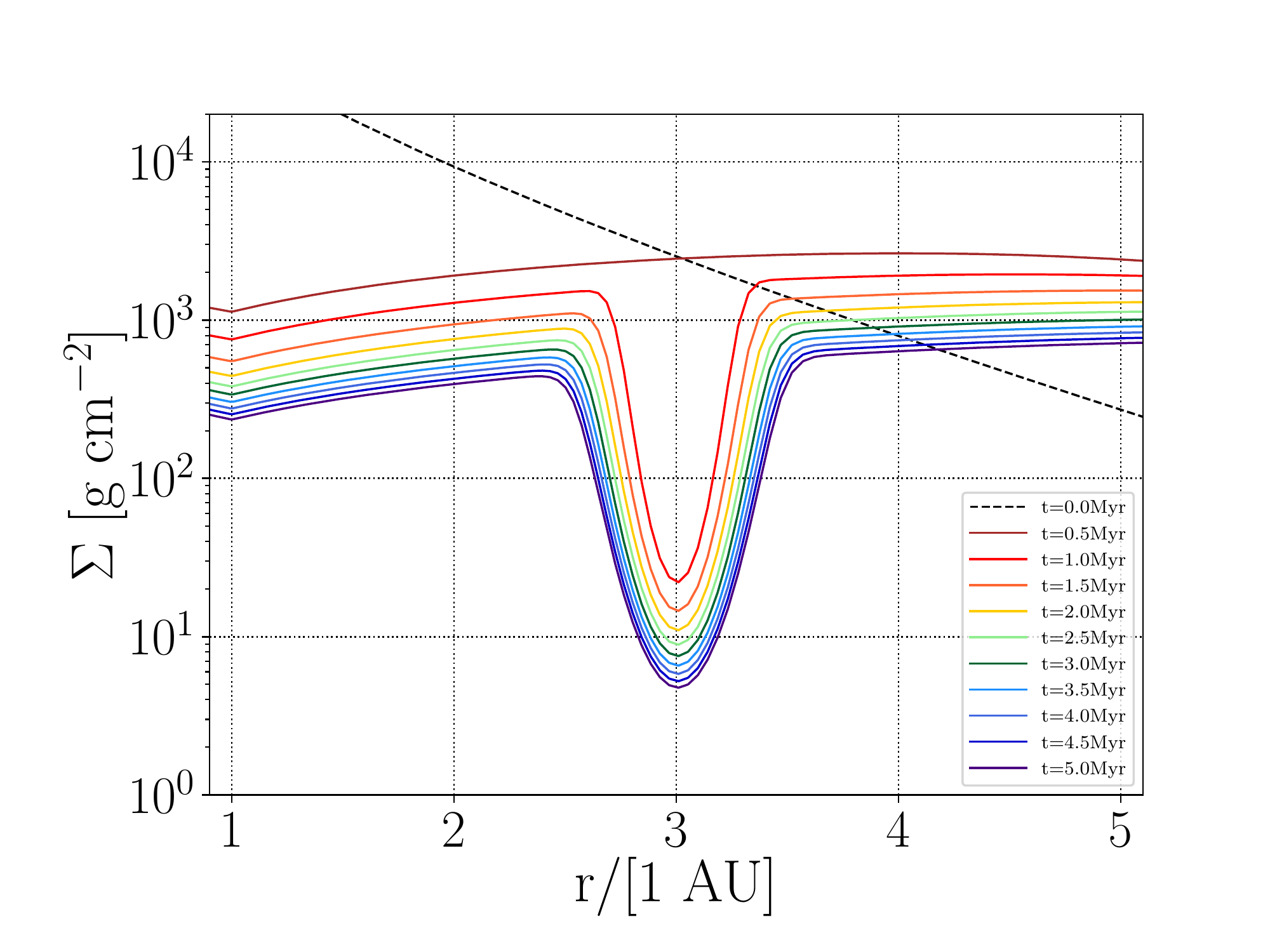}
 \caption{Surface densities at various times in the disk's evolution (colors denote the same times as in Figure 1),
 in the 1 - 5 AU region. Jupiter is assumed to form at 3.0 AU and start to open a gap starting at 0.6 Myr. }
 \label{fig:Sigmazoom}
\end{figure}
\end{centering}

Because chondrites and other planetary materials form in the inner disk, we show in Figure 2 a zoomed-in view
of the 1 - 5 AU region. Note that the axis measures heliocentric distance directly, rather than its logarithm as in
Figure 1.  In our models, the viscosity of gas in Jupiter's feeding zone is increased once it forms, and soon after
0.6 Myr gas in its vicinity viscously spreads rapidly. 

%
%
\begin{centering}
\begin{figure}[ht]
 \centering
 \includegraphics[width=.8\linewidth]{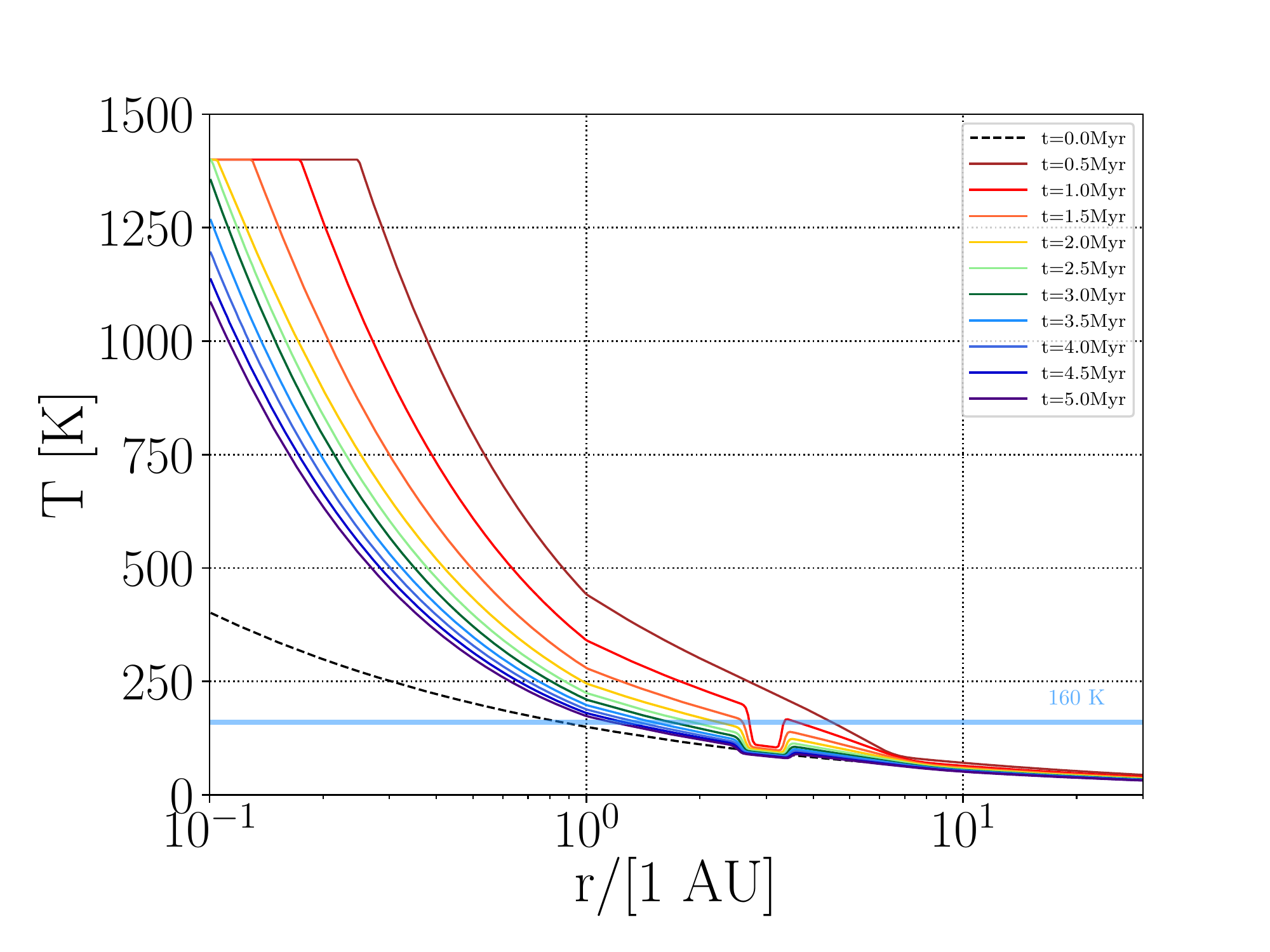}
 \caption{Temperatures at various times in the disk's evolution.  The black dashed curve denotes a passive disk
 temperature profile $T = 160 \, (r / 1 \, {\rm AU})^{-3/7} \, {\rm K}$, for comparison. Colors refer to the same times as in Figure 1.
 The passive disk profile always applies beyond 7 AU, and accretional heating tends to always dominate inside several AU.  
 The horizontal dashed line marks $t = 160 \, {\rm K}$, the approximate temperature at which water ice sublimates. 
 After Jupiter opens a gap at 3 AU at 0.6 Myr, the snow line migrates inward from 4.3 AU, through the 2 - 3 AU region.}
 \label{fig:Temperature}
\end{figure}
\end{centering}

Figure~\ref{fig:Temperature} shows the evolution of $T(r,t)$ in the disk.
The colored curves denote the same times in disk evolution as in Figure~\ref{fig:Sigma}.
An additional dashed curve is drawn depicting the profile $T(t) = 160 \, (r / 1 \, {\rm AU})^{-3/7} \, {\rm K}$,
which is the passively heated disk profile of Chiang \& Goldreich (1997) for a solar luminosity $\approx 0.7 \, L_{\odot}$,
at a time $t \approx 2.5$ Myr.
Inside about 5 AU, temperatures in the disk exceed this value, showing that accretional heating is significant.
Accretional heating remains significant inside 3 AU throughout the disk evolution.
The persistence of accretional heating is ultimately due to the high optical depths and surface densities in the inner
disk, which remain high because our assumed viscosity is low. 
Inside about 1 AU, temperatures are high enough, $T > 1400$ K, to allow CAI formation. 
We call this region, hot enough to vaporize ferromagnesian silicates but allow for growth of more refractory particles,
the ``CAI Factory," following Cuzzi et al.\ (2003). 
We cap the temperature at 1400 K to account for the fact that at higher temperatures the silicates would vaporize, 
greatly decreasing the opacity.

The CAI Factory extends out to a radius $r_{\rm CAI} = 1.40$ AU at $t = 0.01$ Myr, but moves inward as disk mass is lost 
and accretion diminishes: 
$r_{\rm CAI} = 0.37$ AU at $t = 0.25$ Myr, 
$r_{\rm CAI} = 0.25$ AU at $t = 0.50$ Myr, 
$r_{\rm CAI} = 0.22$ AU at $t = 0.75$ Myr, and 
$r_{\rm CAI} = 0.17$ AU at $t = 1.00$ Myr. 
As time evolves, the mass of gas and solids in the CAI Factory diminishes and is located ever closer to the Sun,
so the likelihood of chondrites capturing CAIs from this region decreases, although it does not vanish until 
about $t = 2.5$ Myr, when $r_{\rm CAI}$ decreases below 0.10 AU.

%
%
\begin{centering}
\begin{figure}[ht]
 \centering
 \includegraphics[width=.8\linewidth]{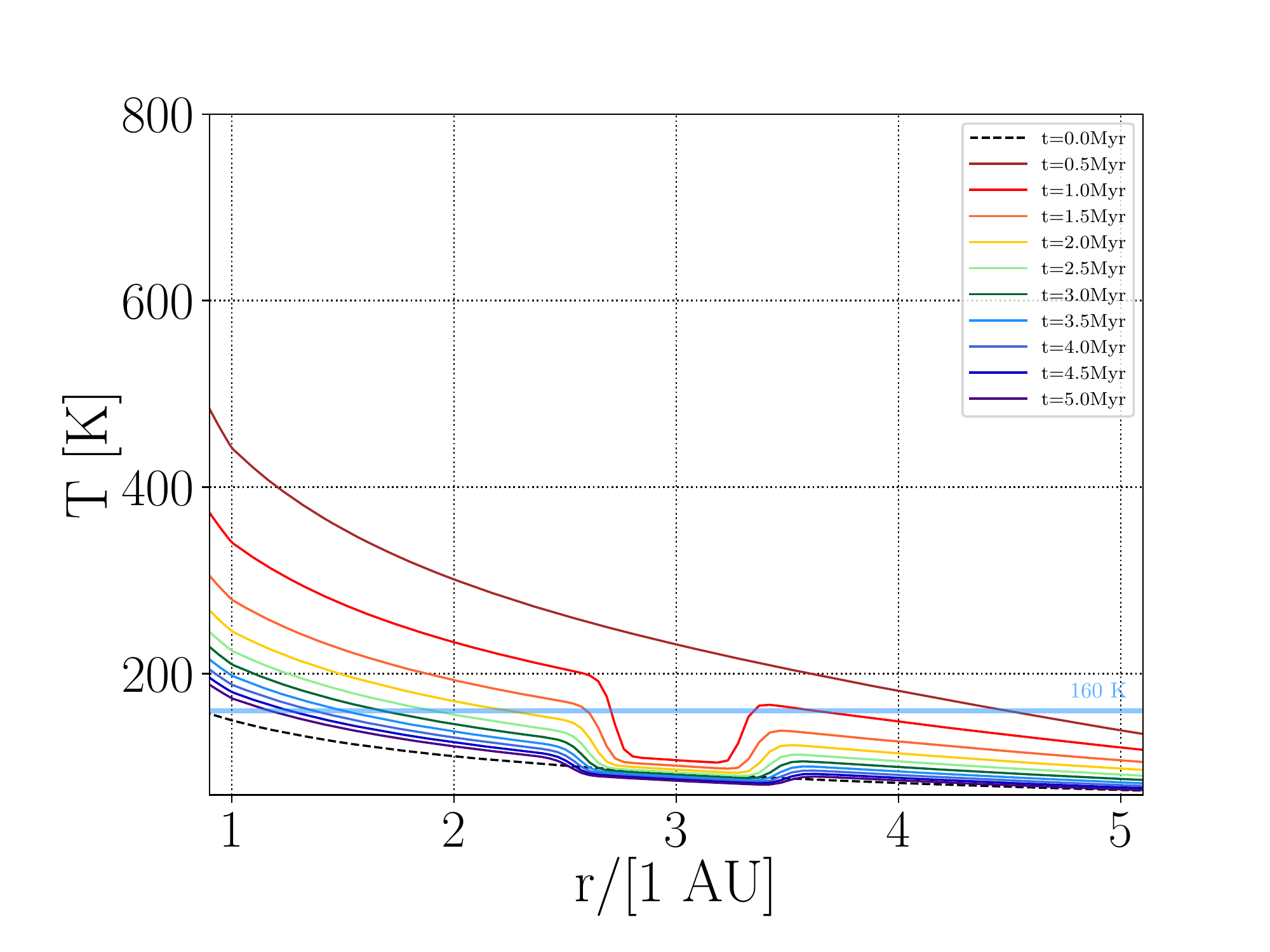}
 \caption{Temperatures at various times in the disk's evolution, in the 1 - 5 AU region.  Colors have the same meanings
  as in Figure 3. Note that both heliocentric distance and temperature are measured on a linear scale.  The dashed
  curve again refers to a profile in a passively heated disk, and the horizontal dashed line marks $T = 160$ K; 
  water ice sublimates at temperatures between 161 and 169 K in our models.  After Jupiter opens a gap at 3 AU at 0.6 Myr, 
  the snow line migrates inward from 2.70 AU at 1.0 Myr, to 2.08 AU at 2.0 Myr. }
 \label{fig:Temperaturezoom}
\end{figure}
\end{centering}

In Figure~\ref{fig:Temperaturezoom} we present a zoomed-in view of the temperature in the 1-5 AU region.
Several behaviors become clearer at this scale. For example, accretional heating is significant throughout
this region inside about 4 AU at $t < 1$ Myr.
Figure~\ref{fig:Temperaturezoom} demonstrates the behavior of the snow line, the radius $r_{\rm snow}$ inside of 
which temperatures are high enough for water ice to sublimate.
The vapor pressure of water at low temperatures ($T < 160 \, {\rm K}$) is 
$\log_{10} \left[ P_{\rm vap}(T) / (1 \, {\rm Pa}) \right] = -(3059 \, {\rm K}) / T + 14.88$
(Mauersberger \& Krankowsky 2003).
For typical conditions, $\Sigma_{\rm g} \approx 1000 \, {\rm g} \, {\rm cm}^{-2}$ and water makes up a fraction 
$\approx 0.57\%$ of the mass, so when 50\% of the water has condensed, 
and $r \approx 3 \, {\rm AU}$, we find $P_{\rm H2O} =  P_{\rm vap}(T)$ at $T = 167 \, {\rm K}$. 
This line moves in from a maximum distance of about 4.7 AU at 0.3 Myr, to about 4.1 AU at 0.6 Myr, when Jupiter forms.
It is natural that Jupiter would form around the snow line (Stevenson \& Lunine 1988), 
and formation just inside the snow line is predicted in some models (Ida \& Guillot 2016; Schoonenberg \& Ormel 2017). 
The chemistry of Jupiter also suggests it accreted carbonaceous material but not abundant ice (Lodders 2004),
inside the snow line. 
Because Jupiter forms inside the snow line, little water exists in the disk interior to Jupiter, 
and as it opens a gap and induces a reversal of the pressure gradient beyond it, it prevents the 
influx of large ($> 100 \, \mu{\rm m}$) grains that might bring in water later as the inner disk cools. 
This is the hypothesis posed by Morbidelli et al.\ (2016) for why the inner disk is so dry,
and our parameters and disk model are consistent with this scenario.
After Jupiter forms and depletes the area around it of gas, and also lets the 2 - 3 AU region drain, the 
snow line moves inward.
At 1 Myr it is at 2.70 AU, reaching 2.08 AU at 2.0 Myr and 1.61 AU at 3.0 Myr.

%
%
\begin{centering}
\begin{figure}[ht]
 \centering
 \includegraphics[width=.8\linewidth]{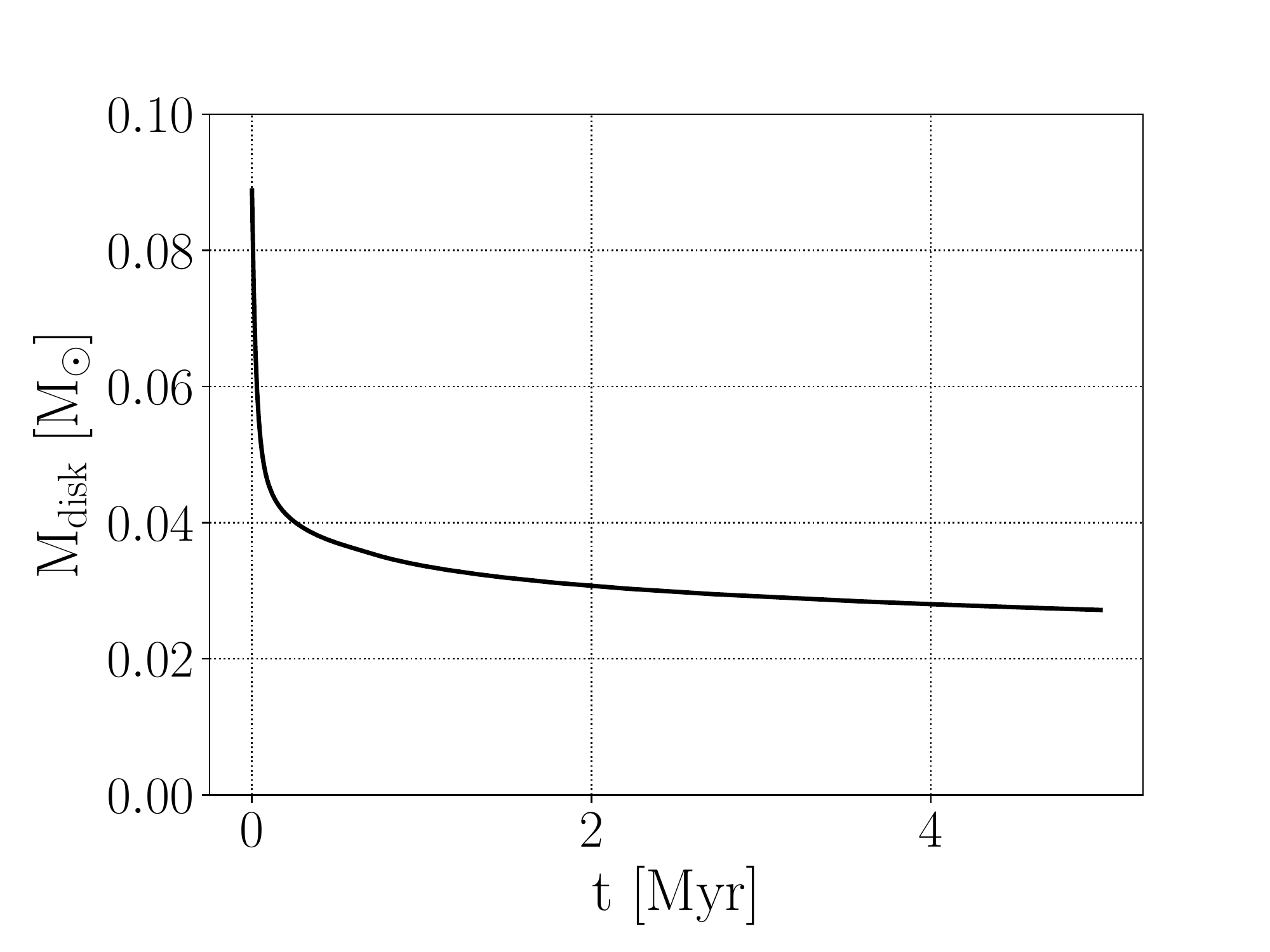}
 \caption{Evolution of the disk mass over time. The first few $\times 10^5$ yr are marked by a very active inner
  disk with high mass accretion rates $> 10^{-7} \, M_{\odot} \, {\rm yr}^{-1}$, falling to 
  $< 10^{-8} \, M_{\odot} \, {\rm yr}^{-1}$ after the inner disk is depleted,
  and $\sim 10^{-9} \, M_{\odot} \, {\rm yr}^{-1}$ by 4 Myr. }
 \label{fig:mass}
\end{figure}
\end{centering}
%

Figure~\ref{fig:mass} shows the mass in the disk as a function of time. 
At early times, the inner disk is very massive and feeds the star at a high accretion rate, 
In the first 
$\approx 0.2$ Myr, 
the disk loses $\approx 0.05 \, M_{\odot}$ at an average mass accretion rate 
$\sim 2 \times 10^{-7} \, M_{\odot} \, {\rm yr}^{-1}$. 
As material is lost to the Sun, the surface density decreases and temperatures and viscosity also decrease.
After about 0.2 Myr, the disk tends to evolve on the viscous timescale of the regions outside the CAI Factory, 
and accretes at slower rates $\approx 10^{-9} - 10^{-8} \, M_{\odot} \, {\rm yr}^{-1}$. 

%
%
\begin{centering}
\begin{figure}[ht]
 \centering
 \includegraphics[width=.8\linewidth]{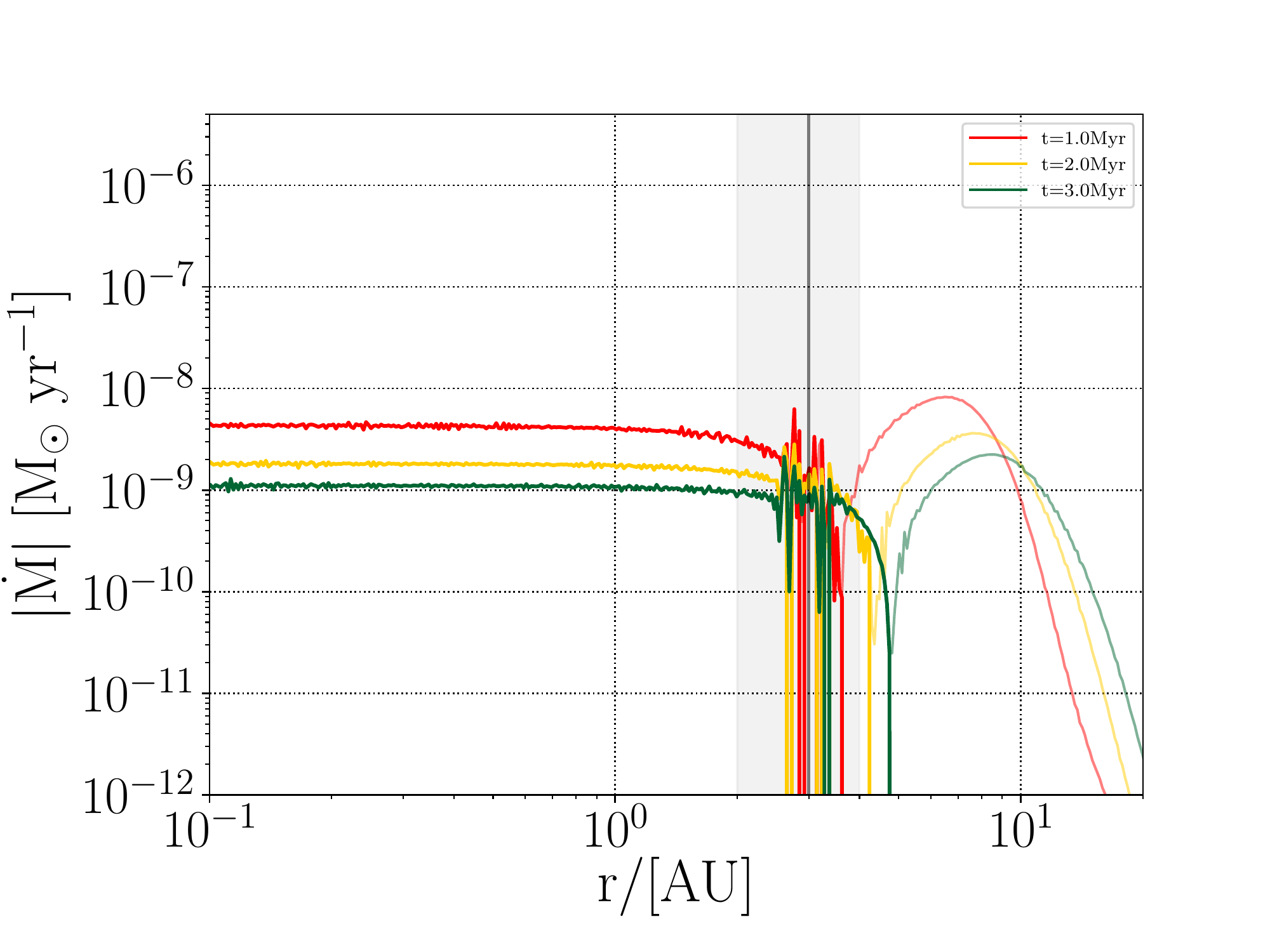}
 \caption{Mass accretion rates in the disk at three different times in the evolution of the disk: 
 1.0 Myr (red), 2.0 Myr (yellow) and 3.0 Myr (green). Thicker portions of each curve denote zones ($r < 2$ AU)
 where the mass accretion rate is inward; lighter portions ($r > 4$ AU) denote where the flow is outward.  
 The shaded region from 2 to 4 AU denotes regions close to Jupiter, at 3 AU.}
 \label{fig:massflow}
\end{figure}
\end{centering}
%

Figure~\ref{fig:massflow} shows the mass flows in the disk as a function of time. 
The initial self-similar disk profile by construction sees mass flow switch from inward accretion to outward 
decretion at the ``transition radius", located at $R_{\rm t} = \left[ 2 (2-\gamma) \right]^{-1/(2-\gamma)} \, R_1$
(Equation 3).
For our case, $\gamma = 15/14$ and $R_1 = 1$ AU, yielding $R_{\rm t} = 0.52$ AU at $t = 0$ Myr.
Inside this radius, mass moves inward, and outside of it it moves outward. 
The transition radius tends to move outward with time in a self-similar disk, and in our simulations lies at about 
2 AU at 0.5 Myr.
Inside of this radius, mass flow is inward and uniform at just over $1 \times 10^{-8} \, M_{\odot} \, {\rm yr}^{-1}$. 
Outside of this radius, mass flow is outward.
After the formation of Jupiter at 0.6 Myr, the two halves of the disk begin to evolve independently. 
The inner disk behaves more or less like an accretion disk with fixed inner and outer boundaries, and evolves to a 
steady-state structure with transition radius in Jupiter's gap, just inside 3 AU.
As the mass is depleted from this region, the surface density decreases in magnitude, but the structure remains the
same. The mass accretion rate onto the Sun decreases steadily, reaching $1 \times 10^{-9} \, M_{\odot} \, {\rm yr}^{-1}$
by 3 Myr.
These mass accretion rates onto the central star are completely consistent with observations of T Tauri star 
accretion rates (Hartmann et al.\ 1998). 
The way we have constructed the gap, flow of gas and small ($\sim 1 \, \mu{\rm m}$) dust across the gap is possible, 
but Jupiter accretes a significant fraction but not all of the material entering the gap.
Meanwhile, the outer disk also behaves more or less like a disk with fixed inner and outer boundaries.
Some mass very close to Jupiter moves inward, feeding it at rates $\sim 10^{-9} \, M_{\odot} \, {\rm yr}^{-1}$, 
but mostly the mass in the outer disk moves radially outward, in a non-uniform fashion.
The steadily outward flow of gas throughout the outer disk suggests that materials formed in the inner disk,
such as fragments of chondrules and CAIs, can easily find themselves mixed into the outer disk, possibly explaining 
the presence of such objects in the Stardust samples (Zolensky et al.\ 2006). 

%
%
\begin{centering}
\begin{figure}[ht]
 \centering
 \includegraphics[width=.8\linewidth]{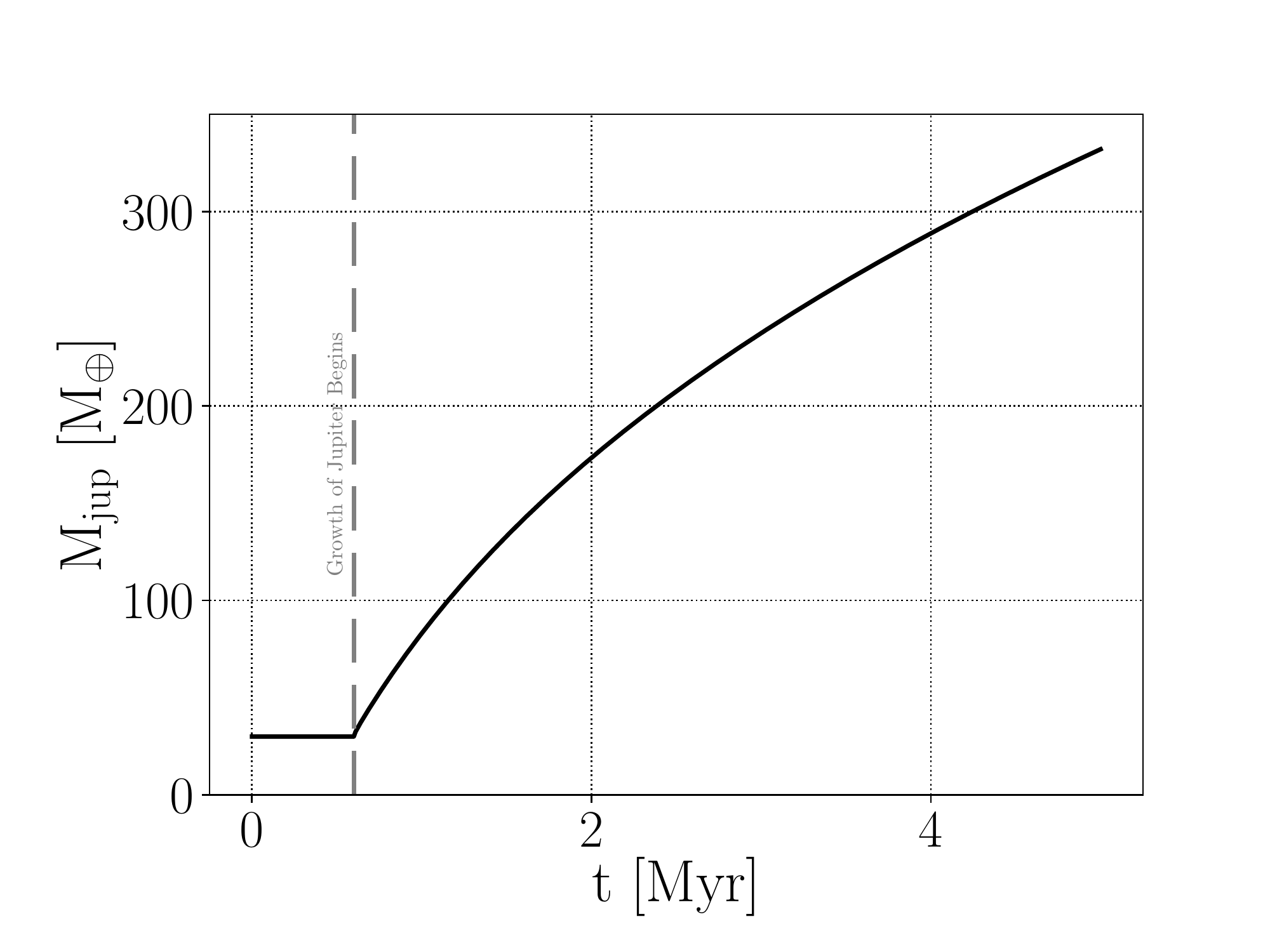}
 \caption{Mass of Jupiter as a function of time. Jupiter is assumed to grow instantaneously to 
  $30 \, M_{\oplus}$ at 0.6 Myr and then start to accrete gas within its Hill radius on a timescale
  $\sim 1 \times 10^5$ yr.  Jupiter grows to its final mass $318 \, M_{\oplus}$ just as the nebula
  presumably dissipates at about 4.5 Myr.}
 \label{fig:jupiter}
\end{figure}
\end{centering}
%

Figure~\ref{fig:jupiter} shows the growth of Jupiter as a function of time. 
After Jupiter is assumed to instantaneously reach $30 \, M_{\oplus}$ at 0.6 Myr, it accretes gas quickly, 
reaching $83.1 \, M_{\oplus}$ by 1 Myr, $173.3 \, M_{\oplus}$ by 2 Myr, $237.4 \, M_{\oplus}$ by 3 Myr,
$288.7 \, M_{\oplus}$ by 4 Myr, and $332.0 \, M_{\oplus}$ (1.04 Jupiter masses) at 5 Myr. 
By the end of the simulation, Jupiter is growing roughly as $t^{0.6}$.
We presume the disk dissipates sometime around 5 Myr, leaving Jupiter with its final mass.

%
%
\begin{centering}
\begin{figure}[ht]
 \centering
 \includegraphics[width=.8\linewidth]{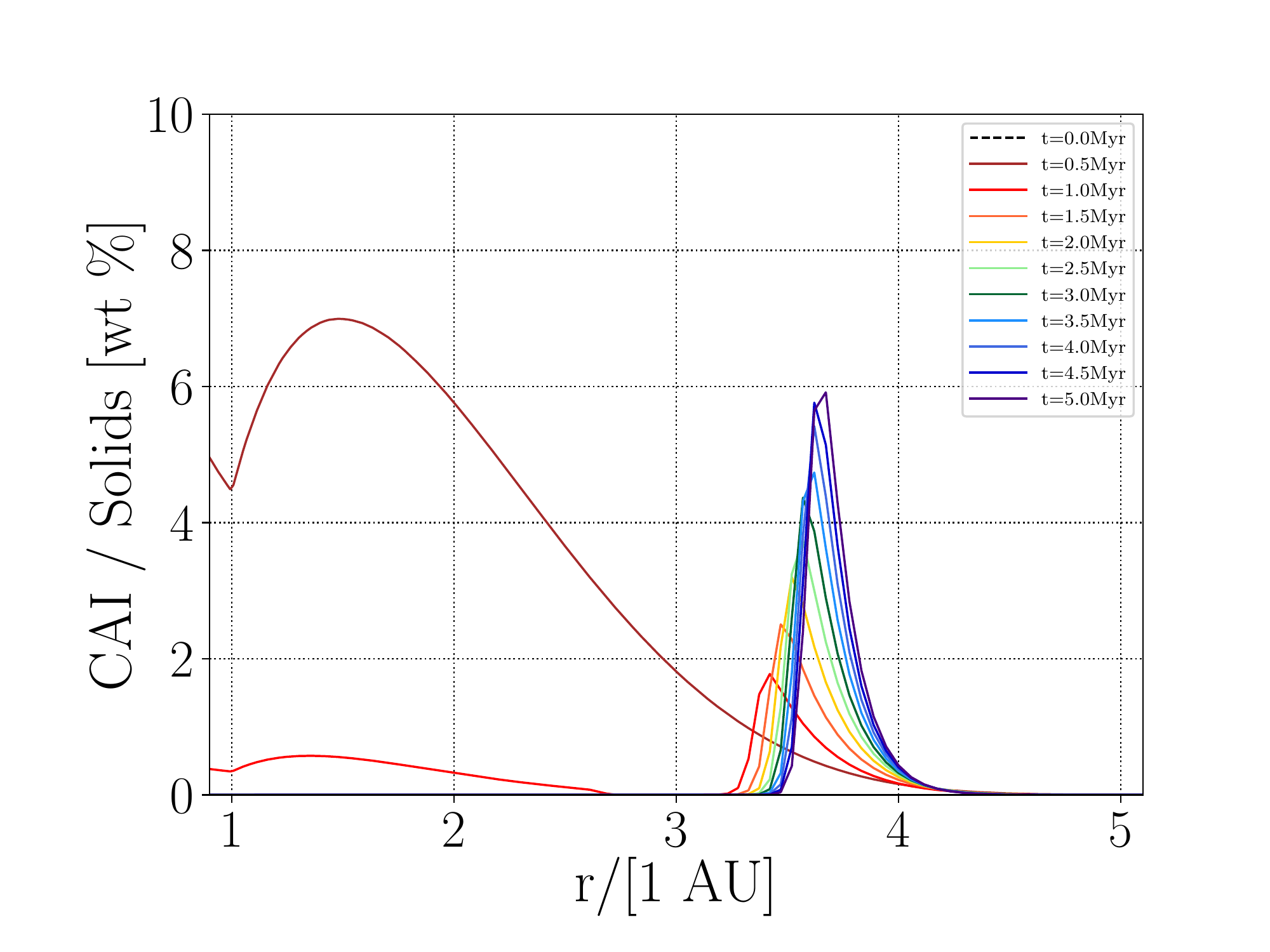}
 \caption{Abundances of CAIs (in wt\%, relative to all rocky solids in meteorites) at various times in nebular evolution, 
  in the 1 - 5 AU region.  Colors refer to the same times as in Figure 1. By 2.5 Myr, the CAI abundance is 
  $< 0.1\%$ beyond 2 AU, except in the pressure bump at 3.5 - 3.75 AU, where it climbs to several percent.  }
 \label{fig:cais}
\end{figure}
\end{centering}
%

Figure~\ref{fig:cais} shows the distribution of CAIs in the disk at various times.  
Note that we report mass fractions, and therefore wt\%. CAI fractions are usually reported in vol\%.
In what follows, we assume that CAIs have the same density as their host chondrites, but if, for example, CAIs
had densities 10\% lower than the bulk chondrite, the volume fraction would be increased by about 10\%, e.g.,
from 3.0\% to 3.3\%.
In all the plots pertaining to refractory and CAI abundances, we assume that all CAIs have radius $2500 \, \mu{\rm m}$.
The disk starts with no CAIs, but immediately after $t = 0$, a fraction of the solid material inside the CAI Factory 
is converted into CAIs that then diffuse throughout the disk.
By 0.5 Myr, CAIs are abundant and have distributed themselves throughout the disk in an approximate power law,
with an abundance (relative to all solids) of 7.1\% at 1.5 AU, falling to 1.9\% at 3 AU, 0.2\% at 4 AU, and 
$< 10^{-4}$ at 5 AU.
The mass of CAIs beyond 3 AU increases steadily until about 0.13 Myr, peaking at $0.32 \, M_{\oplus}$, then falling
to $0.048 \, M_{\oplus}$ by 0.6 Myr, plateauing at that level for the rest of the simulation. 
The concentration of CAIs caught in the pressure trap increases over time as CAIs far out in the disk spiral into the 
trap, and as small solids are lost along with gas from this region. 
The peak concentrations of CAIs in the pressure bump increase from 3.4\% at 2.0 Myr, 4.7\% at 3.0 Myr, and 5.9\% at 4.0 Myr, 
consistent with the total mass of CAIs beyond 3 AU being conserved. 
In contrast, the mass of CAIs inside 3.0 AU at 0.13 Myr is greater, $1.17 \, M_{\oplus}$, but it falls steadily
as CAIs spiral into the Sun, reaching $0.35 \, M_{\oplus}$ at 0.6 Myr, eventually decreasing to $< 2 \times 10^{-6} \, M_{\oplus}$ 
at 2 Myr.
The concentrations at 2 AU drop from 0.36\% at 1 Myr, to 0.02\% at 1.25 Myr. 
It is extremely significant that CAIs are not removed from the pressure bump, even as they are lost from
essentially every other region.  This is essentially the resolution of the CAI storage problem.

%
%
\begin{centering}
\begin{figure}[ht]
 \centering
 \includegraphics[width=.8\linewidth]{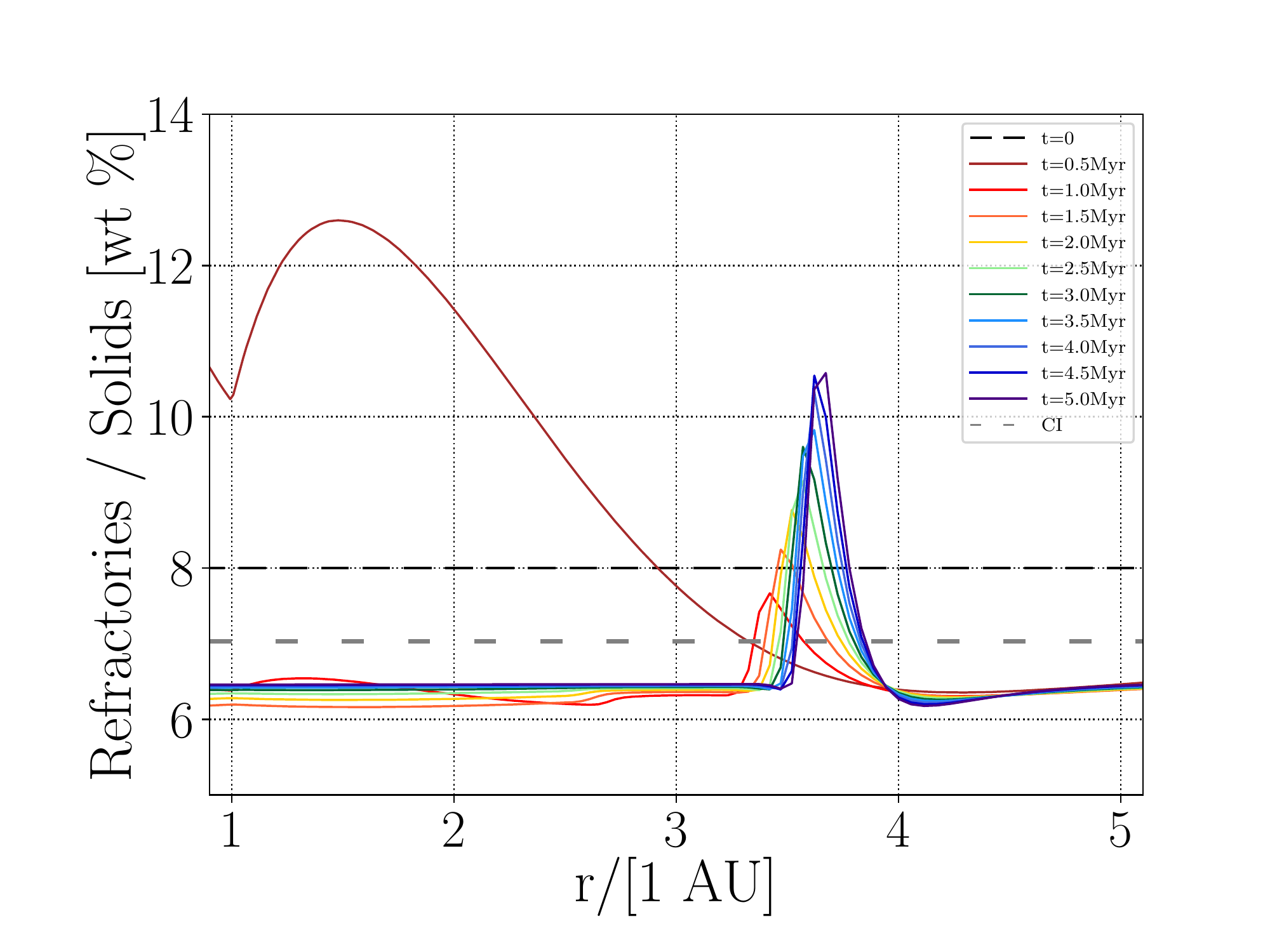}
 \caption{Abundances of refractory materials (relative to all rocky solids) at various times in nebular evolution,
  in the 1 - 5 AU region. Colors refer to the same times as in Figure 1. The assumed abundance of refractories
  in a solar nebula composition, 8\%, is drawn as the dark horizontal dashed line, and the assumed abundance of 
  CI chondrites, 7.03 wt\%, is drawn as the lighter horizontal dashed line.}
 \label{fig:refractories}
\end{figure}
\end{centering}

Figure~\ref{fig:refractories} shows the abundance of refractories at various times in nebular evolution.
This is calculated in each zone by taking 8\% of the mass of the material with original (solar) composition, 
adding the mass of CAIs, and comparing to the mass of all solids. 
Because we have imposed the constraint that only 60\% of the original refractory material can be converted into
large CAIs, a baseline amount of refractories exists throughout the disk.
In the 2 - 3 AU region, much material has been converted into CAIs, and the baseline refractory abundance is about
6.4\%, or about $0.90 \times$ CI. 
In the outer disk, there is less mixing with material that has seen temperatures $> 1400$ K, and the baseline 
amount tends to be slightly higher at 5 AU, eventually climbing to the original 8\% at many tens of AU.
For example, at 1.5 Myr, the refractory abundances rise from 7.07\% at 10 AU, to 7.31\% at 15 AU, with disk edge at 16 AU.
At 3.0 Myr, the refractory abundances rise from 6.77\% at 10 AU, to 7.03\% at 15 AU, to 7.04\% at the disk edge at 19 AU.
At 4.5 Myr, the refractory abundances rise from 6.68\% at 10 AU, to 6.90\% at 15 AU, to 6.93\% at the disk edge at 22 AU.
For reasons we discuss below (\S 5.4.4),
we take CI chondrites to form at about 3.0 Myr at 15 AU, where the refractory 
abundance is 7.03\%, a depletion of about 12\% relative to the solar composition. 
Refractory abundances higher than these baseline amounts are due to inclusion of CAIs.
In the 2 - 3 AU region, at 0.6 Myr, CAIs are still reasonably abundant, especially closer to the Sun.
The refractory abundance varies from about 10.3\% $= 1.3 \times$ CI at 2.0 AU to 8.0\% $= 1.0 \times$ CI at 2.7 AU.  
As time goes on, the CAIs drain from the 2 - 3 AU region and the refractory abundance decreases, and the 
gradient begins to reverse. 
By 2.0 Myr, the refractory abundance varies from 6.27\% $= 0.89 \times$ CI at 2.0 AU, to 6.38\% 
$= 0.91 \times$ CI at 2.7 AU.
From 2 to 3 Myr, the 2 - 3 AU region is marked by depletions of refractories relative to CI chondrites, mostly 
because some of the refractory material has been wrung out of it to make CAIs.
In contrast, the refractory abundance in the pressure bump at 3.6 AU is enhanced over a CI composition,
because of the concentration of CAIs there, 
and the peak refractory abundances there increase over time, from $1.31 \times$ CI at 2.0 Myr, to $1.60 \times$ CI at 4.0 Myr.

\subsection{Convergence Test} 

Because the code is explicit and must satisfy a Courant condition including diffusion, the time to run a simulation
tends to scale as the cube of the number of radial zones.
A run with 300 zones takes about a day to perform on a laptop, but 450 zones takes about 4 days, and 500 zones takes 
about 10 days. 
It is therefore important to assess the number of zones needed for numerical convergence. 
The simulations presented above were run with 450 zones, but we also conducted runs with 200, 300, and 500 zones. 
For the case with 200 zones, numerical diffusion of the CAI fluid out of the pressure bump was significant, and
the CAI abundances were significantly lower (e.g., 1 - 2\% instead of 3 - 4\%).
We saw no qualitative changes in the outcomes with increasing numerical resolution once we used 300 zones and 
at least a dozen zones over the pressure bump.
Most outputs (the position of the snow line, surface densities, etc.) changed only at the percent level when 
increasing the resolution from 300 to 450 to 500 zones. 
The peak CAI abundance in the pressure bump was the output most sensitive to the numerical resolution, but 
the results using 300, 450, and 500 zones are within 10\% of each other.
We therefore consider cases with 450 zones to be sufficiently resolved. 

\subsection{Sensitivity Analysis}

The simulations above make clear that the CAI abundances in CCs are just explained in our model if
there is trapping of CAIs in the pressure bump beyond Jupiter, formed at 0.6 Myr, at 3.0 AU. 
Significant outward transport of CAIs from meridional transport is also a key feature of our model. 
In what follows, we try to assess the sensitivity of our results to each of these effects.
This analysis will refer to a suite of cases we ran just like the above, but with $\alpha = 1.25 \times 10^{-4}$ 
in the inner disk, and a mix of CAIs such that 75\% (by mass) have radii $1600 \, \mu{\rm m}$ and 25\% 
have radii $800 \, \mu{\rm m}$.

{\it Jupiter Opening Gap}:
For the survival of CAIs in the disk, the opening of a gap by Jupiter is paramount.
In our runs in which Jupiter does not form and does not open a gap, 
the surface density profile resembles that of Figures 1 and 2, but without the reduction in $\Sigma$ around Jupiter's location.
The pressure gradient force remains monotonically outward.
Particles are never trapped anywhere in the disk, spiralling in to the Sun instead.
After an initial burst of production of CAIs inside 1 AU by 0.2 Myr, CAIs diffuse and are transported throughout the disk, 
reaching appreciable levels beyond 3 AU by 0.5 Myr.
After that, though, their abundances decrease steadily at all radii.
The evolution inside 3 AU is largely the same with or without Jupiter: CAIs decrease in abundance, largely disappearing by 2 Myr.
Outside 3 AU CAIs disappear even sooner.  At no place in the disk would CAI abundances reach $> 3$ wt\%. 
Significantly, the concentration of CAIs decreases monotonically outward, so that CCs would be predicted
to have significantly fewer CAIs than OCs and ECs.
The opening of a gap by Jupiter, and the subsequent reversal of the pressure gradient, is the key to resolving the CAI storage
problem.

{\it Location of Jupiter's Formation}:
In our canonical model, Jupiter forms at 0.6 Myr at 3.0 AU. 
A formation at 3.0 AU (or a formation at 4-5 AU, followed by migration to 3 AU and then a cessation of migration)
is unexpected, as Jupiter orbits today at 5.2 AU. 
We have also conducted runs with the canonical parameters in which Jupiter formed at 4.0 and 5.2 AU.
Because CAIs are produced inside 1 AU and must be transported by diffusion or meridional transport outward, 
formation of Jupiter farther from the Sun significantly decreases the number of CAIs trapped in the pressure bump
beyond Jupiter.  
In runs where Jupiter forms at 3.0 AU, the CAI abundance (as a fraction of all non-volatile solids) at 3 Myr is 
maximized at about 3.6 AU, where it is about 3.4 wt\% at that time. 
For comparison, forming Jupiter at the same time but at 4.0 AU leads to the CAI abundance at 3 Myr in the pressure 
bump (at about 4.8 AU) to be no higher than 0.4 wt\%.
Forming Jupiter at 5.2 AU leads to the CAI abundance at 3 Myr in the pressure bump (at about 6.3 AU) to be no higher 
than $\sim 0.01$ wt\%.
The CAI abundance in the pressure bump increases by about an order of magnitude for each AU it forms closer to the
Sun than 5.2 AU. 
To comply with meteoritic constraints, the CAI abundance in the pressure bump cannot decrease by more than about 25\%.
Formation of Jupiter farther than about 3.2 AU would lead to too few CAIs in the region where CCs formed.
Formation inside 3.0 AU would lead to more CAIs in CCs, but would also deplete the protoplanetary disk in the asteroid belt 
region. 
Our model considers several meteorite types, including L chondrites and ureilites to form at about 2.6 - 2.8 AU.
We conclude that Jupiter had to form between about 2.9 and 3.2 AU.

{\it Time of Jupiter's Formation}:
In our model, CAIs are produced very early in disk evolution and transported outward very rapidly before they drift relative 
to the gas and spiral inward.
As result, the earlier Jupiter forms, the more CAIs can be trapped in the pressure bump beyond it. 
From their isotopic analyses involving ${}^{182}{\rm W}$, Kruijer et al. (2017) inferred that Jupiter formed at times between 
0.4 and 0.9 Myr. 
Figure 8 illustrates that the CAI fraction at 3.0 AU essentially drops by a factor of 2 every 0.5 Myr. 
Repeating our calculations with Jupiter forming at different times, we find results consistent with this trend. 
If Jupiter formed at 0.4 Myr, the peak CAI abundances at 3.0 Myr would be 4.8 wt\%, instead of 3.0 wt\% if Jupiter forms at 0.6 Myr.
These drop to 2.2 wt\% if Jupiter forms at 0.8 Myr, or 1.7 wt\% if Jupiter forms at 0.9 Myr. 
We conclude that Jupiter must form and open a gap no later than about 0.7 Myr, if meteoritic constraints are to be met. 
Formation at an earlier time would lead to higher CAI abundances and would more easily match meteoritic constraints.

{\it Alpha Profile}:
We have assumed a very particular variation of the turbulence parameter $\alpha$ with heliocentric distance $r$, and the
sensitivity to this parameter must be assessed. 
The runs we have presented assume $\alpha = \alpha_{\rm inner} = 5 \times 10^{-4}$ for $r \leq r_{\rm inner} = 1 \, {\rm AU}$, 
$\alpha = \alpha_{\rm outer} = 1 \times 10^{-5}$ for $r \geq r_{\rm outer} = 10 \, {\rm AU}$, and varying as a power law 
in between, i.e., falling as $\alpha = 5 \times 10^{-4} \, (r / 1 \, {\rm AU})^{-1.699}$ for $1 \, {\rm AU} < r < 10 \, {\rm AU}$.
When cast in this way, a remarkably large parameter space could be explored, varying $\alpha_{\rm inner}$, $\alpha_{\rm outer}$,
$r_{\rm inner}$ and $r_{\rm outer}$.
In addition, we held the values of $\alpha(r)$ to be constant throughout the simulation, but one could imagine varying these 
in time as well. 
It is impossible to exhaustively search all of parameter space. Instead, we tried to find a profile of $\alpha(r)$ that 
conformed with expectations about the disk and matched meteoritic constraints. 

We chose $\alpha_{\rm outer} = 1 \times 10^{-5}$ because mixing in the disk is too pervasive if $\alpha$ has 
higher values than this in the outer disk. Even for this low value of $\alpha$, we predict CI chondrites will be depleted 
by 12\% relative to the Sun; higher values of $\alpha_{\rm outer}$ would lead to greater depletions that would violate 
observational constraints. 
Lower values of $\alpha_{\rm outer}$ are not plausible, as even hydrodynamic instabilities will lead to turbulence at these
levels or greater.

We chose $\alpha_{\rm inner} = 5 \times 10^{-4}$ because it allowed a high fraction of CAIs in the pressure bump and also 
satisfied many meteoritic constraints on the evolution of the snow line.
The greater $\alpha_{\rm inner}$ is, the greater mass of CAIs is transported to beyond 3 AU by the time Jupiter forms.
For the mix of CAI sizes considered above, we have run cases with $\alpha_{\rm inner} = 1.25 \times 10^{-4}$, $1.5 \times 10^{-4}$,
$2 \times 10^{-4}$, and $3 \times 10^{-4}$; the peak CAI mass fraction in the pressure bump at 3 Myr for these $\alpha_{\rm inner}$
are 2.3\%, 3.2\%, 5.2\%, and 9.4\%, respectively. 
The peak abundances of CAIs in the pressure bump region are therefore very sensitive to $\alpha_{\rm inner}$, partly because
the diffusion of CAIs is simply stronger, and partly because the $d \ln ( \Sigma \alpha ) / d \ln r$ term in the meridional 
transport equation (Eq.~20) has greater magnitude.
Other disk properties depend on $\alpha_{\rm inner}$, but not as sensitively. 
Counterintuitively, larger $\alpha_{\rm inner}$ leads to lower temperatures in the 2-3 AU region of inner disk in the 2-3 Myr time
interval when the snow line is sweeping through the asteroid belt region. 
This is because accretional heating depends on both $\Sigma$ and $\alpha$. Larger $\alpha$ would lead to more heating, but it also 
drains the inner disk of mass more rapidly. 
For $\alpha_{\rm inner} = 5 \times 10^{-4}$, the snow line reaches inside 2.1 AU at 2.0 Myr, whereas for 
$\alpha_{\rm inner} = 1.25 \times 10^{-4}$, the snow line would only reach inside 2.1 AU at 4.0 Myr.
To meet the constraints that ECs formed inside the snow line at around 1.9 - 2.1 AU at about 2.0 Myr, the largest permissible
value of $\alpha_{\rm inner} = 5 \times 10^{-4}$.
We set $\alpha_{\rm inner}$ at this highest possible value, both to maximize the number of CAIs in the outer disk
and to allow the snow line to sweep in to 2 AU by 2 Myr, so that OCs can accrete water.
We then chose the largest plausible CAI radius, so that the abundances in the outer disk are brought down to their observed values,
and allowing the inner disk to be drained of CAIs. 
Values of $\alpha_{\rm inner}$ between $1 \times 10^{-4}$ and $5 \times 10^{-4}$ all provide a reasonable fit to the 
constraints. 
At the low end of the range, small CAIs ($\approx 800 \, \mu{\rm m}$ radius) are preferred.
The ratio of CAIs in the pressure bump to the inner disk is barely high enough to match constraints (ratio $> 40$), meaning the
inner disk does not drain of CAIs.  Also, the snow line does not sweep in much into the asteroid belt.
At the high end of the range, the results are as described elsewhere in this paper: large CAIs ($\approx 2500 \, \mu{\rm m}$ radius) 
are preferred. This effectively drains the inner disk of CAIs and still allows the correct value in the pressure bump, but the snow
line moves in past EC-forming region before 2 Myr, meaning ECs would have water, violating an important constraint.

{\it Meridional Transport}: 
Finally, we have conducted a case in which CAIs are not allowed to settle to the midplane.
Instead of preferentially experiencing the midplane gas velocity, which can be outward due to meridional flow, they always
experience the vertically averaged radial velocity of the gas, which is usually inward. 
For example, from Figure 6 the net mass accretion rate is clearly seen to be inward inside 3 AU.
At 0.5 Myr, the average velocity of gas is inward, with inward velocity $-1.5 \, \nu / r$.
In the same region and time, $T$ and $\alpha$ are sharply decreasing functions of $r$, so that 
by Equation 18 the midplane gas velocity is potentially outward rather than inward. 
At 0.5 Myr, between 1 and 2 AU, the average midplane velocity is in fact $u_{\rm g,r} = +0.6 \, \nu / r$. 
Instead of experiencing an average inward velocity as vertically well-mixed micron-sized particles do, CAIs experience
an outward gas velocity carrying them from the CAI Factory out beyond Jupiter. 
The effect of meridional transport is to greatly increase the peak abundance of CAIs in the pressure bump.
The same general behaviors and spatial distribution of CAIs are seen in runs without meridional transport, but instead of 
peak abundances at 3 Myr being 3.5\%, they would be only 1.0\%, a factor of 3 smaller. 

\section{Analysis}

The disk model presented above predicts the surface density, temperatue, refractory and CAI abundances,
at each radius and time in nebular evolution. 
Assuming that planetary bodies represent snapshots in the evolution of the disk, with compositions derived
from material in that time and place, these predictions can immediately translate into information about
the compositions and histories of chondrites and achondrites.
Prompt accretion of material is a commonly invoked concept in meteoritics (Wood 2005), and is given
strong support by models of rapid planet growth (e.g., streaming instability: Johansen et al.\ 2007; 
and pebble accretion: Lambrechts et al.\ 2014) and the size distribution of asteroids (Morbidelli et al.\ 2009). 
We examine whether our model is consistent with the types of meteorites seen in our collections.
First we consider whether the model can simultaneously match the constraints on time and place of formation,
refractory lithophile abundance, and CAI abundance, for 5 types of achondrite (ureilites, HEDs, 
acapulcoite-lodranites, aubrites, and winonaites), and for the known chondrite types.
We then make predictions about the physical conditions present when they formed, and compare those starting
conditions to what is known about the chondrites. 
Finally, we show that aerodynamic sorting of chondrules by turbulent concentration into aggregates, then of 
aggregates into planetesimals by streaming instability, matches the constraints on chondrule sizes, lending
strong support to the idea that chondrites are snapshots in time of the solar nebula.

\subsection{Achondrite Parent Bodies}


Many stony achondrites formed in the first $\sim 2$ Myr of disk evolution.
Evidence strongly suggests these parent bodies were very similar to the parent bodies of chondrites,
but simply had sufficient live ${}^{26}{\rm Al}$ to melt. 
For example, many primitive achondrites contain rare relict chondrules: e.g., the acapulcoites and lodranites
(Schultz et al.\ 1982; McCoy et al.\ 1996), and winonaites (Benedix et al.\ 1998; Farley et al. 2015). 
This strongly suggests that primitive achondrites are simply chondrites that heated to the point of melting but
did not experience melt migration (Weisberg et al.\ 2006).
This interpretation is reinforced by the observation that the acapulcoites and lodranites (Patzer et al.\ 2004)
and the winonaites (Weisberg et al.\ 2006) have compositions intermediate between ECs and H chondrites. 

Given that achondrites probably formed in the same manner as chondrites and resemble chondrites, our
models of refractory element abundances should apply to them as well. 
Unfortunately, the CAI abundances of achondrites are not known (they melted), and even the starting compositions
before melting are difficult to infer. 
We could find only 5 stony achondrites with enough information about refractory abundances and time of accretion
to discuss: ureilites, howardite-eucrite-diogenites (HEDs), acapulcoites-lodranites, aubrites, and winonaites.
For these achondrites we review their bulk composition and refractory abundances, their time of accretion, and
what is known about their formation locations, to see if they are consistent with our disk model. 
In what follows, we assume all CAIs have radii of $2500 \, \mu{\rm m}$.
and we assume that CIs are depleted by 12\% relative to the starting composition. 
Although our model is adjusted to match chondrites, we show that our models are consistent with what is known 
about achondrite formation.

\subsubsection{Ureilites}

The ureilite parent body (UPB) is inferred from Al-Mg and Hf-W systematics to have accreted in
$< 1-2$ Myr (Lee et al.\ 2009), at $< 1.9_{-0.7}^{+2.2}$ Myr (Baker et al.\ 2012), 
at 1.4 Myr (van Kooten et al.\ 2017) or 1.6 Myr (Budde et al.\ 2015), 
although Wilson \& Goodrich (2016) argue that inclusion of melt migration in the thermal models implies 
the UPB accreted closer to 0.6 Myr after CAIs.
Based on their $\eCr$ anomaly and an inferred trend of $\eCr$ vs. heliocentric distance for Earth, Mars 
and Vesta, Yamakawa et al.\ (2010) inferred that the UPB formed at about 2.8 AU. 
This is consistent with their chemical similarities to CV chondrites
(Rubin 1988), which would form on the other side of Jupiter, at 3.6 AU .
The refractory lithophile abundance of the UPB is difficult to constrain, but Goodrich et al.\ (2013) suggest 
that the depletion of Mo and W and volatiles in ureilites could be explained if they accreted an ``excess" CAI component. 
On the other hand, the Almahata Sitta ureilite appears to be depleted in refractory lithophiles (Friedrich et al.\ 2010). 
Assuming the UPB formed at 2.8 AU and about 0.6 Myr, we can infer the conditions of its formation.
In forthcoming work (Desch et al., submitted), we propose that the UPB composition resembles a mix of 
63\% H chondrite, 33\% CV chondrite, and 4\% CI chondrite, not dissimilar to the mix inferred for Vesta
(Righter \& Drake 1997), and implying a refractory enrichment of $1.05 \times$ CI, like that of Vesta.

At 2.8 AU, the local conditions are strongly affected by the presence of Jupiter. 
Already by 0.6 Myr the abundance of CAIs in this region is decreasing, having peaked at 4.0\% at 0.2 Myr, and 
falling to 1.7\% by 0.6 Myr. 
As soon as Jupiter forms (or arrives) at 3.0 AU and starts to open a gap, CAIs are quickly depleted from this region,
with their abundance falling to 0.04\% by 0.7 Myr. 
Refractory abundances in this region also drop rapidly, from $1.39 \times$ CI at 0.3 Myr, to $1.10 \times$ CI abundances at 0.6 Myr,
to $0.89 \times$ CI at 0.7 Myr. 
Assuming the UPB forms at 2.80 AU at 0.603 Myr, i.e., within $3 \times 10^4$ yr of Jupiter's arrival, 
they would have refractory abundances $1.049 \times$ CI, and CAI abundances of 1.3\%, and the local 
surface density and temperature would be $\Sigma = 1920 \, {\rm g} \, {\rm cm}^{-1}$ and $T = 213$ K.
This temperature is below the ``tar line" temperature of $T = 350$ K (Lodders 2004), consistent with accretion and 
retention of carbonaceous material, but is above the snow line temperature; we predict that the UPB would not accrete ice. 
We speculate that the opening of a gap may increase the solids-to-gas ratio and trigger the formation of the UPB.
Formation of ureilites at 2.8 AU and 0.6 Myr is consistent with their bulk compositions, and their slight refractory enrichments
plus moderate CAI abundances.

\subsubsection{Howardite-eucrite-diogenites}

Vesta is a large asteroid orbiting at 2.36 AU, and is the parent body of the howardite, eucrite and diogenite (HED) 
achondrites (McCord 1970; Drake 2002). 
The time of Vesta's accretion is generally thought to have occurred within $\sim 1$ Myr of CAIs
(Schiller et al.\ 2011; Formisano et al.\ 2013; Neumann et al.\ 2014; Touboul et al.\ 2015). 
The thermal modeling of SF14                     suggests an accretion age of $0.8 \pm 0.3$ Myr. 
The initial bulk abundance of Vesta has been modeled as 70\% L chondrite and 30\% CV chondrite material 
(Righter \& Drake 1997). 
This mixture of compositions would be slightly enriched in refractories, by about 
3.5\% relative to CI chondrites.

We find that solid material is enriched in refractories by just this amount at 2.36 AU at 0.765 Myr, 
when the refractory abundance is $1.035 \times$ CI and the CAI abundance is 1.2\%.
The local surface density and temperature would be $\Sigma = 1810 \, {\rm g} \, {\rm cm}^{-2}$, $T = 244$ K.
We predict that Vesta would not have accreted ice. 
It is remarkable that for our standard parameters tuned to explain chondrite abundances, Vesta's composition is consistent with 
formation in place, at 0.77 Myr, in the middle of the time interval calculated by SF14. 
These values are consistent with what is known about Vesta from HEDs. 

\subsubsection{Acapulcoites-Lodranites}

The acapulcoites and lodranites together form a clan, based on similar characteristics.
Acapulcoites have minerals like OCs, are similar to chondrites in bulk composition, and a few 
have relict chondrules (McCoy et al.\ 1996; Mittlefehldt et al.\ 1996; Mittlefehldt \& Lindstrom 1998; Mittlefehldt 2005).
Both acapulcoites and lodranites appear to be depleted in a metal-sulfide component that would
be the first to melt in a heated asteroid, and lodranites appear to be acapulcoites additionally
depleted in a basaltic component.
Rubin (2007) suggests the precursors to the acapulcoite-lodranite clan were similar in composition to
CR chondrites, but more enriched in metal and sulfides. 
Since CR chondrites have refractory abundances $1.02 \times$ CI, this implies
a slight depletion of refractories in acaplucoites relative to CI. 
Mittlefehldt (2014) 
has reviewed the literature on the compositions of the clan's precursor materials, and concludes that they have
an abundance of refractory lithophiles (normalized to Mg, relative to CI), of $0.95 \times$ CI, 
slightly more depleted than H chondrites, which they list as $0.98 \times$ CI, referencing Wasson \& Kallemeyn (1988). 
We consider this value to have been superseded by the dataset of Rubin (2011), for which the 
the H chondrite refractory abundance is $0.899 \times$ CI.
We therefore consider the refractory abundance of acapulcoites to be closer to $0.87 \times$ CI, similar to 
that of EL chondrites. 
This is consistent with recent spectroscopic surveys of the innermost asteroid belt that find asteroids that 
spectroscopically match the acapulcoites (Lucas et al.\ 2017), suggesting a source in the innermost disk near
where ECs formed (i.e., $< 2.1$ AU).
Thermal modeling of the acapulcoite-lodranite parent body suggests it was 270 km radius and formed 
1.66 Myr after CAIs (Henke et al.\ 2014), or had radius 25-65 km and formed at 1.3 Myr (Golabek et al.\ 2014). 
Likewise, the thermal modeling of SF14 suggests an accretion time of $1.3 \pm 0.3$ Myr. 
Based on Hf-W dating, Touboul et al.\ (2009) inferred an accretion at 1.5 - 2.0 Myr.
We infer the acapulcoites formed in the innermost part of the asteroid belt, at roughly 1.3 Myr, 
with a depletion relative to CI of about 13\%. 

In our simulations, such strong depletions are not achieved in many times and places before 2 Myr. 
Between about 1.0 and 1.6 Myr, around the 2 AU region, the refractory abundance undergoes a minimum in time at about 1.3 Myr.
As a function of heliocentric distance, refractory abundances are fairly uniform or slightly decreasing with $r$ between
1.8 and 2.0 AU, then begin to rise with increasing $r$.
The very lowest refractory abundances, $0.88 \times$ CI, occur at 2.0 AU at 1.3 Myr, which is where we favor their formation.
Acapulcoites would have formed with 0.01 wt\% CAIs, in an environment with $\Sigma = 1050 \, {\rm g} \, {\rm cm}^{-2}$ 
and $T = 206$ K. 
Despite the relatively early formation time, we speculate that the acapulcoites would have had a lower abundance of CAIs
than the HED and ureilite parent bodies, and therefore a low fraction of ${}^{26}{\rm Al}$ that might have helped it 
fully differentiate. 
The acaplucoite parent body would not have accreted ice. 

\subsubsection{Aubrites}

Aubrites are spectrally and dynamically linked to asteroids at the inner edge of the asteroid belt, especially 
434 Hungaria at 1.94 AU, but possibly other members of the Hungaria family between 1.8 and 2.0 AU. 
(Gaffey et al.\ 1992; Kelley \& Gaffey 2002; Cuk et al.\ 2014).
Aubrites contain abundant minerals reflecting very reduced conditions, such as sulfides, and are considered 
melted enstatite achondrites, perhaps close in composition to EL6 chondrites.
This would imply a depletion of refractory abundances of the same order as the acapulcoite-lodranites, 
$\approx 0.87 \times$ CI. 
Thermal modeling by SF14 suggests a time of accretion of $1.5 \pm  0.1$ Myr. 

We find that conditions at 1.94 AU at 1.5 Myr are consistent with these constraints. 
Aubrites would form with a refractory abundance is $0.88 \times$ CI and $< 0.01$ wt\% CAIs, in
an environment with $\Sigma = 920 \, {\rm g} \, {\rm cm}^{-2}$ and $T = 197$ K.
The aubrites would not have accreted ice. 

\subsubsection{Winonaites}

Very little is known about the origin of the winonaites, which are primitive achondrites.
Winonaites are believed to derive from the same parent body as type I AB irons (Goldstein et al.\ 2009),
and isotopically type I AB irons appear to derive from the inner disk (Kruijer et al.\ 2017). 
Relict chondrules are seen in some winonaites (Benedix et al.\ 1998; Farley et al. 2015), 
and Hunt et al.\ (2017) have presented geochemical evidence that the winonaites indeed suffered only limited melting. 
The winonaites are intermediate in mineral composition between ECs and H chondrites (Weisberg et al.\ 2006), and
from the data presented by Hunt et al.\ (2017), winonaites appear to be depleted in refractory abundances
to the same degree as EH and H chondrites, i.e., $\approx 0.89 \times$ CI.
Winonaites also can contain more C (0.8 wt\%) than ECs (Hunt et al.\ 2017), suggesting formation further out in the
disk, perhaps at temperatures below the tar line temperature of 350 K.
Hf-W dating of winonaites and type I AB irons show  metal-silicate differentiation took place
at $3.6_{-2.0}^{+2.3}$ Myr or $3.1_{-1.9}^{+2.3}$ Myr (Schulz et al.\ 2009), which in combination
with thermal modeling suggests an accretion time of $\sim 1.8$ Myr (Hunt et al.\ 2017), although
with great uncertainty. 

Our model predicts refractory abundances of $0.89 \times$ CI at 1.8 Myr between radii of 2.0 and 2.4 AU. 
Winonaites would accrete with $< 0.01$ wt\% CAIs in environments with $\Sigma \approx 820 - 930 \, {\rm g} \, {\rm cm}^{-2}$,
and temperatures falling from 177 K at 2.0 AU to 168 K at 2.2 AU, to 160 K at 2.4 AU. 
Based on the overall reduced chemistry of the winonaites, we suspect they formed at temperatures too high for water ice
to condense, i.e., at $r < 2.2$ AU. 
Being below the tar line temperature of about 350 K, these bodies potentially would accrete C.
We favor formation of the winonaite parent body at about 2.1-2.2 AU at 1.8 Myr.

\subsection{Chondrite Formation}

The parent bodies of iron meteorites and achondrites were forming during the first 2 Myr of solar nebula
evolution, and presumably planetesimals would continue to form in a similar fashion as the disk continued to evolve.
A key difference is that planetesimals forming after about 2 Myr would contain less live ${}^{26}{\rm Al}$,
and therefore could remain unmelted to the present day
(Grimm \& McSween 1993; McSween et al.\ 2002; Hevey \& Sanders 2005; Sanders \& Taylor 2005).
Because they did not melt, chondrites retain CAIs, and the CAI abundance in different chondrite classes 
imposes an additional, important constraint on our model.
Here we review what is known about the refractory element abundances and CAI abundances of chondrites, as 
well as what is known about the time and place of their formation. 
We show that our model reproduces most of the observational constraints on chondrites.
Using the estimated times of formation and the refractory abundances, we calculate the heliocentric distances
at which each chondrite parent body could form.

\subsubsection{Enstatite Chondrites}

Enstatite chondrites (ECs) are spectrally associated with the E asteroids that orbit the Sun from approximately
1.9 to 2.1 AU (Gradie \& Tedesco 1982; Gaffey 1993; DeMeo \& Carry 2013).
Because of their very reduced mineralogies, ECs are inferred to have formed in regions with $T > 170 \, {\rm K}$, 
so that they are depleted in water.
Despite forming closer to the Sun and at higher temperatures,
ECs are the most depleted in refractories of any chondrites, $\approx 0.87 - 0.88 \times$ CI,  and 
CAIs are largely absent ($< 0.1$ vol\%) in ECs. 
Having abundant chondrules, ECs can be expected to form after increased chondrule production commenced at 1.5 Myr.
On the basis of Mn-Cr systematics, the accretion time of ECs has been estimated at $\sim 2$ Myr 
(Shukolyukov \& Lugmair 2004).
ECs reached relatively high temperatures, and the thermal modeling by SF14 suggests that ECs accreted relatively early, 
at $1.8 \pm 0.1$ Myr. 

The conditions between 1.9 and 2.1 AU at 1.7 Myr predicted by our model are an excellent match to the formation
conditions of ECs.
Refractory abundances are $0.88 \times$ CI and CAI abundances are $< 0.01$ wt\%.
Between 1.9 and 2.1 AU, surface densities range from $\Sigma = 828$ to $891 \, {\rm g} \, {\rm cm}^{-2}$,
and temperatures between 187 and 177 K.
ECs form well inside the snow line and would not accrete any water (although the snow line would sweep through this
region between 2.0 and 2.5 Myr).

\subsubsection{Rumuruti-type Chondrites}

Very little is known about the R chondrite parent body, but for completeness, we consider where and when it 
might have formed. The thermal modeling of SF14                     suggests it accreted at $2.1 \pm 0.1$ Myr.
Refractory abundances in R chondrites are $\approx 0.97 \times$ CI, with low ($< 0.1$ vol\%) CAI abundances.
The R chondrite parent body shows a higher degree of oxidation than other chondrites, suggesting perhaps it 
accreted with substantial water, at temperatures $< 160$ K.
Another odd feature of R chondrites is their relatively high $\epsilon{}^{54}{\rm Cr} \approx 0.4$, the highest value
among inner solar system objects, and equal to the CK chondrites (Warren 2011).
These suggest formation of the R chondrite parent body far out in the asteroid belt.

In the context of our model, we cannot explain such high refractory abundances without accretion of CAIs.
However, refactory abundances increase with increasing $r$, consistent with formation far out in the disk.
Within the time interval 2.0 - 2.2 Myr, the highest refractory abundance our model predicts is at 2.6 AU,
$0.91 \times$ CI. 
We speculate that if accretion of abundant ice were to somehow introduce extra CI dust into the R chondrite-forming
region, this would enhance the refractory abundances as well as increase $\epsilon{}^{54}{\rm Cr}$; 
but these are effects not included in the model.

We place the time and place of R chondrite formation to be 2.2 Myr, at 2.6 AU. 
The model predicts refractory abundances $0.91 \times$ CI and CAI abundances $< 0.01$ wt\%.
The surface density and temperature at this time and place would be $610 \, {\rm g} \, {\rm cm}^{-2}$ and 123 K.
The R chondrites would have accreted abundant ice.


\subsubsection{Ordinary Chondrites}

Ordinary chondrites are spectrally linked to S-type asteroids that dominate the asteroid belt from about 2.2 to 2.5 AU,
but which extend throughout the asteroid belt (Gradie \& Tedesco 1982; Gaffey 1993; Carry \& DeMeo 2013).
It is hypothesized that the parent body of the H chondrites may be the large asteroid 6 Hebe, that orbits at 2.43 AU
(Binzel et al.\ 1996; Bottke et al.\ 2010). 
Likewise, it is hypothesized that the parent body of the LL chondrites may the large asteroid 8 Flora, that orbits at 
2.20 AU (Vernazza et al.\ 2009; Dunn et al.\ 2013), and L chondrites have been linked with the Gefion family at 2.74 AU 
(Nesvorny 2009).
Unlike ECs, OCs do not have such reduced mineralogies and many must have accreted some water, 
indicating that many probably formed in regions cold enough for ice to condense, with $T < 160 \, {\rm K}$.
They have far less ice than CM, CR and CI chondrites, though, consistent with the idea that they formed in the inner 
disk (as constrained isotopically: Warren 2011), in a region cold enough to condense ice but not as water-rich as the outer disk.
H, L, and LL chondrites have refractory depletions of about $0.89-0.90 \times$ CI, 
and all have relative low CAI abundances $< 0.1$ wt\%, with H chondrites possibly as high as $0.2$ wt\%. 

The H chondrite parent body was considerably thermally metamorphosed, and is inferred to have formed early. 
Henke et al.\ (2013) modeled the thermal evolution of the H chondrite parent body and inferred it formed 
about 2 Myr after CAIs.
SF14, treating all the OCs as arising from the same parent body, derived an accretion time
of $2.1 \pm 0.1$ Myr, very similar to the accretion age derived by Blackburn et al.\ (2017), $2.15 \pm 0.1$ Myr.
OCs must accrete after the chondrules within them formed.
Chondrules in L chondrites have been dated using Al-Mg systematics to have formed 1.6 - 2.2 Myr after CAIs 
(Rudraswami \& Goswami 2007), or using Pb-Pb at 2.5 - 3.0 Myr after CAIs (Bollard et al.\ 2017)
Chondrules in LL chondrites likewise have been dated using Al-Mg systematics to have formed 1.0 - 2.5 Myr (Rudraswami et al.\ 2008),
0.7 - 2.4 Myr (Mostefaoui et al.\ 2002), or 1.5 - 3.0 Myr after CAIs (Villeneuve et al.\ 2009).
Given the potential for alteration and the limitations of the techniques, we do not feel the data demand a maximum time of chondrule 
formation as late as 3.0 or even 2.5 Myr.
Based on Mn-Cr systematics of aqueously produced fayalite, combined with thermal modeling, Doyle et al.\ (2015)
inferred that the L chondrite parent body formed between 1.8 and 2.5 Myr.  
From all this we infer that the OCs accreted at $2.1 \pm 0.1$ Myr.

Our model predicts conditions consistent with each of the OC types.
Within the time interval 2.0 - 2.2 Myr, refractory abundances of $0.899 \times$ CI, consistent with the H chondrite, are
found from 2.2 to 2.5 AU. 
Formation at 2.1 Myr at 2.43 AU yields a refractory abundance $0.899 \times$ CI, but predicts a
CAI abundance $< 0.01$ wt\%.
This is consistent with 6 Hebe being the parent body. 
The surface density would be $843 \, {\rm g} \, {\rm cm}^{-2}$ and the temperature 150 K, suggesting the H chondrite parent
body was in a region cold enough to condense ice, although the inner disk may have been depleted in water by Jupiter blocking
the inward flow of ice (Morbidelli et al.\ 2016).

Within the time interval 2.0 - 2.2 Myr, refractory abundances of $0.904 \times$ CI, consistent with the L chondrite, are
found from 2.5 to 2.6 AU. 
At 2.0 Myr and 2.6 AU, the refractory abundance is $0.905 \times$ CI, consistent with the L chondrite, and the CAI abundance
would be $< 0.01$ wt\%.
The surface density would be $708 \, {\rm g} \, {\rm cm}^{-2}$ and the temperature 131 K, suggesting the L chondrite parent
body was in a region cold enough to condense ice. 
The surface density beyond 2.6 AU decreases rapidly because of the gap opened by Jupiter, so the model does not favor formation
beyond 2.6 AU, but this is encouragingly close to the location of the Gefion family at 2.74 AU. 

Within the time interval 2.0 - 2.2 Myr, refractory abundances are not quite as low as the $0.890 \times$ CI value for the LL
chondrite.
Lower refractory abundances are found at earlier times and closer to the Sun.
At 2.0 Myr and 2.20 AU, the refractory abundance is $0.894 \times$ CI, reasonably consistent with the LL chondrite, and the 
CAI abundance would again be $< 0.01$ wt\%.
This is consistent with 8 Flora being the parent body of LL chondrites. 
The surface density would be $815 \, {\rm g} \, {\rm cm}^{-2}$ and the temperature 162 K, suggesting the LL chondrite parent
body was in a region marginally cold enough to condense ice. 

Given the time of formation, our model predicts a range of radial distances from the Sun that would yield the refractory abundances
of each of the OC types. 
Remarkably, those ranges coincide with the location of 6 Hebe for the H chondrites, 8 Flora for the LL chondrites, and the model
predicts a formation of the L chondrite parent body in the outer asteroid belt near 2.6 AU, reasonably close to the 2.74 AU 
location of the Gefion family. 


%
%

\subsubsection{Carbonaceous Chondrites}

{\it CV and CK Chondrites}: CVs have among the highest mass fractions of CAIs of any chondrites, and so we 
presume they formed in the pressure bump beyond Jupiter at about 3.6 AU.  
Because CK chondrites are strongly associated with CV chondrites, and have similarly high CAI abundances, we assume 
that CK chondrites formed in the same place.
CK chondrules in general are too thermally metamorphosed to be isotopically dated.
The accretion ages of CVs are constrained by ages of chondrules and dated by various isotopic systems.
Amelin \& Krot (2007) report Pb-Pb ages of CV3 Allende chondrules apparently $\approx 0.6 \pm 1.1$ Myr younger 
than Allende CAIs.
The initial ${}^{26}{\rm Al}/{}^{27}{\rm Al}$ ratios in Allende chondrules reported by Bizzarro et al.\ (2004, 2005) 
imply chondrule formation continued to at least 1.4 Myr,
and Nagashima et al.\ (2015) inferred a time of formation for CV3 chondrules from 2.0 to 2.5 Myr after CAIs. 
Budde et al.\ (2016) found Allende chondrules were melted about 2 Myr after CAIs. 
Likewise, Akaki et al.\ (2007) found an Al-rich chondrule that melted at about 1.7 Myr after CAIs, and 
Kawasaki et al.\ (2015) found a type C CAI that had melted at 1.6 Myr after CAI formation.
More recently, Becker et al.\ (2015) analyzed Hf-W isotopes in whole-rock systems composed of chondrules and matrix
and inferred a time of formation at least 2.6 Myr after CAIs. 

The CV parent body has not been identified, but based on paleomagnetism studies of the CV3 chondrite Allende, it is 
inferred that the CV chondrites formed on the surface of a partially differentiated body large enough to sustain a 
core dynamo (Elkins-Tanton et al.\ 2011; Carporzen et al.\ 2011).
This would also explain the close association (based on siderophile element abundances) of CV chondrites with the 
Eagle Station pallasites (Humayun \& Weiss 2011)).
From Hf-W and Al-Mg systematics, metal-silicate fractionation on the Eagle Station pallasite parent body
took place at least 2 Myr after CAIs, 
To differentiate in its interior, such a body would have to be large ($\sim 200$ km in radius) and have accreted by 
about 1.5 Myr (Elkins-Tanton et al.\ 2011), although it is possible that the bulk of the body had formed early, and
the CV chondrite material was swept up at a later time.
In this model, CK chondrites come from the same parent body, at greater depths where they would experience 
greater thermal metamorphism. This would imply that they accreted at the same time, or, if they were swept up, 
perhaps slightly earlier than CV chondrites. 
The thermal models of SF14                     suggest the CK parent body accreted at $2.6 \pm 0.2$ Myr, and
the CV parent body at $3.0 \pm 0.2$ Myr. 
Or, possibly the single parent body started to form at 1.5 Myr and continued to accrete material for $> 1$ Myr. 
Put together, we view it as likely that CV chondrites represent material in the pressure bump at 3.6 AU, at about
2.5 - 3.0 Myr after CAIs, with CK chondrite material accreted at about the same time or slightly earlier,
possibly 2.5 Myr. 

Our model predicts conditions consistent with these constraints. 
As the disk evolves, refractory abundances and CAI abundances increase steadily in the pressure bump region. 
In the pressure bump region, at 3.5 - 3.6 AU, we predict peak CAI mass fractions that increase from 3.4 wt\% at 2.0 Myr, 
to 4.0 wt\% at 2.5 Myr, to 4.7 wt\% at 3.0 Myr, and 5.1 wt\% at 3.5 Myr, to 5.9 wt\% at 4.0 Myr. 
Likewise, we predict peak refractory abundances that increase from $1.31 \times$ CI at 2.0 Myr, to 
$1.38 \times$ CI at 2.5 Myr, to $1.46 \times$ CI at 3.0 Myr, to $1.51 \times$ CI at 3.5 Myr, to $1.60 \times$ CI at 4.0 Myr.
Conditions from 2 to 3 Myr are consistent with CV chondrites, which have CAI mass fractions $\approx 3 - 4$ wt\% and
refractory abundances $1.35 \times$ CI.

At 3.62 AU and 2.4 Myr, the model predicts $\Sigma = 1000 \, {\rm g} \, {\rm cm}^{-2}$ and $T = 114$ K, although 
we speculate heating by spiral shocks launched by Jupiter may have increased the average temperature (Lyra et al.\ 2016).
Chondrites formed at that time and place would have refractory abundances $1.25 \times$ CI and CAI abundances 3.0 wt\%,
consistent with CK chondrites.
At 3.63 AU and 2.8 Myr, the model predicts $\Sigma =  900 \, {\rm g} \, {\rm cm}^{-2}$ and $T = 108$ K. 
Chondrites formed at that time and place would have refractory abundances $1.38 \times$ CI and CAI abundances 3.7 wt\%,
consistent with CV chondrites. 

{\it CO Chondrites}: The COs have elevated refractory abundances and substantial CAI abundances, but not 
as much as CVs and CKs. Given that they have many similarities to CV and CK chondrites, we infer that they
also formed in the pressure bump, but perhaps at a different location within it.
The chondrules in CO chondrites are dated by Al-Mg to have formed as late as about $2.5 \pm 0.3$ Myr after CAIs 
(Kurahashi et al.\ 2008). 
The thermal modeling of SF14 suggests an accretion time of $2.7 \pm 0.2$ Myr.

At 3.71 AU and 2.7 Myr, the model predicts $\Sigma \approx 940 \, {\rm g} \, {\rm cm}^{-2}$ and $T = 108$ K.
Chondrites formed at that time and place would have refractory enrichment $1.11 \times$ CI and CAI abundance 1.9\%, 
consistent with CO chondrites, although the CAI abundance is slightly overpredicted.
We conclude that CO chondrites formed in the same pressure bump as the CV and CK chondrites, at about 3.7 AU, 
at about 2.7 Myr. 
The CV and CK chondrites may have formed on different bodies, or on the single CV parent body hypothesized by 
Elkins-Tanton et al.\ (2011),
while the CO chondrites may have formed on one or more parent bodies elsewhere in the pressure bump region. 
Heating of gas in this region by spiral shocks launched by Jupiter, plus parent-body thermal metamorphism,
may be why CO, CV and CK chondrites lack the water CM and CR and CI chondrites have.

{\it CM Chondrites and Tagish Lake}: 
There are few constraints on the formation of the CM chondrite parent body.
Based on Al-Mg systematics, Kurahashi et al.\ (2008) estimated a time of accretion 3 - 4 Myr. 
Thermal evolution models by SF14                     suggest an accretion time at 3 - 4 Myr.
Mn-Cr systematics have been used to date carbonate formation to about 2 to 5 Myr after CAIs 
(de Leuw et al.\ 2009; Fujiya et al.\ 2013; Jilly et al.\ 2014), and aqueous alteration to
$3.93 \pm 0.23$ Myr after CAIs (Lee et al.\ 2012).
We therefore infer that CM chondrites probably accreted at $\approx 3.5$ Myr.
CM chondrites have refractory abundances $1.13 \times$ CI and CAI abundances $\approx 1.5$ wt\%. 

Tagish Lake is an ungrouped CC with characteristics intermediate between CM and CI chondrites
(Brown et al.\ 2000; Zolensky et al.\ 2002; Blinova et al.\ 2014a,b). 
It is extremely friable and has suffered considerable aqueous alteration.
Its bulk refractory lithophile abundances are similar to those of CM chondrites, 
While chondrules and CAIs are severely altered and rare compared to other CCs, it does contain CAIs.
We could not find modal abundances reported, but in comparison to CM and CR chondrites, we estimate they must be 
a fraction of a percent. 
Its carbonates are isotopically dated to have formed at the same time as CM carbonates.
Thermal modeling suggests it formed at about 
$3.5_{-0.5}^{+0.7}$ Myr (SF14)                    

At 3.76 AU and 3.5 Myr, the model predicts $\Sigma \approx 940 \, {\rm g} \, {\rm cm}^{-2}$ and $T = 99$ K.
Chondrites formed at that time and place would have refractory enrichment $1.13 \times$ CI and CAI abundance 2.2\%, 
consistent with CM chondrites, although the CAI abundance is slightly overpredicted.
We conclude that CM chondrites also formed in the same pressure bump as the other CCs, close to the location where
COs formed, but later than CO chondrites.
If the lower water content in CO chondrites is attributed to heating of the pressure bump region by spiral shocks from
Jupiter, we would conclude that this heating had stopped by the time CM chondrites formed, or did not reach this region. 
Because its refractory abundances and CAI abundances and accretion time are completely consistent with
CM chondrite properites, we must conclude that Tagish Lake formed at or just beyond where the CM chondrites
formed, i.e., at about 3.8 AU.

{\it CR Chondrites}: A distinguishing characteristic of the CR chondrites is that they seem to have formed much 
later than other chondrites, at $\approx \, 4$ Myr after CAIs, based on the ages of chondrules in them (Schrader et al.\ 2017;
Budde et al.\ 2018).
The thermal modeling of SF14 likewise     suggests a late formation 3 - 4 Myr. 
Combining the modeling of Schrader et al.\ (2017) and SF14, we infer a time of formation between 
3.7 and 4.0 Myr after CAIs. 
Also, CR chondrites have only slightly elevated refractory element abundances, about 
$1.02 \times$ CI, and much lower CAI abundances than other CCs, $\approx 0.5 - 1.0$.

At 3.84 AU and 4.0 Myr, the model predicts $\Sigma \approx 730 \, {\rm g} \, {\rm cm}^{-2}$ and $T = 93$ K.
Chondrites formed at that time and place would have refractory enrichment $1.02 \times$ CI and CAI abundance 1.3\%, 
consistent with CR chondrites, although the CAI abundance is slightly overpredicted.
We conclude that CR chondrites also formed in the same pressure bump as the other CCs, but farther from Jupiter,
and at a later time. 
They are the last major chondrite class to form in this region.

{\it CI Chondrites}: 
CI chondrites were substantially aqueously altered, and thermal modeling by SF14 suggests 
the CI chondrite parent body accreted at 3 - 4 Myr to mobilize this water. 
CI chondrites lack large, abundant chondrules and CAIs, and therefore the CI chondrite parent body must have 
formed well beyond where chondrule formation occurred. 
But the constraints on where CI chondrites formed may be even more stringent than this. 
Inner solar system rocky material is depleted in refractory elements, and carbonaceous chondrite material would
also be depleted in refractories if not for the CAIs they contained.
This depletion results from conversion of some of their refractory materials into CAIs, which are mobile and often
lost to the Sun or other parts of the disk.
Material in the CI chondrite-forming region, if it mixed with these materials and did not accrete CAIs, also 
should be somewhat depleted in refractories, relative to the solar nebula starting composition, which is that of the Sun.
By dint of having compositions that apparently match the Sun and therefore primordial composition of the disk, 
mixing of CI chondrite material with these other reservoirs must have been limited, arguing for an origin of
CI chondrites quite far out in the disk.

Limited mixing does not mean zero mixing, though, and it is a robust prediction of our models that chondrites 
formed even beyond 10 AU will be depleted in refractories.
Indeed, evidence that CIs do not represent a pristine composition comes from measurements of D/H ratios,
which show that the water accreted by CI chondrites more closely resembles water in the inner solar nebula
rather than comets or other outer solar nebula objects (Alexander et al.\ 2012, 2017).
This suggests that the CI chondrites formation region did mix with the rest of the disk, and the only question 
is how much. 
The match between CI abundances and the solar photospheric abundances is excellent but not perfect. 
Uncertainties in both solar photospheric abundances and in meteoritic measurements mean that discrepancies of 
on the order of 10\% would not violate observational constraints
(Lodders 2003; Lodders et al.\ 2009; Asplund et al.\ 2009; Palme et al.\ 2014). 
To limit the degree of mixing, CI chondrites must have formed early, and far out in the disk.

By 3 Myr, even with our low assumed value of $\alpha = 1 \times 10^{-5}$, our model predicts that much of the outer disk 
has mixed with more depleted regions of the outer disk just exterior to Jupiter, and even with the inner disk 
(importantly, Jupiter does not necessarily prevent the mixing of gas and small dust across the gap).
At 1.5 Myr, the only regions that are not depleted by at least 10\% from the starting composition lie beyond 11 AU.
By 2.0 Myr, no parts of the disk have escaped being depleted by at least as much. 
We favor formation of CIs at 3.0 Myr and 15 AU.
At this time, from 15 AU out, the disk is uniformly depleted in refractory lithophile elements by 12.2\% relative to the
starting composition, which is the Sun.
We conclude that CI chondrites formed at 15 AU or beyond, at 3.0 Myr, and that the abundances of their refractory 
lithophile elements (Ca, Al, Ti, Sc, rare earths) are depleted relative to the starting composition by about 9\%.
We adopt this composition for the CI chondrites throughout this paper. 
At 15 AU and 3.0 Myr, $\Sigma = 6.2 \, {\rm g} \, {\rm cm}^{-2}$ and $T = 46$ K. 
CI chondrites would accrete essentially no CAIs. 
While the location of the CI chondrites so far out in the disk is provocative, it is an inevitable consequence of 
the compositional match between the CI chondrites and the solar photosphere, and is justified by the lack of 
chondrules and the high water-to-rock ratio of CI chondrites.
Interestingly, Gounelle et al.\ (2006) calculated the trajectory of the CI chondrite Orgueil from historical accounts, 
and found it to match a Jupiter family comet originating beyond the orbit of Jupiter. 
Gounelle et al.\ (2006)  also discuss other evidence linking CI chondrites to comets.
As for the depletion of refractories by 12\%, not only is this allowed by the observational uncertainties, there is some 
evidence for such a depletion, discussed in \S 6.4.


{\it CB/CH/Isheyevo Chondrites}: The CB and CH chondrites (and Isheyevo, which appears to be a hybrid of the two
groups), are classified isotopically as CCs, but they are unlike any other chondrites.
CBs show remarkably high elevations in ${}^{15}{\rm N}$ abundance (Weisberg et al.\ 2001). 
CAIs are present but very rare in two CBs, and absent in the 3 other CBs (Krot et al.\ 2005). 
CH chondrites are metal-rich and are remarkable for containing spherules of metal that show strong evidence for 
condensation from a vapor (Meibom et al.\ 2000; Petaev et al.\ 2001). 
CB chondrites contain abundant igneous spherules of silicate composition that resemble chondrules in other
chondrites, but which have igneous textures (cryptocrystalline and skeletal) that are very unlike the textures
of chondrules in all other chondrite groups (mostly poprphyritic), and indicative of rapid cooling (Krot et al.\ 2001;
Rubin et al.\ 2003). 
Pb-Pb dating of these `chondrules' shows they formed $\approx 4.5$ Myr  after CAIs and the birth of the solar system
(Krot et al.\ 2005; Bollard et al.\ 2015), 
much later than the formation times of other chondrites examined here.
The timing and other characteristics of these chondrites strongly indicates they formed in the vapor plume 
produced by a large impact between two parent bodies, in a dissipating nebula (Krot et al.\ 2005).
The silicate- and metal-rich layers in Isheyevo show sedimentary laminations that suggest sweepup of 
aerodynamically sorted particles in this plume, by the impacted parent body (Garvie et al.\ 2017). 

Because of its impact origin, it is not possible to derive the time of accretion of the CH/CB/Isheyevo parent body:
it has no chondrules formed in the nebula to date, and other isotopic systems have been reset to the time of 
impact. 
It is possible to say that the parent bodies had CAIs (Krot et al.\ 2017), but the original CAI fractions on the 
impactors are not well constrained, nor are their starting refractory abundances.
We therefore cannot analyze the CB/CH chondrites in the framework of our model, but they do constrain the model
nonetheless.
The required energetic impact is probably associated with a dynamical destabilization associated with the migration
of Jupiter (Johnson et al.\ 2016). 
In the context of our model, this suggests that the outward migration of Jupiter from 3 AU to 5.2 AU began shortly 
before the impact that produced CB/CH chondrites at 4.5 Myr.
This supports the picture we have presented here that Jupiter stayed near 3 AU through the formation of the CR 
chondrites at 4.0 Myr, and only afterwards began to migrate outward. 


%
%
\begin{table}
\centering
\caption{Conditions of achondrite and chondrite formation} 
\vspace{0.2in} 
\begin{tabular}{c|cccc|cccc} 
             & \multicolumn{4}{c}{Meteoritic Constraints} & \multicolumn{4}{c}{Model Predictions} \\
Meteorite    & r (AU)   &  formation   & (X/Mg) & CAIs      & r (AU)   &  formation   & (X/Mg) & CAIs   \\
Type         &          &  time (Myr)  & / CI   & (vol\%)   &          &  time (Myr)  & / CI   & (wt\%) \\ 
\hline 
ureilites    &  2.8     & $\sim 0.6$   &  1.05? &           &  2.8     &  0.6         &  1.05  & 1.3    \\ 
HEDs         &  2.36    & $\sim 0.8$   &  1.03  &           &  2.36    &  0.8         &  1.03  & 1.2    \\
acapulcoite- & $< 2.1$  & $\sim 1.3$   &  0.87  &           &  2.0     &  1.3         &  0.88  & 0.0    \\
 lodranites  &          &              &        &           &          &              &        &        \\
aubrites     &  1.94?   & $\sim 1.5$   &  0.88  &           &  1.94    &  1.5         &  0.88  & 0.0    \\ 
winonaites   &          & $\sim 1.8$   &  0.89  &           &  2.4     &  1.8         &  0.89  & 0.0    \\
\hline 
ECs          & 1.9 - 2.1 & 1.7 - 1.9   & 0.87 - 0.88 & $< 0.1$ & 1.9 - 2.1 &  1.7    &  0.88  & 0.0    \\ 
\hline
R            &          & 2.0 - 2.2    &  0.97  &  $< 0.1$  &  2.6      &  2.2         &  0.91  & 0.0   \\ 

\hline
H            &  2.43?   & 2.0 - 2.2    &  0.90  &  0.01 - 0.2 &  2.43   &  2.1         &  0.90  & 0.0   \\
L            &  2.74?   & 2.0 - 2.2    &  0.90  &  $< 0.1$  &    2.6    &  2.0         &  0.90  & 0.0   \\ 
LL           &  2.20?   & 2.0 - 2.2    &  0.89  &  $< 0.1$  &    2.20   &  2.0         &  0.89  & 0.0   \\ 
\hline
CK           & beyond   & 2.4 - 2.8    &  1.24  &  $< 4$    &  3.60    &  2.2         &  1.24  & 2.9          \\
CV           & Jupiter  & 2.8 - 3.2    &  1.35  &  $> 3$    &  3.60    &  2.6         &  1.35  & 3.8          \\
CO           &  "       & 2.5 - 2.9    &  1.11  &  1        &  3.72    &  2.7         &  1.11  & 1.9          \\
CM           &  "       & 3.0 - 4.2    &  1.13  &  1 - 2    &  3.76    &  3.5         &  1.13  & 2.2          \\ 
CR           &  "       & 3.7 - 4.0    &  1.02  &  0.5 - 1  &  3.84    &  4.0         &  1.02  & 1.3          \\
CI           &  "       & 3.0 - 4.0    &  1.00  &  0.0      & $\geq$ 15 &  3.0        &  $\equiv$ 1.00 & 0.0  \\
\hline 
\end{tabular}
\end{table}
%

%
%
\begin{table}
\centering
\caption{Predicted conditions of achondrite and chondrite formation} 
\vspace{0.2in} 
\begin{tabular}{c|cccccc} 
Meteorite    & $r$      & Formation    & $\Sigma$                & $\rho$  &  $T$    & $\alpha$     \\ 
Type         & (AU)     & time (Myr)   & $({\rm g} \, {\rm cm}^{-2})$  
                                                                 & $(10^{-10} \, {\rm g} \, {\rm cm}^{-3})$ 
                                                                           &  (K)    & ($10^{-4}$)  \\
\hline 
ureilites    & 2.8      &  0.6         & 1920                    &  3.8    & 213     & 0.87         \\ 
HEDs         & 2.36     &  0.8         & 1810                    &  4.3    & 244     & 1.16         \\
acapulcoite- & 2.1      &  1.3         & 1050                    &  3.2    & 206     & 1.42         \\ 
 lodranites  &          &              &                         &         &         &              \\ 
aubrites     & 1.94     &  1.5         &  920                    &  3.2    & 197     & 1.62         \\

winonaites   & 2.2      &  1.8         & 1050                    &  3.4    & 160     & 1.31         \\ 
\hline
ECs          & 1.9 - 2.1 & 1.7         & 830 - 890               &  3.1 - 2.9  & 187 - 177 &  1.68 - 1.42 \\ 
\hline 
R            & 2.6      &  2.2         &  610                    &  1.7    & 123     & 0.99         \\  
\hline 
H            & 2.43     &  2.1         &  840                    &  2.4    & 150     & 1.11         \\ 
L            & 2.6      &  2.0         &  710                    &  2.0    & 131     & 0.98         \\ 
LL           & 2.20     &  2.0         &  820                    &  2.6    & 162     & 1.31         \\ 
\hline 
CK           & 3.60     &  2.2         & 1060                    & 1.9     & 118     & 0.57         \\
CV           & 3.60     &  2.6         &  940                    & 1.7     & 111     & 0.57         \\
CO           & 3.72     &  2.7         &  940                    & 1.7     & 108     & 0.54         \\ 
CM           & 3.76     &  3.5         &  790                    & 1.5     &  99     & 0.53         \\
CR           & 3.84     &  4.0         &  730                    & 1.3     &  93     & 0.51         \\ 
CI           & $\geq 15$ & 3.0         & 6.2                     & 0.002   &  46     & 0.10         \\ 
\hline 
\end{tabular}
\end{table}
%
%

\subsection{Aerodynamic Sorting into Planetesimals}

When comparing chondrite compositions to specific times and places in the solar nebula, we are implicitly
assuming that chondrites are snapshots in time of the nebula, meaning that each chondrite parent body 
quickly assembles from material in its local environment.
Here, `quickly' means $\ll 10^{5}$ yr, the timescale on which the nebula evolves, and `local'
means over scales $\ll 0.1$ AU.
The streaming instability (Youdin \& Goodman 2005; Johansen \& Youdin 2007; Johansen et al.\ 2007; Simon et al.\ 2016)
is a mechanism that satisfies these constraints.
In regions of the disk with solids-to-gas ratios slightly higher than the solar ratio, particles with stopping times
$t_{\rm stop} \sim 0.3 \, \Omega^{-1}$ can be concentrated in just a few orbital timescales into large collections 
of particles, large enough to self-gravitate and become planetesimals.
These planetesimals would have a power-law distribution of masses that closely matches the primordial size distribution
of asteroids (Simon et al.\ 2016). 
A challenge for the streaming instability model is that the particles that are abundant in the 
solar nebula---chondrules and CAIs---are 
typically millimeter-sized and are characterized by $\Omega \, t_{\rm stop} < 10^{-3}$, whereas the particles
that are concentrated by streaming instability must be meter-sized to have $\Omega \, t_{\rm stop} \sim 0.3$.
Cuzzi et al.\ (2017) and Simon et al.\ (2018) suggest that turbulent concentration first concentrates chondrules
and small objects into aggregates 10 cm in size or larger; these aggregates then are concentrated by streaming 
instability.
Two neighboring lithologies in the ordinary chondrite NWA 5717 show strong evidence of being two such aggregates 
that were assembled into that parent body (Simon et al.\ 2018).
We consider it very likely that turbulent concentration of chondrules is the first step to chondrite formation.

This presents a severe test of our model. 
Although our model was constructed to explain the refractory abundances and CAI abundances of chondrites, 
one of the powerful aspects of our model is that it predicts the physical conditions ($\Sigma$, $T$, $\rho$, $\alpha$, 
water ice abundance, etc.) in which each chondrite parent body formed (Table 4).
Because of that, we can predict the mean size of chondrule that is concentrated in each of the regions where each of the 
chondrites formed.
We do so using Equation 2, assuming $\rho_{\rm s} = 3.2 \, {\rm g} \, {\rm cm}^{-3}$ (see discussion by Friedrich et al.\ 2014), 
and using other information from Table 4. 
In our analyses above we did not distinguish between EH and EL chondrites. Here we assume that EH chondrites formed closer
to the Sun, at 1.9 AU, while EL chondrites formed farther from the Sun, at 2.1 AU. 
We then compare our predicted sizes of particles concentrated by turbulence to the mean chondrule diameters reported by 
Scott \& Krot (2014) and determined by Friedrich et al.\ (2014) in their careful study.
While CI chondrites do not contain chondrules, they do contain abundant olivine and pyroxene grains up to several hundred 
$\mu{\rm m}$ in diameter (Leshin et al.\ 1997), so we include these as if they were chondrules.
The match is within 10\% for EL, R, H, L, LL, and CR chondrites, and within 25\% for CK and CV chondrites.
While not strictly chondrites, acapulcoites are primitive achondrites with relict chondrules whose sizes
are in the range 0.4 - 0.7 mm (Friedrich et al.\ 2014), and our model predictions are in the middle of that range.
The model predicts larger particles than are observed for EH, CO and CM chondrites.

Given the wide range of conditions that might have pertained in the disk, 
we consider the match between the model predictions and the meteoritic data to be exceptional.
For example, a typical turbulent viscosity parameter might have been $\alpha \sim 10^{-2}$ (Kalyaan et al.\ 2015),
instead of the $\sim 10^{-4}$ we favor. This would have led to size-sorting by turbulence an order of magnitude 
smaller than we have assumed. It is common to assume $\alpha$ is uniform, or even increasing wih heliocentric distance
in the disk as the magnetorotaional instability (MRI) becomes more effective (Kalyaan et al.\ 2015). 
This would have led to smaller chondrules in CCs as compared to OCs.
It is remarkable that the parameters we chose so that refractory abundances could be matched lead to similar sizes of 
chondrules across all chondrite classes, right around the 0.5 mm diameter seen in OCs, with slightly smaller chondrules 
for ECs and RCs, and slightly larger chondrules in CCs. . 
We suspect that wih better data on chondrule densities the fit would improve; for example, EH chondrites have among the 
highest grain densities, $\sim 3.7 \, {\rm g} \, {\rm cm}^{-3}$,  (Consolmagno et al.\ 2008), and if their chondrules 
have the same density, we would predict a chondrule size of 0.37 mm. 
The anomalously small CO and CM chondrules still require an explanation, but our model clearly is consistent with particle 
concentration by turbulence.
This strongly suggests that aggregates of chondrules can form as described by Simon et al.\ (2018), and that these aggregates
of particles are what are concentrated by streaming instability.
This further provides strong support for the model assumption that chondrites represent snapshots in time of the solar nebula.

%
%
\begin{table}
\centering
\caption{Sizes of aerodynamically sorted chondrules} 
\vspace{0.2in} 
\begin{centering}
\begin{tabular}{c|ccc} 
Meteorite    & d (mm)             & d (mm)             & d (mm)        \\ 
Type         & (observed)${}^{a}$ & (observed)${}^{b}$ & (predicted)   \\
\hline
EH           & 0.2               & {\bf 0.23}         & 0.43          \\
EL           & 0.6               & {\bf 0.50}         & 0.48          \\ 
\hline 
R            & 0.4               & {\bf 0.40}         & 0.43          \\  
\hline 
H            & 0.3               & {\bf 0.45}         & 0.51          \\ 
L            & 0.5               & {\bf 0.50}         & 0.47          \\ 
LL           & 0.6               & {\bf 0.55}         & 0.52          \\ 
\hline 
CK           & 0.8               & {\bf 0.90}         & 0.75          \\
CV           & 1.0               & {\bf 0.90}         & 0.70          \\
CO           & 0.15              & {\bf 0.15}         & 0.70          \\ 
CM           & 0.3               & {\bf 0.27}         & 0.64          \\
CR           & 0.7               & {\bf 0.70}         & 0.62          \\ 
CI           & n/a            & {\bf 0.1-0.3}${}^{c}$ & 0.10          \\ 
\hline 
acapulcoites &                   & {\bf 0.4 - 0.7}    & 0.54          \\  
\hline
\end{tabular}
\end{centering}

\noindent 
a. Scott \& Krot (2014); b. Friedrich et al.\ (2014). c. Leshin et al.\ (1997). 
\end{table}
%
%

\subsection{Summary}

\begin{centering}
\begin{figure}[ht]
 \centering
 \includegraphics[width=.8\linewidth]{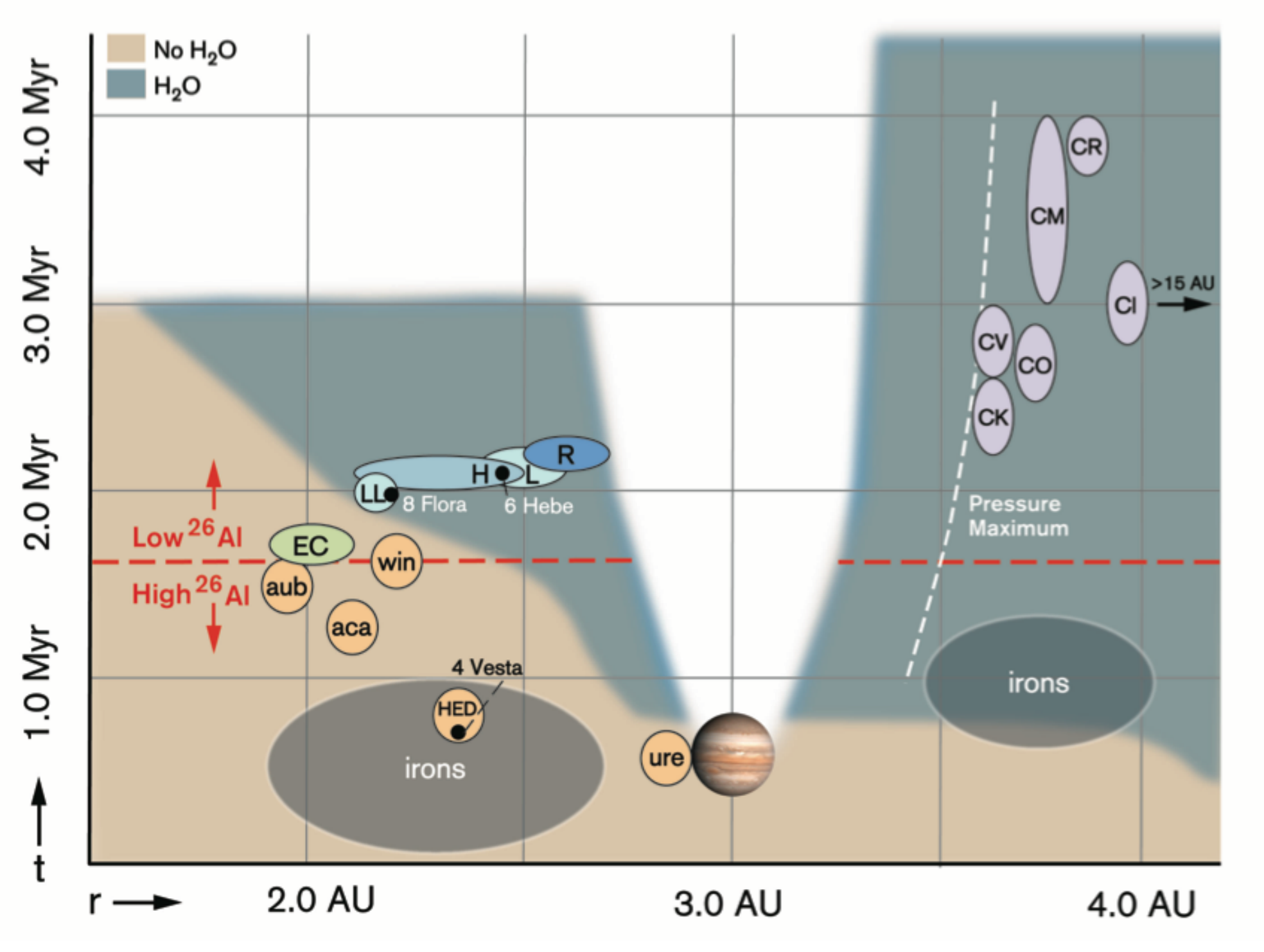}
 \caption{Summary plot of when and where our model predicts different meteorite types formed.  The horizontal
  axis denotes heliocentric distance and the vertical axis denotes different times after CAIs. Brown denotes 
  presence of warm ($T > 160$ K) nebular gas, bluish-brown denotes the presence of cold nebular gas in which
  water ice can condense. Gas is presumed absent inside 3 AU after 2.5 Myr, and in the gap opened up by Jupiter
  at 3 AU starting at 0.6 Myr. The location of the pressure maximum beyond Jupiter is labeled with a white 
  dashed line.  Iron meteorites 
  form at times and places
  denoted by gray ovals. Various achondrites (ureilites ['ure']; howardite-eucrite-diogenites [`HED']; 
  acapulcoite-lodranites [`aca']; aubrites ['aub']; and winonaites [`win']) form at times and places denoted by orange 
  ovals.  These fall below the red dashed line, indicating they formed with a sufficiently high ${}^{26}{\rm Al}$
  content to melt.  Enstatite chondrites (`EC'), ordinary chondrites (`H', `L', `LL'), Rumuruti chondrites (`R') 
  and carbonaceous chondrites (`CK', `CV', `CO', `CM', `CR', `CI') form at times and places denoted by green, blue,
  dark blue, and lavender ovals, respectively.  In our model CI chondrites form at the time indicated, but at or 
  beyond 15 AU.  Black dots denote locations and inferred time of formation for likely parent bodies: 4 Vesta for 
  the HEDs, 8 Flora for the LL chondrites, and 6 Hebe for the H chondrites.  }
 \label{fig:bigkahuna}
\end{figure}
\end{centering}

In Figure~\ref{fig:bigkahuna} we provide a summary plot of where and when we predict each meteorite type to form.
As a function of heliocentric distance, $r$, and time after CAIs, $t$, we plot where and when gas is present,
as defined by $\Sigma > 10^3 \, {\rm g} \, {\rm cm}^{-2}$, and the temperature of the gas, as defined by whether
the gas is cold enough for ice to condense ($T < 160$ K), according to our model.
Jupiter's core ($30 \, M_{\oplus}$) is assumed to form at 0.6 Myr, and Jupiter grows in mass from that time forward, 
opening a gap as it does.
The predicted times and places of accretion of various meteorite parent bodies are plotted as ovals, including all
the chondrites classes and various achondrites. 
We also plot the radial position and predicted time of accretion of several asteroids that may be meteorite parent bodies:
4 Vesta, accepted to be the parent body of the HED achondrites; 6 Hebe, thought to be the parent body of the H chondrites;
and 8 Flora, which may be the parent body of some LL chondrites. 
We also plot the position of the pressure maximum beyond Jupiter, which migrates outward from 3.5 AU at 1 Myr, to 
3.75 AU at 4 Myr. 
We also draw a horizontal line at 1.9 Myr, the approximate time after which parent bodies accrete with too little 
live ${}^{26}{\rm Al}$ to melt.
Achondrites and magmatic iron meteorites form beow this line, and chondrites form above this line. 
Other details are described in the Figure caption.

Certain trends can be extracted from Figure~\ref{fig:bigkahuna}.
Carbonaceous chondrites formed at a variety of locations and times in the disk beyond Jupiter, but the pattern
is that almost all carbonaceous chondrites form near the pressure maximum, where material is concentrated.
We were only able to find sufficient information about initial composition for five achondrites, but they
appear to be distributed throughout all the times (up to 2 Myr) and places (2 to 3 AU)  where achondrites form,
suggesting parent bodies formed throughout this region at all early times.
Just as interesting than where meteorite parent bodies formed, however, is where they did {\it not} form. 
Chondrites appear to represent the tail end of planetesimal formation in this region, with no strong evidence 
that chondrite parent bodies formed in the inner disk after 2.5 Myr.
We assume gas is lost from this region, presumably by photoevaporation from the Sun, starting around 3 Myr.
Clarke et al.\ (2001) have shown that photoevaporation from far-ultraviolet radiation is very rapid ($\sim 1-2 \times 10^5$ yr) 
once accretion rates in the disk drop below a few $\times 10^{-10} \, M_{\odot} \, {\rm yr}^{-1}$,
and even when they are $1 \times 10^{-9} \, M_{\odot} \, {\rm yr}^{-1}$, disk lifetimes are only $\sim 0.4$ Myr
(Alexander et al.\ 2014).  We note that mass accretion rates drop below $1 \times 10^{-9} \, M_{\odot} \, {\rm yr}^{-1}$
at about 3 Myr, so a prediction of our model would appear to be that inner disk gas must vanish by 3.5 Myr at the latest.
The absence of paleofield recorded by angrite basalts at 3.8 Myr suggests there was no gas at that point in the inner disk
(Wang et al.\ 2017).
Similarly, the lack of a paleomagnetic field in the achondrite NWA 7325, almost certainly from the inner disk,
has been interpreted by Weiss et al.\ (2017) as meaning the nebula had dissipated in its locale by 4 Myr. 
Another trend is that accretion of parent bodies seems to have taken longer in the outer disk.
This is highlighted by the fact that magmatic irons formed at $t > 0.9$ Myr, as opposed to $< 0.4$ Myr in the inner disk,
according to Kruijer et al.\ (2017).
And even though some iron meteorites formed in the outer disk, Figure~\ref{fig:bigkahuna} highlights that 
there is an apparent dearth of rocky achondrites formed in the outer disk.

%
%
%

\section{Discussion}

\subsection{Summary}

We have described a 1-D numerical hydrodynamics model we have developed to calculate the distribution of refractory 
elements and CAIs in an evolving protoplanetary disk.
We self-consistently calculate the evolution of the surface density and temperature, assuming a radial profile
of $\alpha$ varying from $5 \times 10^{-4}$ inside 1 AU, to $1 \times 10^{-5}$ outside 10 AU, that is possibly 
consistent with the VSI or other hydrodynamic instabilities. 
The interior region where $T > 1400$ K, the ``CAI Factory", extends initially out to 1.7 AU, 
shrinking over many $\times 10^5$ yr as the disk evolves and loses mass.
A set fraction of the primoridal particles passing through this region are assumed to be thermally processed
and $\approx 8\%$ of that material converted into CAIs.
We track CAIs with a single radius $2500 \, \mu{\rm m}$, following their radial transport due to
advection, drift, and diffusion, and including the effects of meridional flow at the midplane.
We include the formation of Jupiter at 3 AU at 0.6 Myr, and mimic the effects seen in more sophisticated 
numerical simulations of gap opening and accretion.
In our simulations, Jupiter creates at 3.5 - 3.75 AU a persistent local maximum in surface density and 
pressure---the ``pressure bump"---in which particles can be trapped. 
We find that CAIs created in the inner disk are trapped in this region with high efficiency throughout the entire
chondrite formation era, in contrast to the 2 - 3 AU region, from which CAIs spiral in to the Sun.
This resolves the CAI storage problem not just qualitatively, but in quantitative detail.
In the 3.6 - 3.9 AU region we calculate CAI abundances $\approx 1-4$ wt\%, and refractory enrichments 
$1.02 - 1.35 \times$ CI. We are able to identify a specific time and place for each carbonaceous chondrite
type to form in this region.
In the 2 - 3 AU region we calculate vanishing CAI abundances ($< 0.01$ wt\%) and refractory abundances $0.89 - 0.90 \times$ CI
We are able to identify a specific time and place for each ordinary and enstatite chondrite type to form in 
this region, as well as the five achondrites for which we could find sufficient information.
With few exceptions (the model does not explain why R chondrites are less depleted in refractories than other 
chondrites, and it does not fully explain the non-zero abundance of small CAIs in the inner disk), 
the model is consistent with all known data about time and place of formation, and refractory and CAI abundances, for 
all of the chondrite types and five types of achondrites we consider.

Our model builds on previous models that have attempted to explain the variations of refractory lithophile abundances 
and CAI abundances in chondrites (Larimers \& Anders 1970; Wasson 1978; Wood 2005; Rubin 2011; Jacquet et al.\ 2012). 
We especially build on the work of Cuzzi et al.\ (2003), who coined the term the ``CAI Factory", and demonstrated how
CAIs of different sizes could be produced in the inner solar system and distributed throughout an evolving disk.
The importance of meridional transport was pointed out by Ciesla (2010), who modeled the radial transport of CAIs in a disk.
Kuffermeier et al.\ (2017) have recently examined formation of CAIs in a solar nebula collapsing from a molecular cloud,
reminiscent of models by Yang \& Ciesla (2012). 
Our work is most similar to that the model of Yang \& Ciesla (2012), who calculated the production and distribution
of CAIs in an evolving disk, similar to our treatment, but beginning with collapse from a molecular cloud.
None of these works, including that of Yang \& Ciesla (2012), included the effects of Jupiter on the structure and
dynamics of the disk, and so did not identify Jupiter as the solution to the CAI storage problem.
While Yang \& Ciesla (2012) found that the fraction of solid mass that is CAIs was $\approx 1 - 3$\% in the 2 - 5 AU 
region, comparable to the CAI abundances in CK and CV chondrites, 
they did not explain why other chondrites---ECs, OCs, and especially CIs---would be devoid of CAIs.
Scott et al.\ (2017) recently pointed out that CAIs may be trapped in the pressure maximum beyond Jupiter, but
they have not made detailed predictions of the CAI or refractory abundances in any chondrite class, as our 
comprehensive, detailed model has done. 

\subsection{Robustness of the Model}

The model we present here fixes a number of parameters, and is fine-tuned to produce parent bodies in the pressure bump
beyond Jupiter consistent with CCs, with CAI abundances $> 3$ wt\%, and parent bodies in the 2 - 3 AU region consistent
with OCs and ECs, with CAI abundances $< 0.1$ wt\%. 
We have not conducted an extensive parameter study but can discuss the sensitivity of the model to various parameters.

The first set of parameters pertain to the disk itself. 
The disk mass was set at $M_{\rm disk} = 0.089 \, M_{\odot}$, 7 times the minimum-mass solar nebula (Weidenschilling 1977b),
but plausible; our results are not very sensitive to this quantity.
The disk was initialized using the self-similar profile of Hartmann et al.\ (1998), using a compact parameter $R_1 = 1$ AU,
as done by Yang \& Ciesla (2012). 
Our results are sensitive to this parameter: increasing $R_1$ even by a factor of 2 leads to much lower surface densities
and temperatures in the inner disk, reducing the size of the CAI Factory and the abundances of CAIs produced.
The opacity of the disk was set to a uniform value $\kappa = 5 \, {\rm cm}^{2} \, {\rm g}^{-1}$, in the middle of 
the range of values calculated by Semenov et al.\ (2003).
Our results tend to depend on this parameter in a complicated fashion: increasing $\kappa$ can increase temperatures, but
that can lead to faster accretion, reducing the surface density, ultimately decreasing the temperatures at later times.
Our results rely on opacity being roughly within a factor of 2 of our nominal value. 

Other parameters refer to the time and place of Jupiter's formation.
For reasons discussed in \S 4.3, to match meteoritic constraints, Jupiter must form and open a gap before 0.7 Myr, and it
must either form at or quickly migrate to a small range of radii, between 2.9 and 3.2 AU. 
Finally, one of the most important parameters we considered was the profile of turbulence, $\alpha(r)$.
Increasing $\alpha$ in the outer disk by factors of a few mixed the outer disk with refractory-depleted material to an unacceptable
degree, which would demand CI chondrites differ from the Sun by $> 25\%$. 
It would also deplete the region beyond Jupiter of gas too quickly.
We therefore require $\alpha_{\rm outer} = 1 \times 10^{-5}$ in the outer disk.
Retaining the same value $\alpha_{\rm inner} = 1 \times 10^{-5}$ in the inner disk leads to the surface density decreasing too 
slowly, increasing temperatures and CAI abundances in the 2 -3 AU region, and yields CAI abundances in the pressure bump that are
 too low, by factors of 2 or more.
As discussed in \S 4.3, only $\alpha_{\rm inner} = 1 - 5 \times 10^{-4}$ are consistent with the meteoritic constraints. 

Other assumptions pertain to the production and transport of CAIs.
One of the biggest assumptions in our model is that a fixed fraction $\approx 30\%$ of the potentially CAI-forming refactory material 
in the starting composition (8\% of the initial, near-CI composition) is converted into large CAIs, while the remaining 70\% is in 
a form that would be measured as refractory material but which is never converted into CAIs. 
This assumption is ad hoc and fine-tuned to explain the refractory abundances of OCs and ECs.
Another big assumption of the model is that CAIs are all the same radius $2500 \, \mu{\rm m}$.
This is a decent match to the size of particle carrying most the mass of CAIs in CK/CV chondrites, but 
it does not match the smaller sizes of CAIs in OCs or CO chondrites; an additional mechanism is required
to explain the ubiquity of a small mass fraction of very small CAIs.
Our results are significantly changed if we used larger or smaller CAIs. Factor of 2 variations in CAI
abundance in CCs would arise from changing the CAI radius by a comparable factor. 
Finally, we have assumed that meridional flow is present.
For the hydrodynamic instabilities we infer to be transporting angular momentum, this is justified; but if it were absent,
our CAI abundances would decrease by a factor of 3.
These are the parameters that were varied in the model, and the ways in which they were optimized.

\subsection{Model Predictions} 

Despite this fine-tuning, the model makes a number of predictions about meteorites that can be used to test it.

{\it Meteoritical Predictions}:
Other than adjusting the CAI abundance in the pressure bump to be $> 3$ wt\%, and fixing the refractory and CAI abundances so that
OCs would match meteoritic constraints, the model was not adjusted to match any other specific meteorite types.
It is a success of the model that times and places could be assigned to all other chondrites that were consistent with their 
refractory and CAI abundances and the estimated time of accretion, as well as the location of the parent body in a few instances.
For only a few objects, perhaps only the H chondrite parent body if it formed where 6 Hebe orbits, are the refractory abundance, 
the CAI abundance, the orbital distance of the parent body and the time of accretion all known; for all other meteorites, the model
predicts one of the missing parameters.
Our predictions for the HEDs are consistent with formation at 2.36 AU where 4 Vesta orbits, our predictions for H chondrites are 
consistent with formation at 2.43 AU where 6 Hebe orbits, and our predictions for LL chondrites are consistent 
with formation at 2.20 AU where 8 Flora orbits.
We are encouraged that we find the same ordering of OC parent bodies in the asteroid belt (moving out from the Sun, 
LL, then H then L) as Binzel et al.\ (2016), based on spectroscopy and tracing the orbits of near-Earth objects.
Our results suggest that the acapulcoite-lodranite parent body probably formed much closer to the Sun, consistent with 
the conclusion of Lucas (2017), based on spectroscopic surveys of asteroids, that this parent body resides in the inner belt. 
%
%
%
%

We also predict the temperature in the nebula at the time each parent body accreted, which may be testable using compositional information, 
as we have started to do for water or carbon content.
Better predictions of water content could be compared to meteoritic data of water-rock ratios and oxidation states of meteorite parent bodies.
And we also predict the density and surface density and even degree of turbulence in the nebula when each chondrite accreted, which 
we used to test the hypothesis of aerodynamic sorting of chondrules by turbulence.
We find strong support for models in which turbulence sorts chondrules into aggregates, which are then concentrated by streaming
instability into chondrite parent bodies. 
For achondrites, we also make quantitative predictions of the abundances of CAIs they accreted in their starting compositions, 
which may be testable using stable isotopes.

{\it Astronomical Predictions}:
Besides the cosmochemical predictions, our model makes a number of testable predictions about the disk itself.
These can be compared broadly to extant protoplanetary disks, although of course there need not be a currently forming disk that exactly
matches the solar system.
Building on the results of Kruijer et al.\ (2017), we likewise predict rapid growth of Jupiter in $< 1$ Myr, but go further to argue that
Jupiter's core probably reached $\sim 30 \, M_{\oplus}$ by 0.6 Myr, and rapidly accreted gas to reach $\sim 100 \, M_{\oplus}$ 
by about 1.2 Myr. 
Our model also makes testable predictions of the concentration of solids in pressure maxima and of the surface density profile, both 
observable using high-resolution millimeter interferometry observations by the {\it Atacama Large Millimeter Array} (ALMA).
For example, Zhang et al.\ (2017) have inferred a flatter surface density profile $\Sigma(r) \propto r^{-0.9}$ in the 5 - 20 AU region
of the TW Hydrae disk than had been inferred for other disks.
Our model predicts a small amount of turbulence, $\alpha \sim 1 - 5 \times 10^{-4}$, in the inner disk, with implications for 
the measureable mass accretion rate of the disk onto the star. 
Lower values of $\alpha$ would not heat a large region of gas to make sufficient CAIs or transport them out of the CAI Factory, while
higher values of $\alpha$ would deplete the inner disk too rapidly for chondrites to form. 
Our model also robustly predicts a degree of turbulence, $\alpha \sim 10^{-5}$, in the outer disk, or else CI chondrites would be 
too depleted in refractories ($> 10\%$) relative to the starting composition and the Sun. 
This is also testable using ALMA observations, and it is intriguing that low values of $\alpha < 10^{-3}$ have been inferred using
ALMA-measured linewidths in the HD163296 disk (Flaherty et al.\ 2015).

\subsection{Implications} 

{\it Angular Momentum Transport in Disks}: 
Assuming future cosmochemical and astronomical observations lend support to our model, the implications would be profound.
From the astronomical perspective, the strongest implications would be for the degree of turbulence and the constraints on the 
nature of angular momentum transport. 
The question of the origin of angular momentum transport has long been hotly debated, and has seen several recent advances.
While the magnetorotational instability (MRI) was considered for a long time to be the most likely mechanism for transporting 
mass and angular momentum in disks, recent work has shown that non-ideal effects suppress the MRI more than had been anticipated,
and that other magnetic effects, such as magnetically-driven disk winds, are more likely to transport angular momentum than the
MRI (Bai 2016; Bai et al.\ 2016). 
At the same time, one of the greatest recent advances has been the discovery that purely hydroydnamic turbulence is capable of 
transporting angular momentum in disks, via instabilities such as convective overstability (Klahr \& Hubbard 2014; Lyra 2014) 
and the vertical shear instability, or VSI (Nelson et al.\ 2013; Umurhan et al.\ 2016; Richard et al.\ 2016). 
In the outer portions of protoplanetary disks, where magnetohydrodynamic (MHD) instabilities are suppressed, it is likely that 
hydrodynamic instabilities, in particular the VSI, should dominate the turbulence and transport, with values of 
$\alpha \sim 10^{-5} - 10^{-4}$. 

Our results strongly support a low degree of turbulence in the outer disk, with $\alpha \sim 10^{-5}$. 
To have CI chondrites form in the disk at $> 3$ Myr and not mix substantially ($> 10\%$) with refactory-depleted material in the inner disk,
the mixing timescale $t_{\rm mix}$ must exceed 30 Myr, which places tight constraints on $\alpha$:
$t_{\rm mix} = r^2 / (3 \alpha H^2 \Omega)$ 
implies $\alpha < 10^{-5}$. 
This is far lower than the level of turbulence predicted by the magnetorotational instability (MRI), typically thought to be 
$\alpha > 10^{-4}$ at least, and usually $\alpha \sim 10^{-2}$ in the outer portions of disks (Simon et al.\ 2017). 
It is much more characteristic of the values associated with purely hydrodynamic 
instabilities such as the vertical shear instability (VSI), with $\alpha \approx 10^{-5} - 10^{-4}$ (Richard et al.\ 2016). 
Moreover, our inference of slightly higher values of $\alpha$ inside of 10 AU strongly suggests an additional source of turbulence
and transport in the inner disk, most likely associated with magnetic disk winds. 

We find a good match between our inferred $\alpha(r)$ profile, falling almost as $r^{-1}$ between 1 and 10 AU, and the 
$\alpha(r)$ profile derived by Suzuki et al.\ (2016) in this region in their disk modeled as MRI-active in layers away from 
the midplane, capable of launching disk winds.  
Suzuki et al.\ (2016) calculate two components of the stress tensor and two different values of the turbulent viscosity 
parameter $\alpha$: the mass-weighted value of $\alpha_{r \phi}$, relevant to the transport of angular momentum; and the 
mass-weighted value of $\alpha_{\phi z}$, arising from torques from the disk wind, relevant to loss of angular momentum 
carried away by the disk wind.
Both $\alpha_{r \phi}$ and $\alpha_{\phi z}$ independently influence the temperature structure of the disk and also contribute 
to mass accretion onto the star, in ways very similar to the roles of the traditional $\alpha$ (which is most closely related 
to $\alpha_{r\phi}$). 
In the simulations of Suzuki et al.\ (2016), they use as free parameters $\alpha_{r \phi}$, $\alpha_{\phi z}$, and a parameter 
$C_{w,0}$ related to the mass flux due to the disk wind.  
These are varied to test the effects of: i) strong or weak MRI turbulence in the inner disk capable of launching winds; 
ii) strength of the disk wind; and iii) a $\Sigma$-dependent disk wind torque.
We favor their simulation with the MRI-inactive case that incorporates a $\Sigma$-dependent torque associated with a 
vertical magnetic field that remains constant with time, which results in an $\alpha_{\phi z}$ profile that increases with 
decreasing $\Sigma$ in the inner disk (see their Figure 7).
In this simulation, the turbulent viscosity $\alpha_{r \phi}$ is assumed to be due to hydrodynamic instabilities, and constant 
at a level $8 \times$ $10^{-5}$, equivalent to $\alpha = 4 \times 10^{-5}$ using the relation given by the authors, 
(i.e., $\alpha = (\sqrt(2)/3) \alpha_{r \phi}$) for conversion to the usual $\alpha$ used in disk models (Shakura \& Sunyaev 1973). 
This is similar to the value $\alpha = 1 \times 10^{-5}$ we adopted beyond 10 AU.
Inside 10 AU, $\alpha$ is dominated by $\alpha_{\phi z}$.
From a simple analysis of Equations 10 and 17 of Suzuki et al.\ (2016), by simplifying the physical quantities inside the 
derivative of the advection term containing $\alpha_{r \phi}$ in their Equation 10, we find that the contribution of 
$\alpha_{\phi z}$ towards both mass transport and thermal energy is much greater than $\alpha_{r \phi}$. 
Our assumed profiles of $\alpha$ in the inner disk resemble the $\alpha_{\phi z}$ profiles of Suzuki et al.\ (2016) and are 
bracketed by the strong- and weak-wind limiting cases in Figure 7 of Suzuki et al.\ (2016).
The fact that our $\alpha(r)$ profile---necessary for production of abundant CAIs and outward transport past Jupiter---so 
closely matches the profile of Suzuki et al.\ (2016) strongly supports the hypothesis that turbulent viscosity is mediated by 
hydrodynamic instabilities (e.g., VSI) plus transport by disk winds inside of 10 AU. 

{\it Accretion and Migration of Parent Bodies}:
Another important implication of our model is that it is possible to assign a time and place to the formation of 
most chondrite and achondrite parent bodies. 
This implies that the material comprising individual meteorites accreted rapidly, in 
a time $< t_{\rm mix} \sim (1/3) (0.1 \, {\rm AU})^2 \, / (\alpha H^2 \Omega) \sim 10^4$ yr. 
Despite the fact that radial diffusion and radial transport of CAIs are central features to our model, it appears
that during the time a parent body accretes, local regions maintain their individual character.
It is possible, though, that some parent bodies could have accreted over longer timescales, in which case they would
grow from material of a changing refractory composition or CAI abundance.
The CK/CV parent body, presuming it is a single parent body, seems to have grown over a timescale almost $\sim 1$ Myr.
Because the pressure maximum beyond Jupiter is a natural location for solids to accrete, our model supports (but does
not require) the idea
of a single parent body of the Eagle Station pallasites and CK and CV chondrite, as suggested by Elkins-Tanton et al.\ 
(2011), based on paleomagnetism studies. 

It also seems to be the case that migration after formation of parent bodies was limited, as the HEDs seem to match
conditions where 4 Vesta is found today, the H chondrites seem to match conditions where 6 Hebe is found today, 
and the LL chondrites seem to match conditions where 8 Flora is found today.
In general, ECs are best fit by conditions $< 2.1$ AU where E-type asteroids are found, and OCs match conditions in the 
2.2 - 2.6 AU region where S-type asteroids are found.
Clearly Jupiter must have migrated from 3 AU to its current location at 5.2 AU, and during the process some S-type
asteroids and many CC parent bodies must have been scattered into the outer asteroid belt.
Even so, many asteroids do not seem to have migrated, and for those that did, the extent of migration is limited to $< 1$ AU. 

{\it Match of CI Chondrites to the Sun}: 
A very robust, important, and testable prediction of our model is that CI chondrites do {\it not} exactly match solar abundances, 
and should be depleted in refractory elements that condense at temperatures at about 1400 K or greater and are converted into CAI material.
The average value of $({\rm X}/{\rm Mg})_{\rm CI} / ({\rm X}/{\rm Mg})_{\odot}$, we predict,
should be about 0.878 (where X denotes a refractory element). 
This is just at the limits of uncertainties of meteoritic and solar photosphere abundance measurements, 
$\approx 10$\% (Lodders et al.\ 2009; Asplund et al.\ 2009; Palme et al.\ 2014), but we find evidence suggestive of this. 

Lodders et al.\ (2009) reported abundances in the Sun and in CI chondrites (normalized to Si)
 for which uncertainties were $< 0.1$ dex, or about 25\%.
We divide the elements into four groups using their chemical affinities and 50\% condensation temperatures $T_{\rm cond}$,
as listed by Lodders (2003):
refractory lithophiles, with $T_{\rm cond} > 1336$ K, the condensation temperature of Mg 
(V, Be, Ba, Sr, Ca, Nb, Ti, Al, Y, Sc, Hf, Zr, and the rare earth elements); 
moderately volatile lithophiles, with $600 \, {\rm K} < T_{\rm cond} < 1336 \, {\rm K}$
(Cd, Zn, F, Rb, B, Cl, Na, K, Mn, and Si);
refractory siderophiles, with $T_{\rm cond} \geq 1328$ K, the condensation temperature of Fe
(Fe, Co, Ni, Rh, Pt, Os, Ir, Ru, Mo, and W);
and moderately volatile siderophiles, with $600 \, {\rm K} < T_{\rm cond} < 1328 \, {\rm K}$
(S, Sn, Pb, Ge, Ag, Ga, Sb, Cu, P, Cr, and Pd).
We take the mean value of the ratio of CI abundance relative to solar photospheric abundance, weighted by the uncertainties:
$\bar{x} = (\sum_{i} x_i / \sigma_i^2) / (\sum_i 1 / \sigma_i^2)$ and $\sigma_{x} = (\sum_i 1 / \sigma_i^2)^{-1/2}$.

Among the refractory lithophiles, we find a weighted average $0.965 \pm 0.040$, and among the refractory siderophiles
we find an average of $0.933 \pm 0.068$. These are consistent with 3.5 - 6.7\% depletions in refractories in CIs relative
to the Sun, albeit only at the $\sim 1\sigma$ level.
In contrast, we find for the moderately volatile lithophiles a weighted average $1.042 \pm 0.063$, and for the 
moderately volatile siderophiles an average $1.068 \pm 0.038$.
CIs are not depleted in moderately volatile elements, instead appearing to be enriched by about 4.2 - 6.8\%. 
Although they are not statistically significant, these results suggest a depletion in CIs relative to the Sun of 
refractory elements that condense at higher temperatures than Mg, around the 1400 K threshold we assumed.
Unfortunately, the data are not of sufficient precision to directly find the average value of 
$({\rm X}/{\rm Mg})_{\rm CI} / ({\rm X}/{\rm Mg})_{\odot}$, where X is a refractory lithophile element,
because the abundance of Mg itself is too uncertain.
Lodders et al.\ (2009) compute a Mg abundance in CI chondrites that is 0.98 times the abundance in the Sun,
give or take a factor of 1.18. Therefore the refractory abundance of CIs could be anywhere from 0.84  to 1.16 $\times$ CI.
The ratio of refractory lithophile abundances to moderately volatile lithophile abundances suggests this ratio would
be more like 0.93, though,
and the data are certainly consistent with a depletion of refractory elements, both lithophile and siderophile,
at the $\approx 12\%$ level. 
Further refinements of solar photospheric abundances, and cosmochemical measurements of whole-rock elemental
abundances in CI chondrites, at the $< 5 \%$ precision level, could test this prediction. 

{\it Missing Meteorites?}:
By plotting the locations and times of formation of meteorite parent bodies in Figure 10, it is possible to identify 
regions and times from which we do not have meteoritic samples, and guess as to the significance of this lack.
For example, no bodies appear to form beyond Jupiter until the CC irons that form at $\sim 1$ Myr (Kruijer et al. 2017).
This could be because formation timescales are longer in the outer disk, or possibly because Jupiter migrated from 
4 to 3 AU at around $\sim 0.5$ Myr.
We have not plotted any bodies forming beyond 2.4 AU between 1 and 2 Myr, but we attribute this to the fact that we found very few 
examples of achondrites for which we had simultaneous information on time of accretion and pre-melt composition.
It probably is significant that no meteorite parent bodies in the inner disk seem to have formed after 2.2 Myr.
There are no major chondrite groups we have not plotted, so very likely the gas in the inner disk dissipated soon after this time,
even as gas persisted beyond Jupiter until at least 4 Myr.
This inferred structure for the solar nebula is the same as the astronomically observed transition disk stage of 
protoplanetary disk evolution, which happens at a median age of 2 to 3 Myr (Williams \& Cieza 2011).
It should be noted that meteorites do not uniformly sample the asteroid belt, and proximity to a dynamical resonance is
important for delivering asteroid materials to Earth.
For example, the most common meteorite is ordinary chondrites associated with S-type asteroids, but these parent bodies
may be relatively rare in the asteroid belt (Meibom \& Clark 1999). 

Because the model we present is so comprehensive, it makes many other predictions and has many other implications
beyond the ones listed here. 

\section{Areas of Future Research} 

Here we suggest improvements that could be made to the model, and directions for future investigations. 

\subsection{Timing of CAI Formation}

Potentially our model could be used to predict the temporal distribution of CAI formation.
The regions that have $T > 1400$ K and can form CAIs shrink with time, but some CAIs could be formed several
$\times 10^5$ yr after $t = 0$. 
The prediction of a tail of CAI production at late times is supported by Al-Mg systematics of CAIs that show 
an extended time of formation, $\sim 0.2$ Myr (Kita et al.\ 2013),
in some cases up to $\sim 0.7$ Myr (MacPherson et al.\ 2012),
and by suggestive evidence from oxygen isotopes  that some CAIs
continued to interact a (presumably hot) ${}^{16}{\rm O}$-rich reservoir for 2 - 3 Myr (Ushikubo et al.\ 2017).
It would be good to test the model against these constraints. 
Unfortunately, a limitation of our model is that we initialize the disk with a commonly used self-similar profile 
instead of following the formation of the disk from the molecular cloud stage
as Yang \& Ciesla (2012) and Kuffermeier et al.\ (2017) did. 
As a result, most ($> 90\%$) of the CAIs produced in our model are produced at $t=0$, limiting the 
power of the model to predict the total mass of CAIs and the timing of CAI formation. 
Nevertheless, we note that formation of most CAIs in an initial pulse is consistent with dating of CAIs by Al-Mg systematics, 
which shows they mostly formed in a very short interval, perhaps as short as 0.02 Myr (Thrane et al.\ 2006), and that our 
initial conditions, especially the compact disk, resemble the early surface density profiles of Yang \& Ciesla (2012).

\subsection{CAIs in Comets}

One of the most surprising findings of the {\it Stardust} comet sample return mission was the presence of 
fragments of chondrules, AOAs, and CAIs in the comet Wild 2, that must have formed in the inner solar system
(Simon et al.\ 2008; Bridges et al.\ 2012; Joswiak et al.\ 2017).
It is estimated that 0.5 vol\% of the sample return material is CAIs (Joswiak et al.\ 2017). 
This is comparable to the abundance in CM and CR chondrites formed in our model at about 4 AU, despite
the fact that comets are presumed to form much further out in the disk, at tens of AU.
This is corroborated by other analyses that suggest Wild 2 has a very high fraction (50 - 65\%) of 
chondritic material (Westphal et al.\ 2009). 
It is also perplexing that CI chondrites, which resemble cometary material (Gounelle et al.\ 2006),
contain nowhere near 0.5 vol\% CAIs.
Indeed, our model predicts that the large ($2500 \, \mu{\rm m}$ radius) CAIs we track, 
have extremely low abundances beyond about 15 AU.
However, given that the net flow of gas at 4 AU is out of the pressure bump and radially outward, 
if CAIs did fragment in the pressure maximum where carbonaceous chondrites formed, then our model predicts that they
would be carried outward very effectively.
At 1 Myr, the mean radial flow of the gas is outward beyond 3.5 AU, and at 3 Myr it is outward beyond 4.7 AU, so 
whatever material makes it past $\approx 3 - 5$ AU will be swept outward to the comet-forming region.
In our simulations the outward velocity in the outer disk is typically $\sim 4 \, {\rm AU} \, {\rm Myr}^{-1}$, 
so material would be swept out to the comet-forming region at 15 - 30 AU about 2.5 to 7 Myr later.
It is therefore a very natural outcome of the model that comets should contain almost as much CAI material as CM and CR 
chondrites, provided the outer disk survives for 5-10 Myr, even as the inner disk dissipates.
But if CI chondrites formed beyond 15 AU by 3 Myr, there would not have been time for much of the CAI and other fragments 
to reach that region.
The growth of Saturn and the creation of pressure maxima beyond its orbit may have complicated the picture.
We consider the transport of refractory fragments to the comet-forming region to be an important area of future 
research. 

\subsection{Migration, Growth, and Gap Opening by Jupiter}

Our inclusion of Jupiter could be improved by considering dynamical effects.
In our model, Jupiter instantaneously forms as a $30 \, M_{\oplus}$ core at 3.0 AU, at 0.6 Myr.
The early time is justified by the finding of an isotopic dichotomy in the solar nebula and the use of Hf-W dating
(Kruijer et al.\ 2017), and the fast growth of a $30 \, M_{\oplus}$ core is justified by the 
rapid growth timescales of Jupiter in pebble accretion calculations (Kretke \& Levison 2014).
The location of Jupiter's formation is harder to justify, and the details of how it might have gotten there, 
and how it might have migrated to 5.2 AU, demand better modeling.
 
We speculate that Jupiter probably formed at or just inside the snow line, which at times 0.3 to 0.6 Myr is migrating inward,
from 4.7 to 4.1 AU. The mechanism of Ida \& Guillot (2016), in which ice and dust aggregates radially drift inward through
the snow line, increasing the solids-to-gas ratio inside the snow line, seems a promising mechanism for forming Jupiter 
just past 4 AU. 
Once Jupiter's core starts to form, like other cores formed by pebble accretion it would have been of the right size to
undergo rapid type I inward migration.
Bitsch et al.\ (2015) have shown that cores will migrate inward significantly before they reach sizes large enough to open
a gap, at which point they transition to the slower type II migration (Lin \& Papaloizou 1986).
From the simulations of Bitsch et al.\ (2015), migration inward to 3 AU is plausible.
After that, Jupiter's migration follows that of the disk. A curious feature of our simulations is that Jupiter is near the
transition radius of the disk: Figure 6 makes clear that the mass flow of gas interior to Jupiter is predominantly inward,
while exterior to Jupiter it is predominantly outward. 
During type II migration, Jupiter should follow the net flow of the gas (Lin \& Papaloizou 1986), but it is not clear that 
it would migrate inward or outward.  
Of course, we fixed the location of Jupiter at 3.0 AU in our simulations, but it would be worthwhile to look at type II
planetary migration for planets at a disk's transition radius. 
This may explain why Jupiter does not seem to have migrated much between 0.6 and 4 Myr. 
Possibly Jupiter would not migrate until Saturn migrates inward and the two planets enter a resonance, as in the 
Grand Tack model (Walsh et al.\ 2011). 
Eventually Jupiter must migrate by some mechanism outward to 5.2 AU, and our model suggests this occured sometime after the 
CR chondrites formed at 4.0 Myr, but before Jupiter's migration triggered the formation of the CB/CH/Isheyevo chondrites at 4.5 Myr
(Johnson et al.\ 2015). 
During Jupiter's outward migration, we expect it to scatter OCs into the outer asteroid belt, and especially
to scatter CCs formed outside it into the asteroid belt, as predicted by Walsh et al.\ (2011) and Raymond \& Izodoro (2017). 
All of these effects should be included in a self-consistent model in future work. 

Another aspect of our model that deserves better modeling is the growth rate of Jupiter and the way it opens a gap.
To incorporate these aspects, we have parameterized many effects that should be tested. 
Our modeling of Jupiter's growth by letting it accrete a fraction of the gas within its
Hill sphere each timestep is physically plausible and is modeled in a similar way by Kley (1999). 
D'Angelo \& Lubow (2008) and Machida et al.\ (2010) showed that the accretion rate is often controlled 
by the amount of gas in the Hill radius, and a growth timescale $\sim 10^5$ yr was found by Machida et al.\ (2010). 
Still, the opening of a gap may slow the growth of Jupiter (Bryden et al.\ 1999), and the fast-then-slow
growth of Jupiter does not match the growth seen in other simulations (e.g., Pollack et al.\ 1996).
Better modeling of Jupiter's growth is required, as well as its ability to open a gap. 
We increase $\alpha$ to $\sim 10^{-2}$ in the vicinity (few Hill radii) of Jupiter, as in
simulations by Lyra et al.\ (2016). This has the effect of creating a gap of the right width 
(few $\times 0.1$ AU), and creating a pressure bump with parameters similar to those seen in 
Lambrechts et al.\ (2014), but our treatment is a crude approximation of more sophisticated 
numerical simulations.


\subsection{Distribution of Water}

One of the obvious next steps for our modeling, which we've already alluded to, is to calculate the distribution
of water in the protoplanetary disk.
Our temperature modeling calculates the location of the snow line, indicating regions where water can condense as ice
and where it cannot. 
On this basis we predicted that the parent bodies of the ureilites, HED meteorites, aubrites and ECs should not have 
accreted ice, but other chondrites may have.
Our model does not currently predict the abundance of water ice in different regions, but such a calculation could be 
used to test the model further, and to test the ``fossil snow line" hypothesis of Morbidelli et al.\ (2016), which
states that the inner solar system was in many regions cold enough for ice to condense, but depleted in water content
anyway.
Such modeling also would place a constraint on whether the lack of water in CV and CO chondrites is due to thermal alteration 
on the parent body, or if the pressure bump region was devoid of water, e.g., by heating by spiral shocks.
Lyra et al.\ (2016; their Figure 3) find that a 5 Jupiter-mass planet at 5.2 AU can drive spiral shocks with speeds 
up to Mach 2, about $1.5 \, {\rm km} \, {\rm s}^{-1}$.
These shocks are too slow to melt chondrules, but they would be sufficient to heat the gas by hundreds of K.
It is possible the repeated passages through these spiral shocks could drive water out of this region, allowing it 
to be cold-trapped outside of the pressure bump. 
A calculation of water distribution would also allow predictions about the proportions of type I (FeO-poor) vs. type II
(FeO-rich) chondrules in different chondrites, as well as the distribution of mass-independent fractionation oxygen 
isotope anomalies, believed to be carried by water (Yurimoto \& Kuramoto 2004; Lyons \& Young 2005; Sakamoto et al.\ 2007).

\section{Summary}

We have presented a comprehensive model of disk evolution that resolves the CAI storage problem and connects the
time and place of a meteorite parent body's formation with its abundances of refractory lithophile elements and CAIs.
The model is based on the idea that Jupiter's $\sim 30 \, M_{\oplus}$ core accreted early and orbited at about 3 AU,
opening a gap in the disk. This assumption is strongly motivated by the finding of an isotopic dichotomy in the solar 
nebula (Warren 2011; Kruijer et al.\ 2017). 
A pressure maximum would have existed outside the gap opened by Jupiter, into which CAIs would have been concentrated.
We conclude that carbonaceous chondrites accreted in this region.
Interior to Jupiter, CAIs would have been removed by aerodynamic drag. 
We conclude this is where ordinary and enstatite chondrites formed.
Similar ideas also have recently been advanced by Scott et al.\ (2018) and Melosh et al.\ (2018), but we have developed 
a quantitative model that allows us to predict the location of a chondrite's formation given constraints on the time of 
its formation, and its refractory and CAI abundances.
We have applied this model to 11 chondrite types, and 5 achondrite types for which the bulk composition of the parent body
could be ascertained. 
In almost every case we find an excellent match to other information about where the meteorites formed such as 
spectral matches to asteroids and water content. 
We find complete consistency between where H chondrites formed and the asteroid 6 Hebe, between HED chondrites and 4 Vesta,
and between LL chondrites and 8 Flora.
We also used the model to predict the physical conditions where each chondrite parent body formed and the size of particle
that would be optimally concentrated by turbulence. We find a very good match between our model predictions and the
mean chondrule diameters in various chondrites. This suggests that turbulence concentrated chondrules into aggregates
of particles that were then swept up by streaming instability into planetesimals and chondrite parent bodies.
This justifies the implicit assumption of our model that meteorites are snapshots in time of the solar nebula.  

By constraining the disk properties by the need to match the refractory and CAI abundances of meteorites, we have
gained insights into disk evolution and planet formation.
The disk appears to have evolved via hydrodynamic instabilities and magnetic disk winds rather than the MRI, characterized
by low values of $\alpha$ that decrease with heliocentric distance in the 1 - 10 AU region.
The lack of any chondrites formed interior to Jupiter after about 2.5, even as CR chondrites formed at $\sim$ 4 Myr 
beyond Jupiter, suggests the solar nebula was a transition disk (Williams \& Cieza 2011) between 3 and 4 Myr, 
with a very-early formed Jupiter.
Observations can test whether these aspects are found in extant protoplanetary disks, and whether these effects are 
universal and whether they act as we have modeled them.
We also gained insights into meteorites, especially with the constraint that CI chondrites cannot chemically match the Sun,
and must be depleted in refractory elements by about 12\%. 
Our model makes many other predictions that can be tested. 
We hope our model serves as an example of how to combine meteoritic data with astrophysical modeling.
If the model proves to be robust, then it could be a valuable tool for using information about refractory lithophile 
and  CAI abundances to infer the time and place of a meteorite's formation, compelling the chondrites to confess their secrets.

\acknowledgments

The results reported herein benefited from collaborations and/or information exchange within NASA's 
Nexus for Exoplanet System Science (NExSS) research coordination network sponsored by NASA’s Science Mission Directorate
(Grant NNX15AD53G, PI Steve Desch).

We thank Kath Bermingham, Richard Binzel, Martin Bizzarro, Aaron Boley, Jeff Cuzzi, Emilie Dunham, Katherine Kretke, Sasha Krot, Prajkta Mane, Larry Nittler, Devin Schrader, and Kevin Walsh for helpful conversations. 

We thank an anonymous referee for their very helpful review, which improved this manuscript.

\end{document}